\documentclass[useAMS,usenatbib]{mn2e}

\pdfoutput=1

\usepackage{tabularx}
\usepackage{graphicx}
\usepackage[section]{placeins}
\usepackage{epstopdf}
\usepackage{array}
\usepackage{amsmath}
\usepackage{amssymb}
\usepackage{wrapfig}
\usepackage{upgreek}
\usepackage{color}
\usepackage{longtable}
\usepackage{supertabular}
\usepackage{units}
\usepackage{float}
\usepackage[labelfont=bf,labelsep=period]{caption}
\usepackage{capt-of}

\def \arcsec {$^{\prime\prime}$}
\def \arcmin {$^\prime$}
\def \um {$\upmu$m}

\def \th {$^{{\rm th}}$}
\def \co12 {$^{12}$CO}
\def \co13 {$^{13}$CO}
\def \co {C$^{18}$O}
\def \nh {N$_{2}$H$^{+}$}
\def \sun {$_{\odot}$}

\title[First SCUBA-2 observations of Ophiuchus]{The JCMT Gould Belt Survey: First results from the SCUBA-2 observations of the Ophiuchus molecular cloud and a virial analysis of its prestellar core population}
\author[K. Pattle et al.]{K. Pattle$^{1}$, D. Ward-Thompson$^{1}$, J.M. Kirk$^{1}$,  G.J. White$^{2,3}$,  E. Drabek-Maunder$^{4}$, \newauthor J. Buckle$^{5, 6}$, S.F. Beaulieu$^{7}$, D.S. Berry$^{8}$, H. Broekhoven-Fiene$^{9}$, M.J. Currie$^{8}$, \newauthor M. Fich$^{7}$, J. Hatchell$^{10}$, H. Kirk$^{11}$, T. Jenness$^{8, 12}$, D. Johnstone$^{8, 11, 9}$, \newauthor J.C. Mottram$^{13}$, D. Nutter$^{14}$, J.E. Pineda$^{15, 16, 17}$, C. Quinn$^{14}$, C. Salji$^{5, 6}$, \newauthor S. Tisi$^{7}$, S. Walker-Smith$^{5, 6}$, J. Di Francesco$^{11, 9}$, M.R. Hogerheijde$^{13}$, Ph. Andr\'{e}$^{18}$, \newauthor P. Bastien$^{19}$, D. Bresnahan$^{1}$, H. Butner$^{20}$, M. Chen$^{9}$, A. Chrysostomou$^{21}$, \newauthor S. Coude$^{19}$, C.J. Davis$^{22}$, A. Duarte-Cabral$^{10}$, J. Fiege$^{23}$, P. Friberg$^{8}$,\newauthor R. Friesen$^{24}$, G.A. Fuller$^{16}$, S. Graves$^{5, 6}$, J. Greaves$^{25}$, J. Gregson$^{2, 3}$, M. J. Griffin$^{14}$, \newauthor W. Holland$^{26, 27}$, G. Joncas$^{28}$, L.B.G. Knee$^{11}$, V. K\"{o}nyves$^{18,29}$, S. Mairs$^{9}$, \newauthor K. Marsh$^{14}$, B.C. Matthews$^{11, 9}$, G. Moriarty-Schieven$^{11}$, J. Rawlings$^{30}$, J. Richer$^{5, 6}$, \newauthor D. Robertson$^{31}$, E. Rosolowsky$^{32}$, D. Rumble$^{10}$, S. Sadavoy$^{33}$, L. Spinoglio$^{34}$, \newauthor H. Thomas$^{8}$, N. Tothill$^{35}$, S. Viti$^{30}$, J. Wouterloot$^{8}$, J. Yates$^{30}$, M. Zhu$^{36}$\\
\\
$^{1}$Jeremiah Horrocks Institute, University of Central Lancashire, Preston, Lancashire, PR1 2HE, UK\\
$^{2}$Department of Physical Sciences, The Open University, Milton Keynes MK7 6AA, UK\\
$^{3}$The Rutherford Appleton Laboratory, Chilton, Didcot, OX11 0NL, UK.\\
$^{4}$Imperial College London, Blackett Laboratory, Prince Consort Rd, London SW7 2BB, UK\\
$^{5}$Astrophysics Group, Cavendish Laboratory, J J Thomson Avenue, Cambridge, CB3 0HE\\
$^{6}$Kavli Institute for Cosmology, Institute of Astronomy, University of Cambridge, Madingley Road, Cambridge, CB3 0HA, UK\\
$^{7}$Department of Physics and Astronomy, University of Waterloo, Waterloo, Ontario, N2L 3G1, Canada\\
$^{8}$Joint Astronomy Centre, 660 N. A`oh\={o}k\={u} Place, University Park, Hilo, Hawaii 96720, USA\\
$^{9}$Department of Physics and Astronomy, University of Victoria, Victoria, BC, V8P 1A1, Canada\\
$^{10}$Physics and Astronomy, University of Exeter, Stocker Road, Exeter EX4 4QL, UK\\
$^{11}$NRC Herzberg Astronomy and Astrophysics, 5071 West Saanich Rd, Victoria, BC, V9E 2E7, Canada\\
$^{12}$Department of Astronomy, Cornell University, Ithaca, NY 14853, USA\\
$^{13}$Leiden Observatory, Leiden University, PO Box 9513, 2300 RA Leiden, The Netherlands\\
$^{14}$School of Physics and Astronomy, Cardiff University, The Parade, Cardiff, CF24 3AA, UK\\
$^{15}$European Southern Observatory (ESO), Garching, Germany\\
$^{16}$Jodrell Bank Centre for Astrophysics, Alan Turing Building, School of Physics and Astronomy, University of Manchester,\\ Oxford Road, Manchester, M13 9PL, UK\\
$^{17}$Institute for Astronomy, ETH Zurich, Wolfgang-Pauli-Strasse 27, CH-8093 Zurich, Switzerland\\
$^{18}$Laboratoire AIM CEA/DSM-CNRS-Universit\'{e} Paris Diderot, IRFU/Service d'Astrophysique, CEA Saclay, F-91191 Gif-sur-Yvette, France\\
$^{19}$Universit\'e de Montr\'eal, Centre de Recherche en Astrophysique du Qu\'ebec et d\'epartement de physique, \\ C.P. 6128, succ. centre-ville, Montr\'eal, QC, H3C 3J7, Canada\\
$^{20}$James Madison University, Harrisonburg, Virginia 22807, USA\\
$^{21}$School of Physics, Astronomy \& Mathematics, University of Hertfordshire, College Lane, Hatfield, Herts, AL10 9AB, UK\\
$^{22}$Astrophysics Research Institute, Liverpool John Moores University, Egerton Warf, Birkenhead, CH41 1LD, UK\\
$^{23}$Department of Physics \& Astronomy, University of Manitoba, Winnipeg, Manitoba, R3T 2N2 Canada\\
$^{24}$Dunlap Institute for Astronomy \& Astrophysics, University of Toronto, 50 St. George St., Toronto ON M5S 3H4 Canada\\
$^{25}$Physics \& Astronomy, University of St Andrews, North Haugh, St Andrews, Fife KY16 9SS, UK\\
$^{26}$UK Astronomy Technology Centre, Royal Observatory, Blackford Hill, Edinburgh EH9 3HJ, UK\\
$^{27}$Institute for Astronomy, Royal Observatory, University of Edinburgh, Blackford Hill, Edinburgh EH9 3HJ, UK\\
$^{28}$Centre de recherche en astrophysique du Qu\'ebec et D\'epartement de physique, de g\'enie physique et d'optique,\\ Universit\'e Laval, 1045 avenue de la m\'edecine, Qu\'ebec, G1V 0A6, Canada\\
$^{29}$Institut d'Astrophysique Spatiale, CNRS/Universit\'{e} Paris-Sud 11, F-91405 Orsay, France\\
$^{30}$Department of Physics and Astronomy, UCL, Gower St, London, WC1E 6BT, UK\\
$^{31}$Department of Physics and Astronomy, McMaster University, Hamilton, ON, L8S 4M1, Canada\\
$^{32}$Department of Physics, University of Alberta, Edmonton, AB T6G 2E1, Canada\\
$^{33}$Max Planck Institute for Astronomy, K\"{o}nigstuhl 17, D-69117 Heidelberg, Germany\\
$^{34}$IFSI - INAF, via Fosso del Cavaliere 100, 00133 Roma, Italy\\
$^{35}$University of Western Sydney, Locked Bag 1797, Penrith NSW 2751, Australia\\
$^{36}$National Astronomical Observatory of China, 20A Datun Road, Chaoyang District, Beijing 100012, China}

\begin{document}

\date{}

\pagerange{\pageref{firstpage}--\pageref{lastpage}} \pubyear{2015}

\maketitle

\label{firstpage}

\begin{abstract}
In this paper we present the first observations of the Ophiuchus molecular cloud performed as part of the James Clerk Maxwell Telescope (JCMT) Gould Belt Survey (GBS) with the SCUBA-2 instrument.  We demonstrate methods for combining these data with previous HARP CO, \emph{Herschel}, and IRAM \nh\ observations in order to accurately quantify the properties of the SCUBA-2 sources in Ophiuchus.  We produce a catalogue of all of the sources found by SCUBA-2.  We separate these into protostars and starless cores.  We list all of the starless cores and perform a full virial analysis, including external pressure.  This is the first time that external pressure has been included in this level of detail.  We find that the majority of our cores are either bound or virialised.  Gravitational energy and external pressure are on average of a similar order of magnitude, but with some variation from region to region.  We find that cores in the Oph A region are gravitationally bound prestellar cores, while cores in the Oph C and E regions are pressure-confined.  We determine that \nh\ is a good tracer of the bound material of prestellar cores, although we find some evidence for \nh\ freeze-out at the very highest core densities.  We find that non-thermal linewidths decrease substantially between the gas traced by \co\ and that traced by \nh, indicating the dissipation of turbulence at higher densities.  We find that the critical Bonnor-Ebert stability criterion is not a good indicator of the boundedness of our cores.  We detect the pre-brown dwarf candidate Oph B-11 and find a flux density and mass consistent with previous work.  We discuss regional variations in the nature of the cores and find further support for our previous hypothesis of a global evolutionary gradient across the cloud from southwest to northeast, indicating sequential star formation across the region.
\end{abstract}

\begin{keywords}
stars: formation -- dust, extinction -- ISM: kinematics and dynamics -- ISM: individual objects: L1688 -- ISM: individual objects: L1689 -- submillimetre: ISM
\end{keywords}

\section{Introduction}

The Ophiuchus molecular cloud is a nearby \citep[$139\pm 6$\,pc,][]{mamajek2008}, well-studied (Wilking et al. 2008), site of low-mass star formation \citep{wilking1983}. It consists of two submillimetre-bright central regions, L1688 and L1689, each of which has extensive filamentary streamers \citep[see, e.g.,][]{loren1989a}.  Ophiuchus is considered to be the nearest site of clustered star formation \citep{wilking1983,motte1998}.  Star formation in Ophiuchus is heavily influenced by the nearby Sco OB2 association \citep{vrba1977}, the centre of which is at a distance of $11\pm 3$\,pc from Ophiuchus \citep{mamajek2008}.  The southwest/northeast-aligned filamentary streamers from each of the central regions are thought to be due to the effects of this association \citep{vrba1977,loren1989a}.  The L1688 cloud shows a much more active star formation history than the neighbouring L1689 cloud to the east, supporting this scenario (\citealt{nutter2006} -- hereafter NWA06).

Ophiuchus is a part of the Gould Belt, a ring of molecular clouds and OB associations $\sim1\,$kpc in diameter and inclined $\sim 20^{\circ}$ to the Galactic Plane \citep{herschel1847,gould1879}.  The Gould Belt is considered a `laboratory' for the study of low-mass star formation, as most of the low-mass star forming regions within 500\,pc of the Earth are associated with it.  This has led to surveys aimed at mapping substantial fractions of the Gould Belt being undertaken using the JCMT \citep{scuba2survey}, the \emph{Herschel} Space Observatory \citep{andre2010}, and the \emph{Spitzer} Space Telecope \citep{evans2009}.

In this paper we report on the SCUBA-2 first results for Ophiuchus from the JCMT Gould Belt Survey (GBS) and compare them to HARP CO J=3$\to$2 observations from the JCMT GBS, as well as to data from other Gould Belt surveys.  We study the starless core population of Ophiuchus, in particular investigating the stability of the cores against gravitational collapse in order to identify the prestellar \citep[i.e. gravitationally bound;][]{wardthompson1994} subset of the population of starless cores. There have been many previous wide-field millimetre and submillimetre studies of the starless core population in the L1688 cloud (e.g. \citealt{motte1998} -- hereafter MAN98; \citealt{johnstone2000}; \citealt{simpson2008} -- hereafter S08; \citealt{enoch2008}; \citealt{simpson2011}).

This paper is laid out as follows: in Section 2, we discuss the observations and data reduction.  In Section 3 we discuss data processing, including techniques for combining SCUBA-2 and \emph{Herschel} data.  In Section 4 we present our catalogue of sources, discussing source extraction and characterisation of sources using continuum and line data.  In Section 5 we discuss the energy balance and stability of the starless cores among our sources.  In Section 6 we discuss how the properties of our starless cores vary with region.  Section 7 summarises the conclusions of this paper.

\section{Observations}

\subsection{SCUBA-2}
\label{sec:s2dr}

\begin{figure*}
\centering
\includegraphics[width=0.9\textwidth]{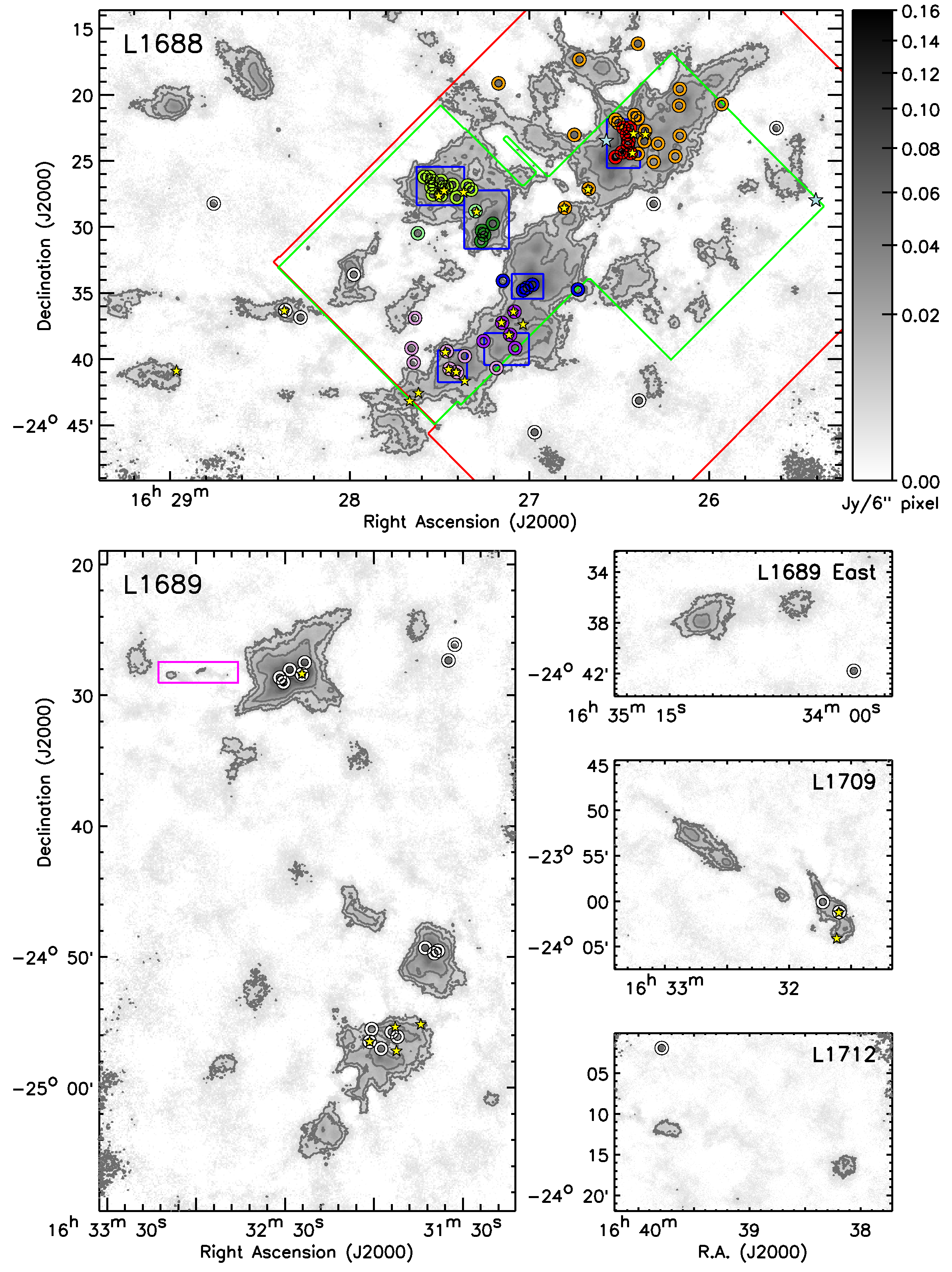}
\caption{850-\um\ flux density data, shown in square root scaling, for each of the sub-regions of Ophiuchus (see text for details).  $^{12}$CO data are available in the area outlined in red; $^{13}$CO and \co\ data are available in the area outlined in green; \nh\ data are available in the areas outlined in blue.  The CO outflow associated with IRAS 16293-2422 is marked in magenta.  Open circles mark the sources we extract from the 850-\um\ data (see text for details of colour coding).  Yellow stars mark the embedded protostars \citep{enoch2009}.  Blue stars mark the B stars HD 147889 and S1.}
\label{fig:850data}
\end{figure*}

\begin{figure*}
\centering
\includegraphics[width=0.9\textwidth]{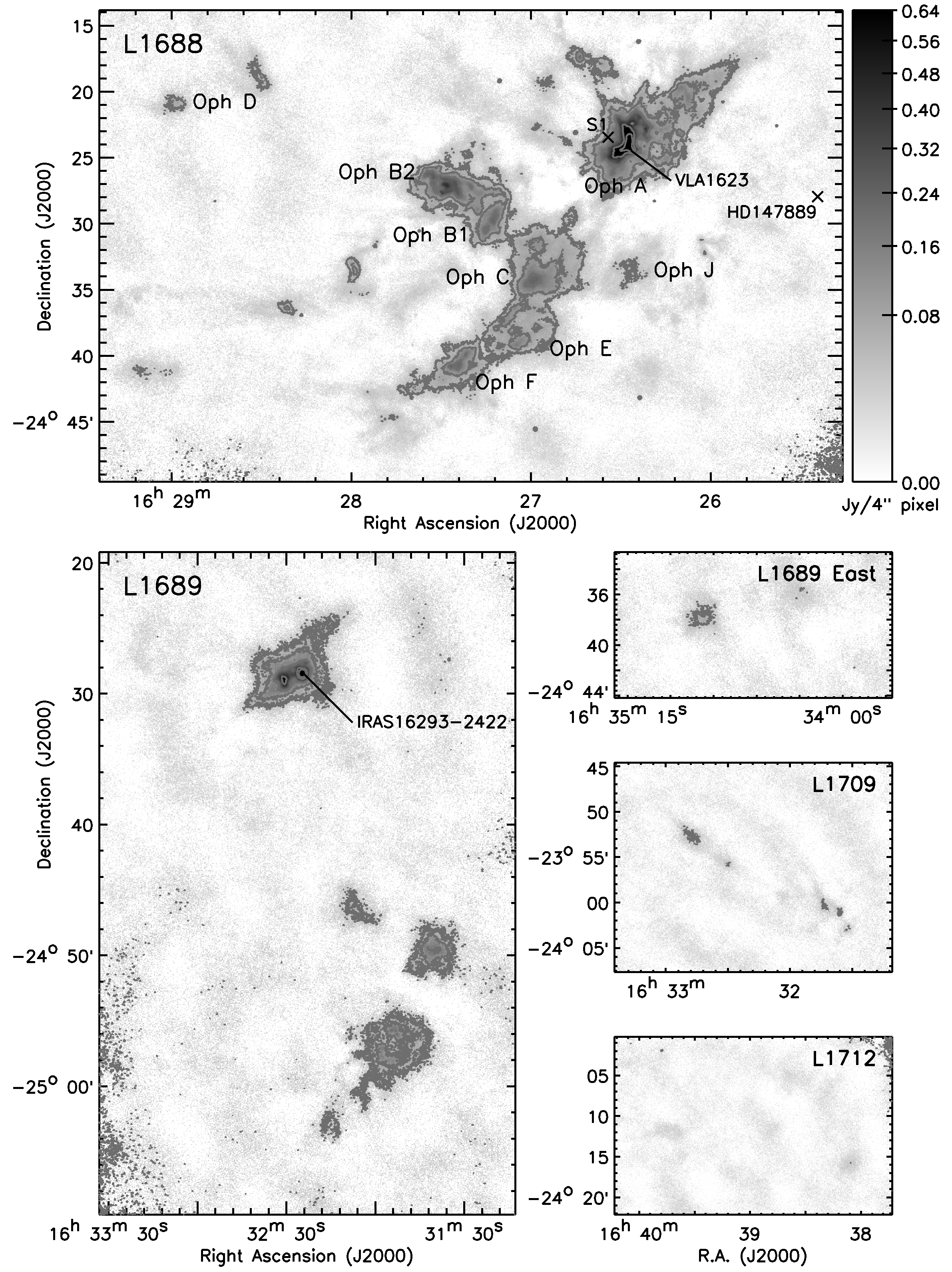}
\caption{450-\um\ flux density data, shown in square root scaling, for each of the sub-regions of Ophiuchus (see text for details).  The B stars HD 147889 and S1 are marked, along with the Class 0 protostars VLA 1623 and IRAS 16293-2422.  The sub-regions of the L1688 cloud are labelled.}
\label{fig:450data}
\end{figure*}

The SCUBA-2 \citep{holland2013} observations presented here form part of the JCMT Gould Belt Survey \citep[GBS,][]{scuba2survey}.  Continuum observations at 850~\um\ and 450~\um\ were made using fully sampled 30\arcmin\ diameter circular regions \citep[PONG1800 mapping mode,][]{bintley2014} at resolutions of 14.1\arcsec\ and 9.6\arcsec\ respectively.  Larger regions were mosaicked with overlapping scans.  The new SCUBA-2 data are shown in Figures \ref{fig:850data} and \ref{fig:450data}, for the regions of the map with significant emission.  The full maps, along with the variance arrays, are shown in Figures A1--A4 in Appendix A.

The data were reduced using an iterative map-making technique \citep[\emph{makemap} in {\sc smurf},][]{chapin2013}, and gridded to 6\arcsec\ pixels at 850~\um\, and 4\arcsec\ pixels at 450~\um, as part of the Internal Release 1 GBS data set.  The iterations were halted when the map pixels, on average, changed by $<$0.1\% of the estimated map rms. The initial reductions of each individual scan were coadded to form a mosaic from which a mask based on signal-to-noise ratio was produced for each region.  The final mosaic was produced from a second reduction using this mask to define areas of emission. Detection of emission structure and calibration accuracy are robust within the masked regions, and are uncertain outside of the masked region. The mask used in the reduction can be seen in Figure A5 in Appendix A.

A spatial filter of 10\arcmin\ is used in the reduction, which means that flux recovery is robust for sources with a Gaussian FWHM less than 2.5\arcmin. Sources between 2.5\arcmin\ and 7.5\arcmin\ in size will be detected, but both the flux and the size are underestimated because Fourier components with scales greater than 5\arcmin\ are removed by the filtering process. Detection of sources larger than 7.5\arcmin\ is dependent on the mask used for reduction.  The mask introduces further spatial filtering, as after all but the final iteration of the map-maker, all emission outside the region enclosed by the mask is suppressed.  The recovery of extended structure outside of the masked regions (shown in Figure A5 in Appendix A) is limited.

The data are calibrated in Jy/pixel, using aperture Flux Conversion Factors (FCFs) of 2.34 and 4.71 Jy/pW/arcsec$^{2}$ at 850~\um\ and 450~\um, respectively, derived from average values of JCMT calibrators \citep{dempsey2013}, and correcting for the pixel area. The PONG scan pattern leads to lower noise levels in the map centre and overlap regions, while data reduction and emission artifacts can lead to small variations in the noise level over the whole map.

Four overlapping subsections of the L1688 region were each observed four times between 2012 May 6 and 2012 July 4 in very dry (Grade 1; $\tau_{225\, {\rm GHz}}\!<\! 0.05$) weather.  Three overlapping subsections of the L1689 region were each observed six times between 2012 June 10 and 2013 June 30 in dry (Grade 2; $\tau_{225\,{\rm GHz} }\!<\! 0.08$) weather.  One section of the L1709 region was observed six times in Grade 2 weather between 2013 July 18 and 2013 July 27, as was one section of the L1712 region between 2013 July 28 and 2013 July 29.  We found a typical 1$\sigma$ noise level of 1.73\,mJy/6\arcsec\ pixel in the 850-\um\ SCUBA-2 data and 14.9\,mJy/4\arcsec\ pixel in the 450-\um\ SCUBA-2 data.

\subsection{HARP}

The HARP \citep[Heterodyne Array Receiver Programme;][]{buckle2009} receiver contains an array of 16 heterodyne detectors, arranged in a $4\times 4$ footprint on the sky.  HARP is used in conjunction with the ACSIS \citep[AutoCorrelation Spectrometer and Imaging System;][]{buckle2009} backend.  The L1688 region of Ophiuchus was observed as part of the JCMT GBS \citep{scuba2survey}, in three isotopologues of the CO J$=\!3\!\to\!2$ transition: $^{12}$CO, $^{13}$CO and \co, at a resolution of 14\arcsec.  These data are presented elsewhere \citep{white2015}.  The region of the SCUBA-2 map for which $^{12}$CO data are available (an area approximately 2050\arcsec$\times$2500\arcsec, centred on L1688) is outlined in red on Figure~\ref{fig:850data}, while the region for which both $^{13}$CO and \co\ data are available (two overlapping regions, each with an area approximately 1000\arcsec$\times$1000\arcsec) is outlined in green.

\subsection{\emph{Herschel} Space Observatory}

The \emph{Herschel} Space Observatory was a 3.5m-diameter telescope, which operated in the far-infrared and submillimetre regimes \citep{herschel}.  The observations for this paper were taken simultaneously with the Photodetector Array Camera and Spectrometer, PACS \citep{PACS}, and the Spectral and Photometric Imaging Receiver, SPIRE \citep{SPIRE,swinyard2010} using the combined fast-scanning (60\arcsec/s) SPIRE/PACS parallel mode.  Of these data sets, we used the three highest-resolution: PACS 70-\um, at 6\arcsec$\times$12\arcsec; PACS 160-\um, at 12\arcsec$\times$16\arcsec and SPIRE 250-\um, at 18\arcsec.  The data used in this paper were taken as part of the \emph{Herschel} Gould Belt Survey -- hereafter HGBS \citep{andre2010}.  HGBS Ophiuchus data are presented elsewhere (\citealt{roy2014}; Ladjalate et al. 2015.).  We use them here for comparison with the SCUBA-2 data.  These data, with Observation IDs 1342205093 and 1342205094, were reduced using HIPE version 5.1.  The SCUBA-2 pipeline was applied to the \emph{Herschel} observations in order to make the data sets comparable, as discussed in Section~\ref{sec:filtering}.  This process removes large-scale structure from the \emph{Herschel} observations, removing the necessity of applying background-correction offsets to the \emph{Herschel} observations.

\subsection{IRAM}

Archival N$_{2}$H$^{+}$ J$=\!1\!\to\!0$ data are also used (\citealt{difrancesco2004}; \citealt{andre2007}).  These observations were carried out with the IRAM 30m telescope at Pico Veleta, Spain, in 1998 June, 2000 July, and 2005 June.  The FWHM of the IRAM beam at 3mm is $\sim\!26$\arcsec.  For the purposes of improving signal-to-noise, we binned the data to a 15\arcsec\ pixel grid.  The regions of the area mapped with SCUBA-2 for which IRAM data are available are outlined in blue on Figure~\ref{fig:850data}.

\section{Data Processing}

Figures \ref{fig:850data} and \ref{fig:450data} show the new SCUBA-2 data.  Figures \ref{fig:rgb} and \ref{fig:rgb2} in Appendix B show the SCUBA-2 data compared to the \emph{Herschel} data.

\subsection{CO contamination}

SCUBA-2 850-\um\ data may be substantially contaminated by the CO J$=\!3\!\to\!2$ transition \citep{drabek2012} which, with a rest wavelength of 867.6\,\um, is covered by the SCUBA-2 850-\um\ filter, which has a half-power bandwith of 85\um\ \citep{holland2013}.  \citet{drabek2012} estimate that the contribution to the measured 850-\um\ continuum emission from CO is generally $\leq\!20\%$, but can reach $\sim\!80\%$ in outflow-dominated regions.  Some CO contamination in the 850-\um\ data is expected for L1688, primarily due to the the bright and extended outflow from the Class 0 protostar VLA1623 \citep{andre1993}.

In this region, CO contamination was corrected for by re-reducing each of the 850-\um\ observations with the integrated $^{12}$CO data added to the SCUBA-2 bolometer time series as a negative signal.  The contribution of CO emission to the total observed flux density in L1688 was found to be 4.6\%.  The fractional CO contamination varies significantly across L1688.  In the dense centres of Oph A, B, C and F the CO contamination fraction is typically $<1$\%, while in Oph E, located along the same line of sight as the edge of the outflow from VLA 1623, the contamination reaches up to 10\%.  However, in the brightest regions of CO emission from the outflow from VLA 1623 and the PDR associated with HD 147889 -- both regions of low 850-\um\ continuum emission -- the contamination fraction reaches $\sim 100\%$.  HARP CO data are only available for the central L1688 region; other regions cannot be corrected for in the same manner.  However, it is only in L1688 that there is likely to be substantial contamination, and as even in L1688 the mean contribution of the CO emission is less than 5\%, dropping to $<1\%$ in the dense, 850-\um-bright regions in which the majority of our sources lie, it is unlikely that measured 850-\um\ flux densities outside of this region are significantly affected.

As a caveat, we note that a CO outflow can be seen in the 850-\um\ data of L1689, to the east of the northernmost part of the region.  This outflow, marked in magenta on Figure~\ref{fig:850data}, was previously identified as submillimetre condensation SMM21 by NWA06, and is likely to be the outflow known to be associated with the protostar(s) IRAS 16293-2422 \citep{mizuno1990}.  This indicates that there is likely to be some CO contamination associated with IRAS 16293-2422 in the L1689 North region.  Flux densities, and hence masses, in this region may be overestimated as a result.

\subsection{Spatial filtering}
\label{sec:filtering}

SCUBA-2 is not sensitive to spatial scales greater than 600\arcsec.  In order to make SCUBA-2 and \emph{Herschel} observations comparable, the large-scale structure must be removed from the \emph{Herschel} observations.  To accomplish this, the SCUBA-2 pipeline was applied to the \emph{Herschel} observations following the method described by \citet{sadavoy2013}, in which the \emph{Herschel} data are added to the SCUBA-2 bolometer time series, and the reduction process, as described in Section~\ref{sec:s2dr}, is repeated, including the application of the mask shown in Figure A5 to the \emph{Herschel} data.  The original SCUBA-2 reduction of the data is then subtracted from the \emph{Herschel}+SCUBA-2 map, leaving the spatially-filtered \emph{Herschel} signal.  The aim of this procedure is to treat the \emph{Herschel} data as if it were SCUBA-2 data.  In order to achieve this, it is necessary to minimise the effect of the \emph{Herschel} data on the mapmaking process by treating it as a small perturbation to the SCUBA-2 signal (the input \emph{Herschel} data is scaled appropriately).  In this way, differences in areas of significant emission, noise levels and beam size between the SCUBA-2 and \emph{Herschel} maps do not distort the final, filtered, map, or prevent the mapmaking process from converging.

This spatial filtering removes the need to use Planck data to determine global background levels for the \emph{Herschel} data sets \citep[see, e.g.,][]{planckoffsets}, as all large-scale structure is removed from the filtered maps, leaving no background signal in emission-free regions.  The filtering process was repeated once for each SCUBA-2 observing position for which there was corresponding \emph{Herschel} data, and the resulting spatially-filtered maps were combined to form a mosaic.  The only region in the SCUBA-2 mosaic of Ophiuchus not covered by \emph{Herschel} is L1712.

\subsection{Common-resolution convolution kernels}
\label{sec:kernels}

SCUBA-2 450-\um\ flux densities have previously been seen to show an excess over the values predicted from interpolation of the \emph{Herschel} 350-\um\ and 500-\um\ bands \citep{sadavoy2013}.  This discrepancy was also seen in our data when they were brought to a common resolution using the published beam models.  The hypothesis that the apparent 450-\um\ excess was caused by the approximation of the SCUBA-2 450-\um\ beam secondary component as a Gaussian \citep{dempsey2013} led to the construction of a set of convolution kernels from the \emph{Herschel} and SCUBA-2 beam maps following the method of \citet{aniano2011}, which we summarise here.  This method works from beam maps rather than published beam models, and involves constructing a convolution kernel $K(A\Rightarrow B)$  that maps point spread function (PSF) $A$ onto the lower-resolution PSF $B$:
\begin{equation}
  {\rm PSF}_{B}=K(A\Rightarrow B)\ast {\rm PSF}_{A}
  \label{eq:convolution}
\end{equation}
In principle, $K(A\Rightarrow B)$ is derived using
\begin{equation}
  K(A\Rightarrow B)={\rm FT}^{-1}\left(\frac{{\rm FT}({\rm PSF}_{B})}{{\rm FT}({\rm PSF}_{A})}\right)
  \label{eq:kernel1}
\end{equation}
where FT represents the Fourier Transform operator and FT$^{-1}$ the Inverse Fourier Transform.  However, in practice the division by ${\rm FT}($PSF$_{A})$ leads to $K(A\Rightarrow B)$ being dominated by noise, unless the high spatial frequency (i.e. high wavenumber $k$) components of PSF $A$ are filtered.  Firstly, high-frequency noise is filtered from both PSFs using a filter $\phi$ which takes the form
\begin{equation}
\phi(k)= \begin{cases} 1 & {\rm for\ }k\le k_{\alpha} \\ 
  {\rm exp}\left[-\left(1.8249\times\frac{k-k_{\alpha}}{k_{\beta}-k_{\alpha}}\right)^{4}\right] & {\rm for\ }k_{\alpha}<k\le k_{\beta} \\
  0 & {\rm for\ }k_{\beta}<k \end{cases}
\label{eq:phi}
\end{equation}
where $k_{\alpha}=0.9k_{\beta}$ and $k_{\beta}=8\pi/$FWHM where FWHM is the FWHM of the instrument primary beam.  Hereafter, FT$_{\phi}=\phi\times$FT.  The highest-frequency components of PSF $A$ are further filtered: Equation~\ref{eq:kernel1} becomes
\begin{equation}
  K(A\Rightarrow B)={\rm FT}^{-1}\left(\frac{{\rm FT}_{\phi}({\rm PSF}_{B})}{{\rm FT}_{\phi}({\rm PSF}_{A})}\times f_{\textsc{a}}\right)
  \label{eq:kernel2}
\end{equation}
and the filter $f_{\textsc{a}}$ takes the form
\begin{equation}
f_{\textsc{a}}(k)= \begin{cases} 1 & {\rm for\ }k\le k_{\textsc{l,a}} \\ 
  \frac{1}{2}\left[1+\cos\left(\pi\times\frac{k-k_{\textsc{l,a}}}{k_{\textsc{h,a}}-k_{\textsc{l,a}}}\right)\right] & {\rm for\ }k_{\textsc{l,a}}<k\le k_{\textsc{h,a}} \\
  0 & {\rm for\ }k_{\textsc{h,a}}<k \end{cases}
\label{eq:f_a}
\end{equation}
where $k_{\textsc{h,a}}$ is the highest wavenumber at which FT(PSF$_{A}$) is appreciable:
\begin{equation}
  {\rm FT}({\rm PSF}_{A})(k_{\textsc{h,a}})=0.005\times{\rm FT}({\rm PSF}_{A})_{{\rm max}}
  \label{eq:k_ha}
\end{equation}
and $k_{\textsc{l,a}}=0.7k_{\textsc{h,a}}$.  Prior to constructing the convolution kernel, the PSFs are centroided, resampled to a common grid of $3645\times3645$ 0.2\arcsec\ pixels, and circularly averaged.  The SCUBA-2 and SPIRE beams are already approximately circular and are largely unchanged by this circular averaging.  The PACS beam, which is substantially elliptical, is more affected, and the convolution process may produce some slight distortion in the convolved 160-\um\ map.  However, as both the circular averaging process and the convolution process conserve flux, and as the PACS 160-\um\ beam (12\arcsec$\times$16\arcsec) is smaller than the SPIRE 250-\um\ beam (18\arcsec) along both its major and minor axes, the total flux measured inside each aperture at 160\um\ will be accurate.  It should also be noted that all of the SCUBA-2, SPIRE and PACS instruments scan in more than one direction on the sky while taking an observation, and hence the beam pattern is rotated several times within each observation.  This means that the beam pattern is to some extent circularly averaged even before the convolution is applied.

\begin{figure}
\centering
\includegraphics[width=0.47\textwidth]{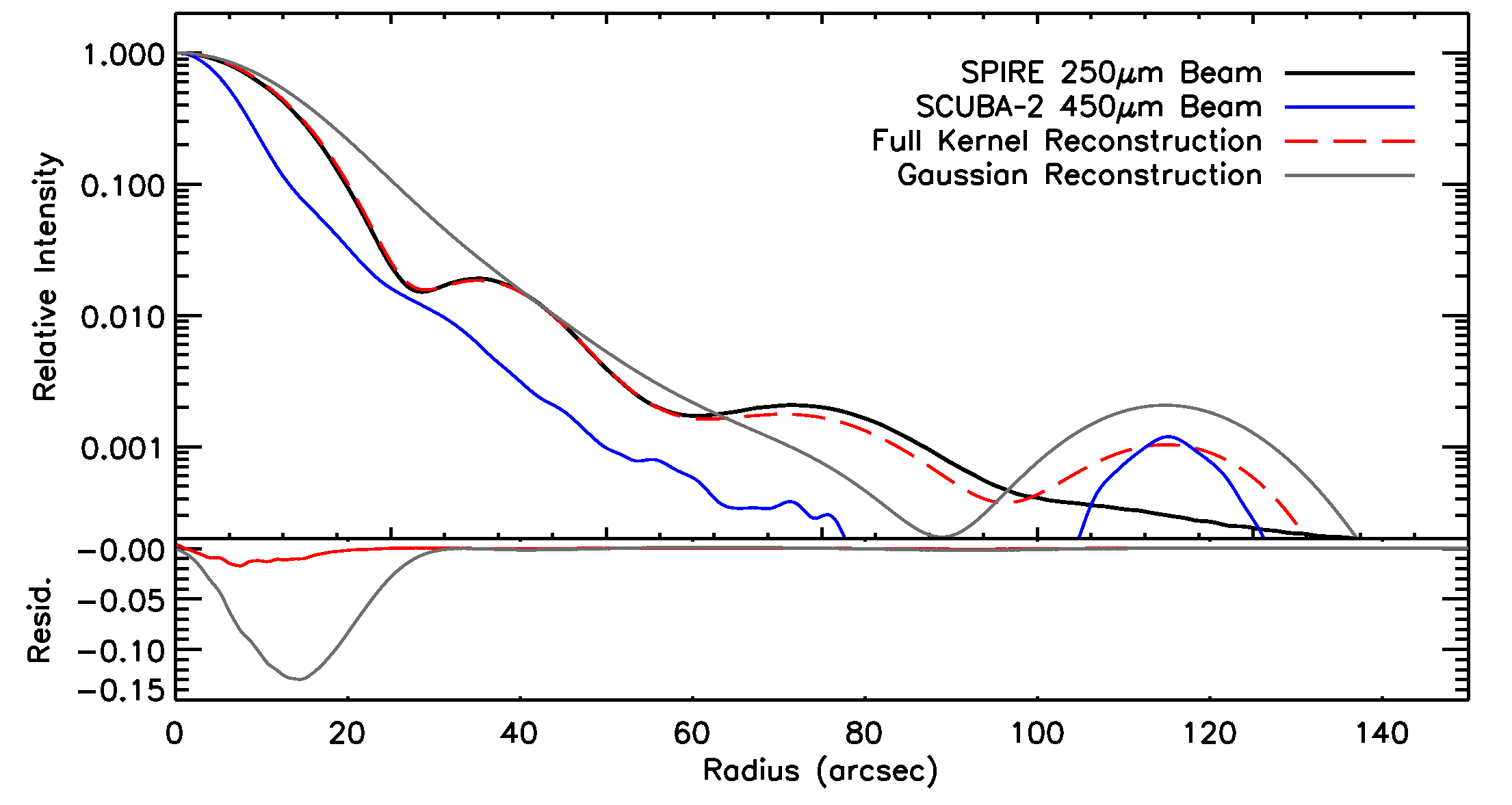}
\caption{Convolution of the SCUBA-2 450-\um\ beam (blue) to the \emph{Herschel} 250-\um\ beam (black).  The red dashed line shows the result of the convolution kernel, while the grey solid line shows the result of convolution by a Gaussian beam model.}
\label{fig:450to250}
\end{figure}

Figure \ref{fig:450to250} shows the result of convolving the maps with these kernels to the lowest-resolution wavelength band being considered (\emph{Herschel} 250-\um); this caused a marked reduction in the discrepancy between the 450-\um\ flux density and the \emph{Herschel} flux densities.  This then allowed the 450-\um\ data to be used in spectral energy distribution (SED) fitting, as discussed below.


\section{Results}

\subsection{Source extraction}
\label{sec:extraction}

Source extraction was performed on the CO-subtracted SCUBA-2 850-\um\ map of L1688, and the non-CO-subtracted SCUBA-2 850-\um\ map of the remainder of the field.  Sources were identified using the curvature-based CuTEx algorithm \citep{cutex} in its detection mode.  CuTEx identifies sources through signal in the second derivatives of the input map, effectively removing background and large-scale structure from the map, and leaving the sharp changes in gradient associated with compact sources.  CuTEx was chosen after extensive testing of various different methods as the algorithm best able to break apart the emission in crowded regions of the map (Oph A and Oph B), and which was in the most agreement with previous studies.

\begin{table*}
\centering
\caption{Results from multiple Gaussian fitting.  Sources 1-70 are from the CO-subtracted section of L1688; 71-93 are from the remainder of L1688, L1689, L1709 and L1712.  Position angles are measured east of north. FWHMs are as measured, without deconvolution.  Sources are named following the conventions of MAN98/S08 for L1688, and NWA06 for L1689.}
\label{tab:sources}
\setlength{\tabcolsep}{3pt}
\begin{tabular}{c c c cc c c c c c c c c}
\hline
Source & Full Name & Source & RA 16$^{\rm h}$: & Dec $-24^{\rm h}$: & FWHM & Angle & $F_{\nu(850\upmu{\rm m})}^{peak}$ & $F_{\nu(850\upmu{\rm m})}^{total}$ & IR &  & S08/ & \\
Index & JCMTLSG & Name & (J2000) & (J2000) & (\arcsec) & ($^{\circ}$) & (Jy/pix) & (Jy) & assn & Type & NWA06 & Region \\
\hline
1 & J162627.4-242352&SM1 & 26:27.36 & 23:52.8 & 20.4$\times$16.2 & 178.7 & 0.651 & 6.762 & S1 & C & SM1 & A \\
2 & J162627.1-242334&SM1N & 26:27.12 & 23:34.8 & 19.6$\times$15.5 & 170.0 & 0.546 & 5.215 & S1 & C & SM1N & A \\
3 & J162629.3-242425&SM2 & 26:29.28 & 24:25.2 & 29.0$\times$17.2 & 139.0 & 0.279 & 4.389 & S1 & C & SM2 & A \\
4 & J162626.4-242428&VLA 1623 & 26:26.40 & 24:28.8 & 20.0$\times$18.9 & 100.0 & 0.465 & 5.555 & Y & P & VLA 1623 & A \\
5 & J162626.6-242233&A-MM5 & 26:26.64 & 22:33.6 & 36.2$\times$18.0 & 106.2 & 0.074 & 1.519 & S1 & C? & A-MM5? & A \\
6 & J162627.6-242302&A-MM6 & 26:27.60 & 23:02.4 & 30.9$\times$22.1 & 169.6 & 0.209 & 4.474 & S1 & C? & A-MM6 & A \\
7 & J162628.8-242233&A-MM7 & 26:28.80 & 22:33.6 & 28.3$\times$19.2 & 24.3 & 0.113 & 1.929 & S1 & C? & A-MM7 & A \\
8 & J162631.4-242446&A-MM8 & 26:31.44 & 24:46.8 & 27.2$\times$17.7 & 88.3 & 0.105 & 1.589 & S1 & C & A-MM8 & A \\
9 & J162621.8-242334&A-MM1 & 26:21.84 & 23:34.8 & 26.5$\times$19.2 & 3.6 & 0.026 & 0.424 & N? & C? & A-MM1? & A$^{\prime}$ \\
10 & J162624.0-242150&A-MM4 & 26:24.00 & 21:50.4 & 27.2$\times$17.7 & 88.3 & 0.035 & 0.525 & N & C & A-MM4 & A$^{\prime}$ \\
11 & J162625.2-242136&A-MM4a & 26:25.20 & 21:36.0 & 14.3$\times$15.7 & 100.0 & 0.027 & 0.191 & N & C & - & A$^{\prime}$ \\
12 & J162645.1-242306&A-MM9 & 26:45.12 & 23:06.0 & 17.1$\times$16.0 & 80.0 & 0.063 & 0.544 & Y & P & A-MM9 & A$^{\prime}$ \\
13 & J162621.6-242247&A-MM10 & 26:21.60 & 22:48.0 & 17.8$\times$19.1 & 174.9 & 0.085 & 0.911 & Y & P? & A-MM10 & A$^{\prime}$ \\
14 & J162640.3-242710&A-MM15 & 26:40.32 & 27:10.8 & 17.3$\times$15.8 & 79.5 & 0.028 & 0.241 & Y & P & A-MM15 & A$^{\prime}$ \\
15 & J162643.4-241724&A-MM18 & 26:43.44 & 17:24.0 & 29.7$\times$22.4 & 71.0 & 0.059 & 1.230 & N & C & A-MM18 & A$^{\prime}$ \\
16 & J162624.0-241612&A-MM19 & 26:24.00 & 16:12.0 & 17.6$\times$16.5 & 80.0 & 0.070 & 0.640 & N & P & A-MM19 & A$^{\prime}$ \\
17 & J162610.3-242052&A-MM24 & 26:10.32 & 20:52.8 & 17.3$\times$15.8 & 79.5 & 0.035 & 0.306 & Y & P & A-MM24 & A$^{\prime}$ \\
18 & J162556.2-242045&A-MM25 & 25:56.16 & 20:45.6 & 17.3$\times$15.8 & 99.5 & 0.016 & 0.139 & Y & P & A-MM25 & A$^{\prime}$ \\
19 & J162610.1-241937&A-MM30 & 26:10.08 & 19:37.2 & 22.8$\times$14.5 & 41.8 & 0.024 & 0.247 & N & C & A-MM30 & A$^{\prime}$ \\
20 & J162630.5-242212&A-MM31 & 26:30.48 & 22:12.0 & 31.9$\times$19.9 & 80.7 & 0.035 & 0.691 & N? & C? & - & A$^{\prime}$ \\
21 & J162624.0-242432&A-MM32 & 26:24.00 & 24:32.4 & 22.4$\times$14.3 & 26.1 & 0.030 & 0.304 & Y & P? & - & A$^{\prime}$ \\
22 & J162617.3-242345&A-MM33 & 26:17.28 & 23:45.6 & 20.9$\times$15.6 & 175.1 & 0.021 & 0.218 & Y & P & - & A$^{\prime}$ \\
23 & J162631.4-242157&A-MM34 & 26:31.44 & 21:57.6 & 30.0$\times$20.6 & 90.9 & 0.038 & 0.736 & S1 & C? & - & A$^{\prime}$ \\
24 & J162648.2-242837&A-MM35 & 26:48.24 & 28:37.2 & 17.3$\times$15.8 & 99.5 & 0.007 & 0.065 & Y & P & - & A$^{\prime}$ \\
25 & J162710.3-241911&A-MM36 & 27:10.32 & 19:12.0 & 17.3$\times$15.8 & 79.5 & 0.036 & 0.313 & Y & P & - & A$^{\prime}$ \\
26 & J162611.5-242443&A2-MM1 & 26:11.52 & 24:43.2 & 25.7$\times$16.7 & 109.1 & 0.018 & 0.246 & N & C & A2-MM1 & A$^{\prime}$ \\
27 & J162618.7-242508&A2-MM2 & 26:18.72 & 25:08.4 & 16.8$\times$16.0 & 78.8 & 0.016 & 0.134 & N & C & - & A$^{\prime}$ \\
28 & J162610.1-242309&A3-MM1 & 26:10.08 & 23:09.6 & 29.4$\times$20.1 & 94.9 & 0.025 & 0.474 & N? & C? & A3-MM1 & A$^{\prime}$ \\
29 & J162712.2-242949&B1-MM3 & 27:12.24 & 29:49.2 & 26.9$\times$19.2 & 136.6 & 0.048 & 0.779 & N & C & B1-MM3 & B1 \\
30 & J162715.1-243039&B1-MM4a & 27:15.12 & 30:39.6 & 26.2$\times$19.2 & 114.9 & 0.050 & 0.796 & N & C & B1-MM4 & B1 \\
31 & J162715.8-243021&B1-MM4b & 27:15.84 & 30:21.6 & 19.5$\times$12.9 & 38.4 & 0.021 & 0.165 & N & C & - & B1 \\
32 & J162716.1-243108&B1-MM5 & 27:16.08 & 31:08.4 & 25.1$\times$17.6 & 98.3 & 0.033 & 0.462 & N & C & B1-MM5 & B1 \\
33 & J162718.0-242851&B1B2-MM2 & 27:18.00 & 28:51.6 & 40.1$\times$14.4 & 107.7 & 0.018 & 0.324 & Y & P? & B1B2-MM2 & B1B2 \\
34 & J162737.2-243032&B1B2-MM3 & 27:37.20 & 30:32.4 & 17.6$\times$19.4 & 177.7 & 0.014 & 0.156 & Y & P & - & B1B2 \\
35 & J162719.4-242714&B2-MM2a & 27:19.44 & 27:14.4 & 27.1$\times$18.2 & 26.8 & 0.028 & 0.441 & N & C & B2-MM2 & B2 \\
36 & J162720.6-242656&B2-MM2b & 27:20.64 & 26:56.4 & 29.6$\times$17.4 & 172.2 & 0.032 & 0.524 & N & C & - & B2 \\
37 & J162724.2-242750&B2-MM4 & 27:24.24 & 27:50.4 & 14.3$\times$15.7 & 80.0 & 0.052 & 0.365 & N & C & B2-MM4 & B2 \\
38 & J162725.7-242652&B2-MM6 & 27:25.68 & 26:52.8 & 32.6$\times$18.0 & 156.2 & 0.077 & 1.412 & N & C & B2-MM6 & B2 \\
39 & J162727.6-242703&B2-MM8a & 27:27.60 & 27:03.6 & 27.2$\times$16.6 & 97.8 & 0.060 & 0.844 & Y & P? & B2-MM8 & B2 \\
40 & J162728.6-242703&B2-MM8b & 27:28.56 & 27:03.6 & 39.1$\times$17.7 & 152.6 & 0.043 & 0.929 & Y & P? & B2-MM8 & B2 \\
41 & J162729.5-242634&B2-MM9 & 27:29.52 & 26:34.8 & 34.5$\times$20.6 & 150.1 & 0.072 & 1.607 & N & C & B2-MM9 & B2 \\
42 & J162729.5-242739&B2-MM10 & 27:29.52 & 27:39.6 & 33.2$\times$18.0 & 141.6 & 0.084 & 1.571 & Y & P & B2-MM10 & B2 \\
43 & J162733.4-242616&B2-MM13 & 27:33.36 & 26:16.8 & 34.9$\times$14.3 & 38.2 & 0.083 & 1.298 & N & C & B2-MM13 & B2 \\
44 & J162732.4-242634&B2-MM14 & 27:32.40 & 26:34.8 & 36.6$\times$19.1 & 23.3 & 0.080 & 1.764 & N & C & B2-MM14 & B2 \\
45 & J162732.6-242703&B2-MM15 & 27:32.64 & 27:03.6 & 25.9$\times$16.3 & 112.9 & 0.071 & 0.945 & N & C & B2-MM15 & B2 \\
46 & J162735.0-242616&B2-MM16 & 27:35.04 & 26:16.8 & 14.3$\times$15.7 & 100.0 & 0.076 & 0.536 & N & C & B2-MM16 & B2 \\
47 & J162732.2-242735&B2-MM17 & 27:32.16 & 27:36.0 & 32.7$\times$20.7 & 144.3 & 0.044 & 0.928 & N & P? & - & B2 \\
48 & J162659.0-243426&C-MM3 & 26:59.04 & 34:26.4 & 28.8$\times$19.5 & 117.0 & 0.041 & 0.718 & N? & C & C-MM3 & C \\
49 & J162701.0-243440&C-MM6a & 27:00.96 & 34:40.8 & 24.5$\times$14.3 & 151.7 & 0.022 & 0.242 & N? & C & C-MM6 & C \\
50 & J162702.2-243451&C-MM6b & 27:02.16 & 34:51.6 & 28.3$\times$19.2 & 48.3 & 0.018 & 0.311 & N? & C & C-MM6 & C \\
51 & J162643.9-243447&C-MM11 & 26:43.92 & 34:48.0 & 17.8$\times$19.1 & 74.9 & 0.025 & 0.271 & Y & P? & C-MM11 & C \\
52 & J162708.9-243408&C-MM13 & 27:08.88 & 34:08.4 & 17.8$\times$15.6 & 175.1 & 0.009 & 0.078 & Y & P & - & C \\
53 & J162704.8-243914&E-MM2d & 27:04.80 & 39:14.4 & 28.1$\times$15.8 & 148.4 & 0.037 & 0.522 & N & C & E-MM2d & E \\
54 & J162709.1-243719&E-MM6 & 27:09.12 & 37:19.2 & 23.2$\times$19.2 & 155.7 & 0.035 & 0.489 & Y & P? & E-MM6 & E \\
55 & J162705.0-243628&E-MM7 & 27:05.04 & 36:28.8 & 20.9$\times$19.5 & 80.0 & 0.025 & 0.318 & Y & P? & E-MM7 & E \\
56 & J162706.5-243813&E-MM9 & 27:06.48 & 38:13.2 & 17.6$\times$16.7 & 80.0 & 0.020 & 0.185 & Y & P? & E-MM9 & E \\
57 & J162715.4-243842&E-MM10 & 27:15.36 & 38:42.0 & 17.3$\times$15.8 & 79.5 & 0.018 & 0.153 & Y & P & E-MM10 & E \\
58 & J162721.6-243950&F-MM1 & 27:21.60 & 39:50.4 & 14.3$\times$15.7 & 100.0 & 0.033 & 0.232 & N & C & F-MM1 & F \\
59 & J162724.2-244102&F-MM2b & 27:24.24 & 41:02.4 & 14.3$\times$15.7 & 100.0 & 0.018 & 0.125 & Y & P & F-MM2b & F \\
60 & J162726.6-244048&F-MM3 & 27:26.64 & 40:48.0 & 17.1$\times$19.5 & 100.0 & 0.047 & 0.498 & Y & P & F-MM3 & F \\
61 & J162727.6-243928&F-MM4 & 27:27.60 & 39:28.8 & 19.8$\times$19.1 & 175.1 & 0.030 & 0.360 & Y & P & F-MM4 & F \\
62 & J162739.4-243914&F-MM5 & 27:39.36 & 39:14.4 & 17.3$\times$15.8 & 99.5 & 0.016 & 0.139 & Y & P & F-MM5 & F \\
63 & J162711.0-244044&F-MM10 & 27:11.04 & 40:44.4 & 21.3$\times$13.6 & 150.1 & 0.011 & 0.101 & Y & P? & - & F \\
\end{tabular}
\end{table*}
\addtocounter{table}{-1}
\begin{table*}
\centering

\caption{\emph{- continued}}
\setlength{\tabcolsep}{4pt}
\begin{tabular}{c c c cc c c c c c c c c}
\hline
Source & Full Name & Source & RA 16$^{\rm h}$ & Dec $-24^{\rm h}$ & FWHM & Angle & $F_{\nu(850\upmu{\rm m})}^{peak}$ & $F_{\nu(850\upmu{\rm m})}^{total}$ & IR &  & S08/ & \\
Index & JCMTLSG & Name & (J2000) & (J2000) & (\arcsec) & ($^{\circ}$) & (Jy/pix) & (Jy) & assn & Type & NWA06 & Region \\
\hline
64 & J162738.6-244019&F-MM11 & 27:38.64 & 40:19.2 & 17.6$\times$15.9 & 80.0 & 0.009 & 0.080 & Y & P & - & F \\
65 & J162738.2-243657&F-MM12 & 27:38.16 & 36:57.6 & 17.6$\times$15.9 & 171.5 & 0.008 & 0.072 & Y & P & - & F \\
66 & J162618.7-242819&J-MM1 & 26:18.72 & 28:19.2 & 17.6$\times$15.9 & 80.0 & 0.023 & 0.207 & Y & P & J-MM1 & J \\
67 & J162537.9-242233&J-MM7 & 25:37.92 & 22:33.6 & 17.3$\times$15.8 & 99.5 & 0.021 & 0.178 & Y & P & J-MM7 & J \\
68 & J162623.5-244311&J-MM8 & 26:23.52 & 43:12.0 & 17.3$\times$15.8 & 86.4 & 0.051 & 0.444 & Y & P & - & J \\
69 & J162658.3-244536&J-MM9 & 26:58.32 & 45:36.0 & 17.6$\times$15.5 & 93.6 & 0.049 & 0.422 & Y & P & - & J \\
70 & J162758.6-243339&H-MM1 & 27:58.56 & 33:39.6 & 29.2$\times$18.5 & 38.6 & 0.050 & 0.845 & N & C & - & 88 \\
71 & J162816.3-243653&H-MM2 & 28:16.32 & 36:54.0 & 17.6$\times$15.9 & 10.0 & 0.018 & 0.160 & Y & P & - & 88 \\
72 & J162821.4-243621&H-MM3 & 28:21.36 & 36:21.6 & 21.8$\times$19.1 & 105.1 & 0.036 & 0.473 & Y? & P? & - & 88 \\
73 & J162845.1-242815&D/H-MM1 & 28:45.12 & 28:15.6 & 17.6$\times$16.4 & 80.0 & 0.016 & 0.141 & Y? & P & - & 88 \\
74 & J162739.1-235819&88N SMM1 & 27:39.12 & 58:19.2 & 19.4$\times$13.3 & 32.8 & 0.008 & 0.066 & Y? & ? & - & 88 \\
75 & J163157.1-245714&SMM 8 & 31:57.12 & 57:14.4 & 28.3$\times$19.2 & 65.7 & 0.037 & 0.638 & N & C & SMM 8 & 89S \\
76 & J163201.0-245641&SMM 9 & 32:00.96 & 56:42.0 & 18.8$\times$16.8 & 92.8 & 0.049 & 0.483 & Y & P & SMM 9 & 89S \\
77 & J163151.6-245620&SMM 11 & 31:51.60 & 56:20.4 & 28.9$\times$19.2 & 82.8 & 0.029 & 0.517 & Y & P & SMM 11 & 89S \\
78 & J163153.5-245558&SMM 12 & 31:53.52 & 55:58.8 & 22.8$\times$14.5 & 158.2 & 0.036 & 0.378 & N? & C & SMM 12 & 89S \\
79 & J163200.2-245544&SMM 13 & 32:00.24 & 55:44.4 & 14.3$\times$15.7 & 86.2 & 0.025 & 0.179 & N & C & SMM 13 & 89S \\
80 & J163137.7-244947&SMM 16a & 31:37.68 & 49:48.0 & 29.2$\times$18.5 & 161.4 & 0.021 & 0.363 & N & C & SMM 16 & 89S \\
81 & J163138.9-244958&SMM 16b & 31:38.88 & 49:58.8 & 14.3$\times$15.7 & 80.0 & 0.019 & 0.137 & N & C & SMM 16 & 89S \\
82 & J163142.0-244933&SMM 16c & 31:42.00 & 49:33.6 & 28.1$\times$16.1 & 109.9 & 0.026 & 0.365 & N & C & SMM 16 & 89S \\
83 & J163355.7-244203&SMM 17 & 33:55.68 & 42:03.6 & 17.8$\times$16.3 & 15.1 & 0.017 & 0.152 & Y & P & SMM 17 & 89E \\
84 & J163228.8-242909&SMM 19 & 32:28.80 & 29:09.6 & 14.3$\times$15.7 & 80.0 & 0.154 & 1.093 & N? & C? & SMM 19 & 89N \\
85 & J163222.6-242833&SMM 20 & 32:22.56 & 28:33.6 & 21.2$\times$19.0 & 79.5 & 1.489 & 18.846 & Y? & P & SMM 20 & 89N \\
86 & J163230.0-242847&SMM 22 & 32:30.00 & 28:48.0 & 23.5$\times$14.3 & 44.7 & 0.058 & 0.611 & N? & C & - & 89N \\
87 & J163226.6-242811&SMM 23 & 32:26.64 & 28:12.0 & 25.6$\times$21.4 & 23.6 & 0.003 & 0.046 & N & C & - & 89N \\
88 & J163221.6-242739&SMM 24 & 32:21.60 & 27:39.6 & 22.1$\times$18.8 & 74.3 & 0.023 & 0.295 & N? & C & - & 89N \\
89 & J163133.4-242735&SMM 25 & 31:33.36 & 27:36.0 & 17.6$\times$16.1 & 80.0 & 0.018 & 0.161 & Y? & P & - & 89N \\
90 & J163131.2-242624&SMM 26 & 31:31.20 & 26:24.0 & 17.6$\times$15.5 & 80.0 & 0.013 & 0.110 & Y? & P & - & 89N \\
91 & J163135.5-240126&1709 SMM1 & 31:35.52 &  1:26.4 & 21.0$\times$15.9 & 82.6 & 0.073 & 0.772 & Y? & P & - & 09 \\
92 & J163143.4-240017&1709 SMM2 & 31:43.44 &  0:18.0 & 18.1$\times$16.8 & 93.5 & 0.023 & 0.217 & N? & C & - & 09 \\
93 & J163945.4-240202&1712 SMM1 & 39:45.36 &  2:02.4 & 17.6$\times$17.4 & 80.0 & 0.037 & 0.353 & - & P & - & 12 \\
\hline
\end{tabular}
\end{table*}
\setlength{\tabcolsep}{6pt}

CuTEx identified 70 sources in the CO-subtracted L1688 region and 23 sources in the rest of the observed field: 4 in the remainder of L1688, 7 in L1689 North, 8 in L1689 South, 1 in L1689 East, 2 in L1709 and 1 in L1712.  All but one of our sources are within the masked areas described in Section~\ref{sec:s2dr} and shown in Figure~A5.  Source 74/L1688N SMM1, which lies outside the mask, is the known protostellar object DoAr 4 (see Table~\ref{tab:protostars} and discussion on source classification below).

Of the 70 sources in the CO-subtracted L1688 region, 46 were uniquely associated with a source in the S08 catalogue.  A source is considered to be uniquely identified in the S08 catalogue if its FWHM area overlaps with that of an S08 source, and if it is the only source in our catalogue to do so.  The S08 catalogue identifies 93 sources in SCUBA observations of L1688, of which 91 are within the CO-subtracted SCUBA-2 field.  In Oph A, all of our sources have a unique counterpart in the S08 catalogue.  In Oph B2 we identify 13 sources while S08 identify 12.  The discrepancies between the two catalogues are mostly in low signal-to-noise regions and are likely to be due in part to the different source-finding criteria used (see discussion on completeness in Section~\ref{sec:completeness}).

Of the sixteen sources in L1689, 13 were uniquely identified sources in the NWA06 catalogue, while the remaining 3 sources were substructure within NWA06 SMM16.

The sources identified by CuTEx were characterised using a custom multiple-Gaussian fitting code, which models the flux density of sources in crowded regions by fitting a two-dimensional Gaussian to each of a set of associated sources simultaneously.  This method uses the source positions and sizes provided by CuTEx as initial input to the fitting routine \emph{mpfit} \citep{mpfit}, along with the model:
\begin{equation}
F(x,y)=a+bx+cy+\sum_{n=1}^{N}A_{n}e^{-\frac{1}{2}\left(\left(\frac{x_{n}^{\prime}}{\sigma_{x,n}}\right)^{2}+\left(\frac{y_{n}^{\prime}}{\sigma_{y,n}}\right)^{2}\right)}
\end{equation}
where
\begin{subequations}
\begin{align}
x_{n}^{\prime}=(x-x_{0,n})\cos(\theta_{n})-(y-y_{0,n})\sin(\theta_{n}) \\
y_{n}^{\prime}=(x-x_{0,n})\sin(\theta_{n})+(y-y_{0,n})\cos(\theta_{n})   
\end{align}
\end{subequations}
and $N$ is the number of sources to be fitted simultaneously.

\begin{table*}
\centering
\caption{Protostellar sources in Ophiuchus, with alternate identifications and classes, where known.  EESG09 -- \citealt{enoch2009}; WGA08 -- \citealt{wilking2008}; AM94 -- \citealt{andre1994}; DoAr -- \citealt{dolidze1959}; VSSG -- \citealt{vrba1975}; WL -- \citealt{wilking1983}; YLW -- \citealt{young1986}, LFAM -- \citealt{leous1991}; GY -- \citealt{greene1992}; GWAYL -- \citealt{greene1994}; ISO-Oph -- \citealt{bontemps2001}; EESG09 Oph-emb -- \citet{enoch2009}; EDJ2009 -- \citealt{evans2009}.  Note that in WGA08 classifications, Arabic numerals indicate a class determined from an IRAC SED while Roman numerals indicate a class determined from a 3.6-24\um\ spectral index. F indicates a flat spectrum.}
\label{tab:protostars}
\begin{tabular}{c c c c c}
\hline
Source & Source &  &  & Class \\
Index & ID & Alternate ID & Class & Reference \\
\hline
4 & VLA 1623 & EESG09 Oph-emb 3 & 0 & EESG09 \\
12 & A-MM9 & GY 116, VSSG 28, ISO-Oph 67 & 2,II  & WGA08 \\
13 & A-MM10 & LFAM 1, ISO-Oph 31 & F,- & WGA08 \\
14 & A-MM15 & GY 91, ISO-Oph 54, EESG09 Oph-emb 22 & I & EESG09 \\
16 & A-MM19 & YLW 32 ISO-Oph 40 & II & AM94 \\
17 & A-MM24 & ISO-Oph 17 & 2,II & WGA08 \\
18 & A-MM25 & DoAr 20, YLW 25, ISO-Oph 6 & II & AM94 \\
21 & A-MM32 & GY 21, LFAM 3, ISO-Oph 37 & F,F & WGA08 \\
22 & A-MM33 & ISO-Oph 21 & 1,I & WGA08 \\
24 & A-MM35 & GY 128, ISO-Oph 7, EESG09 Oph-emb 23 & I & EESG09 \\
25 & A-MM36 & SR 21(A?), YLW 8(A?), ISO-Oph 110 & 2,- & WGA08 \\
33 & B1B2-MM2 & YLW 12A/B?, ISO-Oph 124/125?, EESG09 Oph-emb 11 & I & EESG09 \\
34 & B1B2-MM3 & YLW 46, GY 304, ISO-Oph 159 & 2,- & WGA08 \\
39 & B2-MM8a & GPJ2008 8 &   &   \\
40 & B2-MM8b & YEE2006 20 &   &   \\
42 & B2-MM10 & GY 279, ISO-Oph 147, EESG09 Oph-emb 26 & I & EESG09 \\
47 & B2-MM17 & WLY 1-17? &   &   \\
51 & C-MM11 & WL 12, YLW 2, ISO-Oph 65 & 1,- & WGA08 \\
52 & C-MM13 & WL10, GY 211, ISO-Oph 105 & 2,II & WGA08 \\
54 & E-MM6 & WL 15, ISO-Oph 108, EESG09 Oph-emb 16 & I & EESG09 \\
55 & E-MM7 & GY 197, ISO-Oph 99, EESG09 Oph-emb 6 & 1,I & WGA08 \\
56 & E-MM9 & GY 205, ISO-Oph 103, EESG09 Oph-emb 12 & I & EESG09 \\
57 & E-MM10 & WL 20W/E?, GY 240A/B? ISO-Oph 121 & -,-/2,- & WGA08 \\
59 & F-MM2b & GY 263, EESG09 Oph-emb 12 & I & EESG09 \\
60 & F-MM3 & GY 265, ISO-Oph 141, EESG09 Oph-emb 14 & I & EESG09 \\
61 & F-MM4 & GY 269, ISO-Oph 143, EESG09 Oph-emb 13 & I & EESG09 \\
62 & F-MM5 & GY 314, ISO-Oph 166 & 2,F & WGA08 \\
63 & F-MM10 & GY 224, ISO-Oph 112 & F,F & WGA08 \\
64 & F-MM11 & GY 312, ISO-Oph 165 & 1,I & WGA08 \\
65 & F-MM12 & YLW 47, GY 308, ISO-Oph 163 & 2,II & WGA08 \\
66 & J-MM1 & YLW31, VSSG 1, ISO-Oph 24 & F,II & WGA08 \\
67 & J-MM7 & ISO-Oph 2 &   &   \\
68 & J-MM8 & DoAr 25, YLW 34, ISO-Oph 38 & II & AM94 \\
69 & J-MM9 & DoAr 29, ISO-Oph 88 & II & AM94 \\
71 & H-MM2 & YLW 58, ISO-Oph 196 & II & AM94 \\
72 & H-MM3 & EDJ2009 954, EESG09 Oph-emb 1 & 0 & EESG09 \\
73 & D/H-MM1 & DoAr40 & II & AM94 \\
74 & 88N SMM1 & DoAr 33 & II? & AM94 \\
76 & SMM 9 & GWAYL 6, ISO-Oph 209, EESG09 Oph-emb 10 & I & EESG09 \\
77 & SMM 11 & GWAYL 5?, ISO-Oph 204? LDN 1689 IRS 5? &   &   \\
83 & SMM 17 & EDJ2009 1013 &   &   \\
85 & SMM 20 & IRAS 16293-2422B, EESG09 Oph-emb 2 & 0 & EESG09 \\
89 & SMM 25 & DoAr 44 & II & AM94 \\
90 & SMM 26 & EDG2009 984 &   &   \\
91 & 1709 SMM1 & GWAYL 4, EDJ2009 989, EESG09 Oph-emb 17 & I & EESG09 \\
93 & 1712 SMM1 & IRAS 16367-2356, EDJ2009 989 &   &   \\
\hline
\end{tabular}
\end{table*}

Sources are considered to be neighbours if they are separated by less than twice the FWHM of the larger source.  Groups to be fitted simultaneously are defined such that each source in a group is a neighbour to at least one other source in the group, and no source has any neighbours outside of the group, with isolated sources considered as being in a one-member group.  For each group, the local background is fitted as an inclined plane with coefficients $a$, $b$ and $c$, while for each Gaussian, the quantities $A$, peak flux density, $x_{0}$ and $y_{0}$, central coordinates, $\sigma_{x}$ and $\sigma_{y}$, semi-major and semi-minor axes, and $\theta$, position angle, are fitted.  In order to accurately fit $6N+3$ parameters for each group, \emph{mpfit} was constrained such that for each source, $A>0$, $\Delta x_{0}$ and $\Delta y_{0}\leq 6''$, $\Delta\sigma_{x}$ and $\Delta\sigma_{y}\leq 10\%$, and $\Delta\theta\leq 5^{\circ}$, where $\Delta$ signifies the amount that the quantity is allowed to vary from its initial value supplied by CuTEx.  The fitted quantities do not hit the borders of the allowed parameter space.  CuTEx detects signal in the second derivatives of the input map, and hence can determine source sizes and orientations accurately, as it is sensitive to changes in gradient.

Our sources are listed in Table~\ref{tab:sources}.  In L1688, we continue the naming convention introduced by MAN98 and used by S08, while in L1689 we continue the naming convention of NWA06.  For each source, we list: the index of the source; the name of the source using the official IAU naming convention; the name by which we refer to the source in the text; central right ascension and declination; position angle of the ellipse fitted to the source measured east of north; major and minor FWHMs; best-fit model peak flux density and total 850-\um\ flux density of the background-subtracted source; whether the source has associated emission in the \emph{Herschel} 70-\um\ data (a listing of `S1' indicating that the IR emission at the source location is likely to be due to the reflection nebula associated with the star S1); our evaluation of whether the source is starless or protostellar (`C' indicating a starless core and `P' indicating a protostellar source; classification criteria and question-marked sources are discussed below); the source's identity in the S08 or NWA06 catalogues (if relevant); and the region in which the source is located.  Our sources are marked on Figure~\ref{fig:850data} as open circles, coloured according to region: red for the central Oph A region, (defined as the region contiguous with the prestellar source SM1 where $F_{\nu(850\upmu{\rm m})}^{peak}>0.6\,$Jy/6\arcsec\,pixel); orange for the more diffuse material around Oph A, hereafter referred to as Oph A$^{\prime}$; dark green for Oph B1; light green for Oph B2; blue-green for the intermediate region Oph B1B2; blue for Oph C; dark purple for Oph E; light purple for Oph F; and white for all other regions.  This identification of region by colour is used throughout the rest of this paper, except that cores marked in white in Figure~\ref{fig:850data} are elsewhere marked in black.

We judge a source as being a starless core or protostellar by considering whether its morphology appears to be point-like or extended at 850\um, whether it has associated 70-\um\ emission \citep[see, e.g.][]{konyves2010}, and the shape of its spectral energy distribution.  The first two criteria are of the most importance, as in principle a protostellar source detectable at 850\um\ should have a point-like morphology at both 850\um\ and 70\um.  The SED shape should then confirm the identification.  However, in practice, each of these criteria has limitations.  While a point-like morphology is a good indicator of an unresolved protostellar source, an extended morphology at 850-\um\ does not preclude the presence of a protostar, deeply embedded or otherwise confused with emission from cold gas along the same line of sight.  Extended emission from warm gas may confuse identification of protostars by the presence of 70-\um\ emission at their position, particularly in the reflection nebula associated with the star S1 (70-\um\ associations likely to be caused by this reflection nebula are noted in Table~\ref{tab:sources}).  Similarly, a rising SED at short wavelengths indicates a high-temperature object, possibly a protostellar envelope, but may also be caused by the presence of warm material along the line of sight not directly associated with the source.  In order to clarify these identifications, we also investigated whether there is a previously-identified protostar present within one 850-\um\ JCMT beam size (14.1\arcsec) of each of our source positions.  This criterion was generally used only to confirm the identification made using the observational criteria listed above.  However, in some cases it became necessary to use the presence or absence of a previously-identified protostar as the deciding criterion when classifying a source, particularly in crowded regions with substantial IR contamination.  Previously-known protostars were located using the SIMBAD astronomical database \citep{simbad}.  Those sources we identify as protostellar are listed in Table~\ref{tab:protostars}, with alternative identifications and, where known, their evolutionary class.  Source 93/L1712 SMM1, for which \emph{Herschel} data are not available, was catagorised as protostellar based on its 850-\um\ morphology and identification with the protostar IRAS 16367-2356 (see Table~\ref{tab:protostars}).

For the majority of our sources, a consistent classification can be made from each of our criteria.  However, where this is not the case, our classifications in Table~\ref{tab:sources} are followed by question marks.  In the case of a `C?' listing, this indicates that while all other criteria suggest that this is a starless core, there is some 70-\um\ emission at the location of the source.  In the case of a `P?' listing, this indicates that while the source can be identified with a known protostar, one or more of the selection criteria -- typically, a non-point-like morphology -- suggests that the source might be extended.  A queried classification indicates a slight conflict between our classification criteria, rather than substantial doubt about the nature of the source.

Hereafter, `source' refers to any object in our catalogue, regardless of its classification;  `protostar' refers to an object in our catalogue identified either as a pre-main-sequence star or as containing an embedded protostellar source (those sources listed as `P' or `P?' in Table~\ref{tab:sources}); and `core' refers exclusively to those objects in our catalogue identified as starless cores (`C' or `C?' in Table~\ref{tab:sources}).

\subsection{Source completeness}
\label{sec:completeness}

CuTEx detects sources through signal in the second derivatives of the original map. As a result, source detection is a function of both peak flux density and source FWHM, with sharply peaked sources being recovered better than extended sources with the same peak flux density.  To test the completeness of our set of sources, we injected 50 identical Gaussian sources at random positions in the CO-subtracted SCUBA-2 850-\um\ map of L1688, and attempted to recover these with CuTEx.  We repeated this process for various source sizes and peak flux densities.  For each source size and peak flux density, we repeated the source injection and recovery process ten times, and took the completeness fraction to be the mean fraction of sources recovered.

\begin{figure*}
\centering
\includegraphics[width=0.8\textwidth]{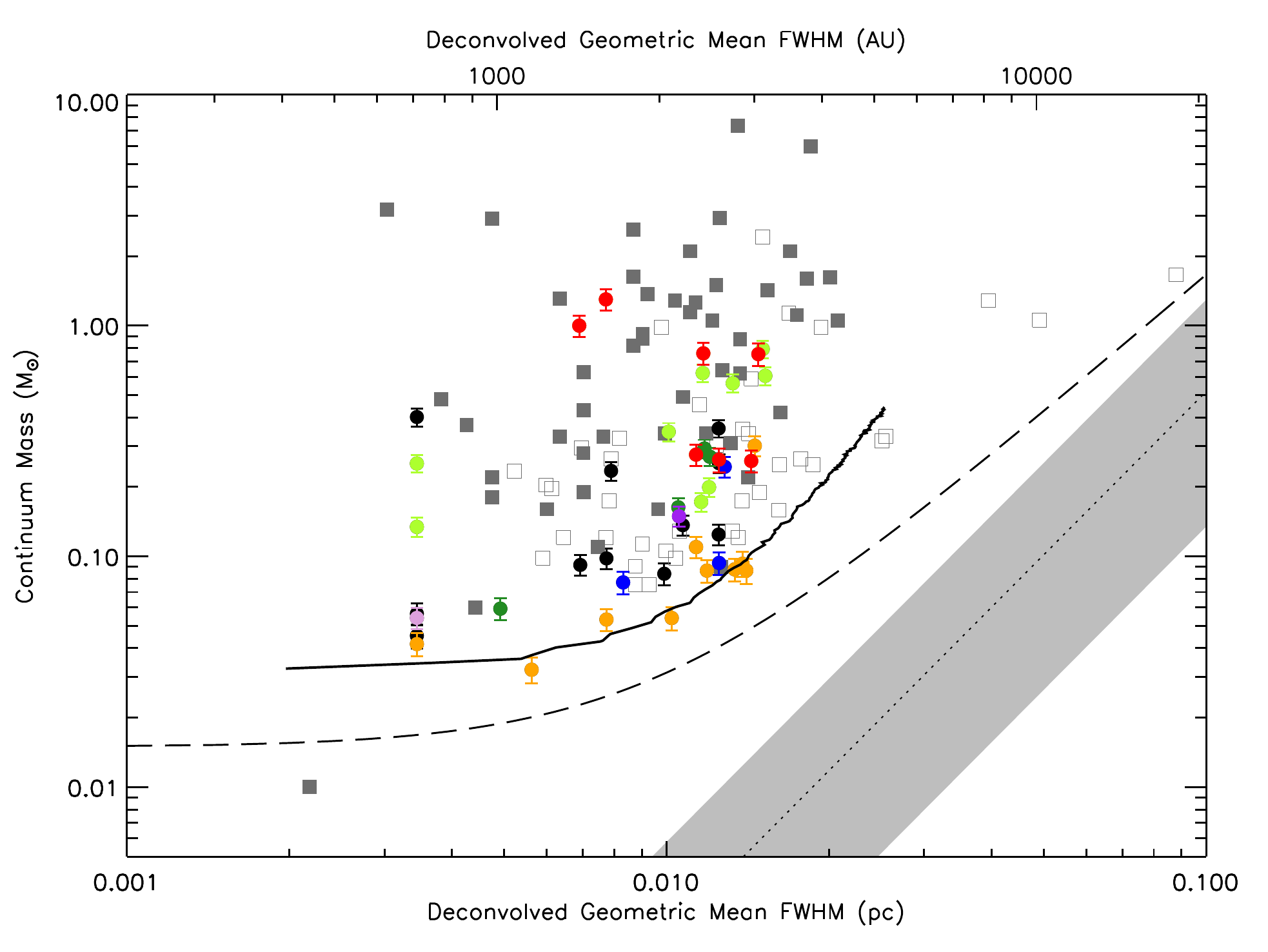}
\caption{Comparison of the masses of our starless cores, calculated from the continuum data, with their deconvolved radii.  Circles with error bars: this study.  Open squares: MAN98.  Filled squares: S08.  Grey band: $M_{\textsc{co}}\propto R_{\textsc{co}}^{2.35}$ relation \citep{elmegreen1996}.  Solid line: 80\% completeness limit.  Dashed line: $5\sigma$ sensitivity limit.  Both limits assume a temperature of 13.5\,K.  Red symbols are cores in Oph A; orange, Oph A$^\prime$; dark green, Oph B1; light green, Oph B2; blue, Oph C; dark purple, Oph E; light purple, Oph F; black, elsewhere in the cloud.}
\label{fig:m/r_plot}
\end{figure*}

For our mean non-deconvolved source FWHM of 19.7\arcsec, CuTEx recovered 50\% of injected sources with a peak flux density of 0.011\,Jy/6\arcsec\ pixel, and 80\% with a peak flux density of 0.020\,Jy/6\arcsec\ pixel.  At our mean source temperature of $\sim 13.5$\,K, these peak flux densities are equivalent to masses of 0.040\,M$_{\odot}$ (50\%) and 0.051\,M$_{\odot}$ (80\%). (See Section~\ref{sec:characterisation} for a discussion of determination of temperatures and derivation of masses.)  The 80\% completeness limit at 13.5\,K as a function of deconvolved source FWHM is shown as a solid line on Figure~\ref{fig:m/r_plot}, below.

The completeness limit in crowded regions of emission will be slightly higher and less certain than the completeness limit in sparsely populated regions, as in crowded regions tightly-packed or superimposed sources must be separated.  In regions of the SCUBA-2 850-\um\ map where $F_{\nu}>10\sigma$, we found a 50\% mass completeness limit of $0.047\pm0.005$\,M$_{\odot}$ at 13.5\,K, approximately consistent with, but slightly more uncertain than, the completeness limit across the map as a whole.  We note that completeness is likely to vary somewhat across the map, and that the completeness limits given in the paragraph above and shown on Figure~\ref{fig:m/r_plot} are average values.

\subsection{Source characterisation from continuum data}
\label{sec:characterisation}

\begin{figure}
\centering
\includegraphics[width=0.47\textwidth]{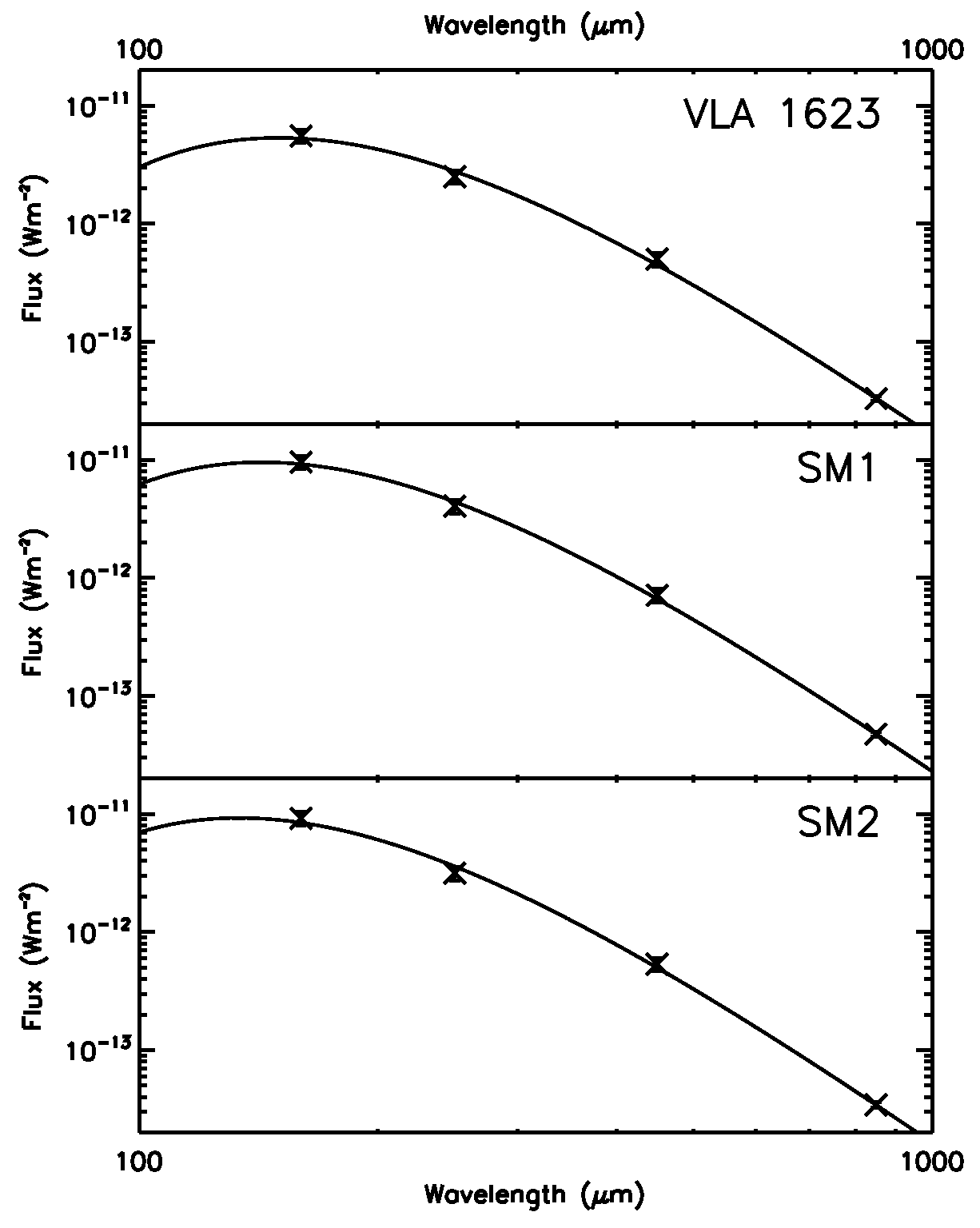}
\caption{Example SED fits for sources VLA1623, SM1 and SM2.}
\label{fig:seds}
\end{figure}

Table~\ref{tab:continuum} lists the properties of our set of sources derived from SCUBA-2 and \emph{Herschel} continuum data.  The deconvolved FWHMs of the sources were determined using the SCUBA-2 850-\um\ equivalent beam size of 14.1\arcsec\ \citep{dempsey2013}.  The equivalent radius of each source was calculated as the geometric mean of the two deconvolved FWHMs.

The data at 160\,\um, 450\,\um\ and 850\,\um\ were convolved to the 250-\um\ resolution of 18\arcsec\ using the convolution kernels described above. Flux densities were measured from the spatially filtered \emph{Herschel} 160-\um\ and 250-\um\ data and the two sets of SCUBA-2 data using elliptical apertures with major and minor axis diameters of twice the measured (i.e. non-deconvolved) major and minor FWHMs of each of the sources (enclosing 99.5\% of the total flux density in a Gaussian distribution).  The resulting SED of each source was fitted with a modified blackbody distribution, in order to determine the mean line-of-sight dust temperature of our sources. The monochromatic flux density $F_{\nu}$ is given at frequency $\nu$ by
\begin{equation}
\lambda F_{\lambda}=\nu F_{\nu}=\nu\Omega fB_{\nu}(T)\left(1-e^{-\left(\frac{\nu}{\nu_{c}}\right)^{\beta}}\right),
\label{eq:sed}
\end{equation}
where $B_{\nu}(T)$ is the Planck function at dust temperature $T$, $\Omega$ is the solid angle of the aperture, $f$ is the filling factor of the source in the aperture, $\nu_{c}=6\,$THz is the frequency at which the optical depth is taken to become unity \citep{wardthompson2002a}, and $\beta$ is the dust emissivity index, here taken to be 2.0.  Figure~\ref{fig:seds} shows three example SEDs.  This process allows determination of the average temperature of the material within the aperture.  There will be be some line-of-sight confusion between cold dust associated with the source (which will itself not be isothermal) and warmer foreground and background emission, possibly leading to an overestimation of source temperatures.  However, the spatial filtering introduced by the SCUBA-2 data reduction process should reduce the contamination by extended emission.  In crowded regions in which sources overlap significantly, the measured flux densities may be contaminated by emission from neighbouring sources.  We emphasise that the temperatures reported here are line-of-sight averages.

Masses were calculated from the best-fit model 850-\um\ flux densities and dust temperatures of our sources following the \citet{hildebrand1983} formulation
\begin{equation}
M=\frac{F_{\nu(850\upmu{\rm m})}D^{2}}{\kappa_{\nu(850\upmu{\rm m})}B_{\nu(850\upmu{\rm m})}(T)},
\label{eq:mass}
\end{equation}
where $F_{\nu(850\upmu{\rm m})}$ is the modelled total flux density at 850-\um, $D$ is the distance to Ophiuchus \citep[$139\pm 6$\,pc;][]{mamajek2008}, $B_{\nu(850\upmu{\rm m})}(T)$ is the Planck function, and $\kappa_{\nu(850\upmu{\rm m})}$ is the dust opacity, as parameterised by \citet{beckwith1990}: $\kappa_{\nu}=0.1(\nu/10^{12}{\rm Hz})^{\beta}$\,cm$^{2}$g$^{-1}$ (assuming a standard dust-to-gas ratio of 1:100).  Again, the dust emmisivity index $\beta$ was taken to be 2.0. 

For the protostellar sources in our catalogue, the temperatures, and hence the masses, determined from the dust emission are those of the protostellar envelope, and not of the protostar itself.  The modified blackbody model used to fit temperatures is applicable only to envelope-dominated sources; the temperatures and masses determined for the Class II protostars in our catalogue (listed in Table~\ref{tab:protostars}) may not be representative.

The mean volume density for each source was calculated assuming that the third axis of each source is the geometric mean of its major and minor axes.  Then, number density $n$ is calculated as 
\begin{equation}
  n=\frac{M}{\mu m_{\textsc{h}}}\frac{1}{\frac{4}{3}\pi R^{3}},
  \label{eq:density}
\end{equation}
where $R$ is the equivalent deconvolved radius, as defined above. Similarly, the column density $N$ of each source is calculated as
\begin{equation}
  N=\frac{M}{\mu m_{\textsc{h}}}\frac{1}{\pi R^{2}},
  \label{eq:coldensity}
\end{equation}
and in both cases, the mean molecular weight $\mu$ is taken to be 2.86, assuming that the gas is $\sim 70$\% H$_{2}$ by mass \citep{csar}.

One of our sources, SMM 23, located in the centre of L1689N, has a very low best-fit peak flux density, 0.003 Jy/6\arcsec\ pixel. This is due to SMM 23 being located between SMM 20/IRAS 16293-2422 and SMM 19, the brightest and second-brightest sources in L1689N respectively, leading to the majority of flux at SMM 23's position being assigned to the two nearby bright sources in the fitting process.  We consider SMM 23 to be robustly detected by CuTEx, and so determine its temperature and mass.  However, due to its properties being poorly constrained by the fitting process, we exclude SMM 23 from all subsequent analysis, leaving 46 starless cores for further analysis.

\subsection{Source mass distribution}

Figure \ref{fig:m/r_plot} shows the distribution of mass with size for the starless cores (those objects marked `C' and `C?' in Table~\ref{tab:sources}) in our sample, compared with previous studies of the same region: MAN98 (with their masses and radii scaled to account for their assumption of a distance of 160\,pc) and S08.  Our cores are comparable in size to those found in previous studies.  The masses of the cores in our sample are comparable to those found by MAN98, while the masses found by S08 are typically higher.

The grey band shown in Figure \ref{fig:m/r_plot} indicates the behaviour expected for transient, gravitationally unbound CO clumps \citep{elmegreen1996}.  Gravitationally bound prestellar cores are expected to occupy the upper part of the mass/size diagram \citep{motte2001}, being overdense compared to transient, unbound structure.

Before \emph{Herschel}, there was discussion of whether starless and prestellar cores are two different populations, separated in the mass/size plane \citep[see, e.g.][and references therein]{PPVch3}.  More recent studies have found cores occupying intermediate locations in the mass/size plane \citep{konyves2010,csar}, indicating that prestellar and unbound starless cores are all part of the same population.  Our cores are restricted to the `prestellar' region in which previous studies have found the starless cores in L1688 to lie (MAN98, S08).  The limit on our ability to recover faint sources is the CuTEx completeness limit.  The 80\% completeness limit, as a function of source size (at a temperature of 13.5\,K) is shown as a solid line on Figure~\ref{fig:m/r_plot}.  However, the 5$\sigma$ sensitivity limit of the SCUBA-2 850-\um\ data (again for a temperature of 13.5\,K), shown as a dashed line on Figure~\ref{fig:m/r_plot}, is such that regardless of our choice of source extraction algorithm, we are not sensitive to material occupying the `unbound' regions of the mass/size plane.

\begin{figure}
\centering
\includegraphics[width=0.49\textwidth]{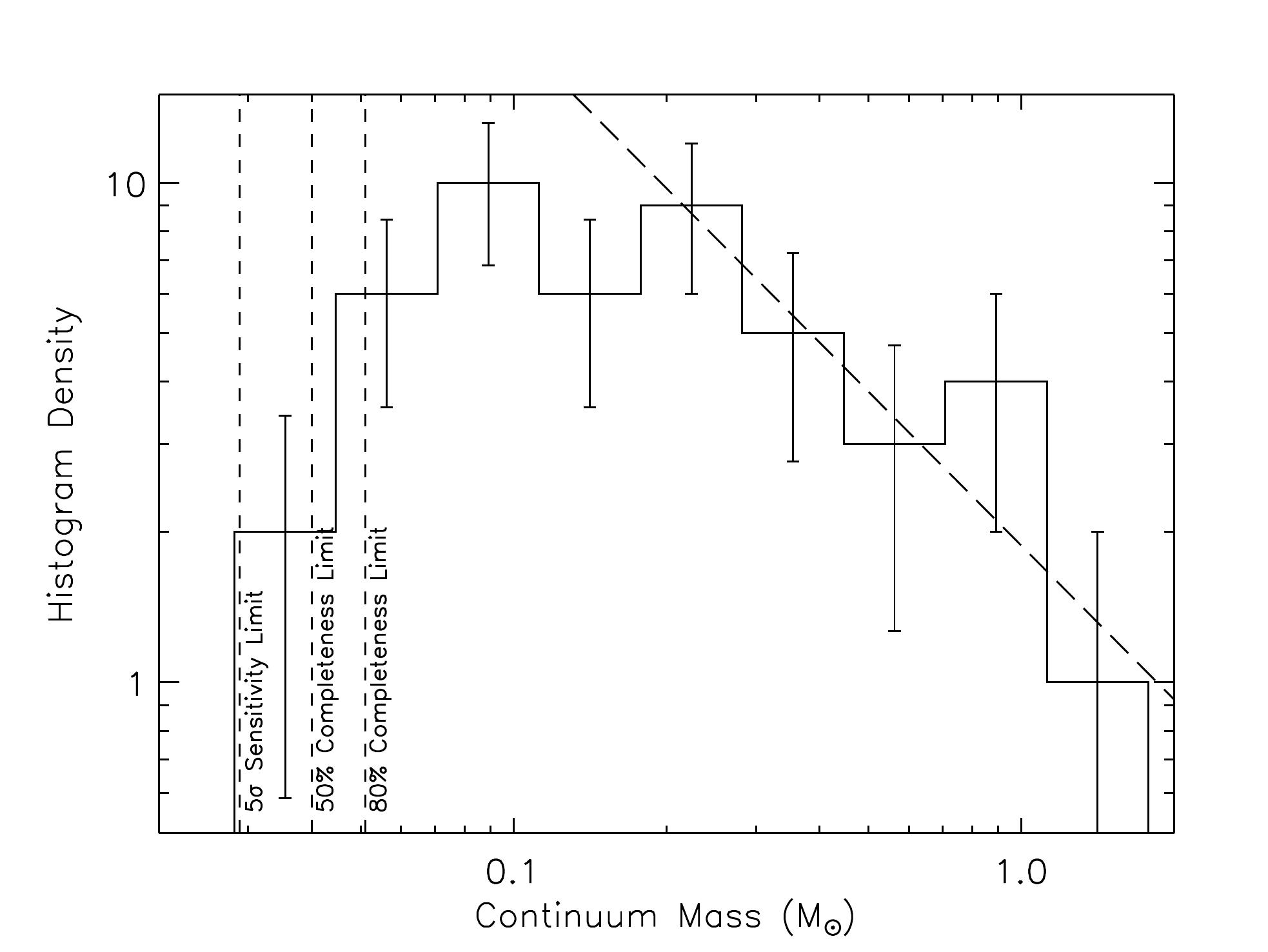}
\caption{Core mass distribution, with best-fitting power-law index $\alpha=2.0\pm 0.4$ for cores with masses $>0.2\,$M$_{\odot}$ plotted as a dashed line.  The $5\sigma$ sensitivity limit and 50\% completeness limits for a temperature of 13.5\,K are also shown.}
\label{fig:mass_hist}
\end{figure}

Figure \ref{fig:mass_hist} shows the mass distribution of our cores.  The mass distribution is consistent with the log-normal + power-law distribution expected for core mass functions (CMFs -- \citealt{chabrier2003}), and previously seen in Ophiuchus by MAN98 and S08.  We fitted a function of the form $N\propto M^{-\gamma}$ to the mass distribution, and found that, for bins centred on masses greater than or equal to 0.2 M$_{\odot}$, the best-fitting power-law index was $\gamma=1.0\pm0.4$, equivalent to a CMF power-law index of $\alpha=\gamma+1=2.0\pm0.4$.

\begin{figure}
\centering
\includegraphics[width=0.49\textwidth]{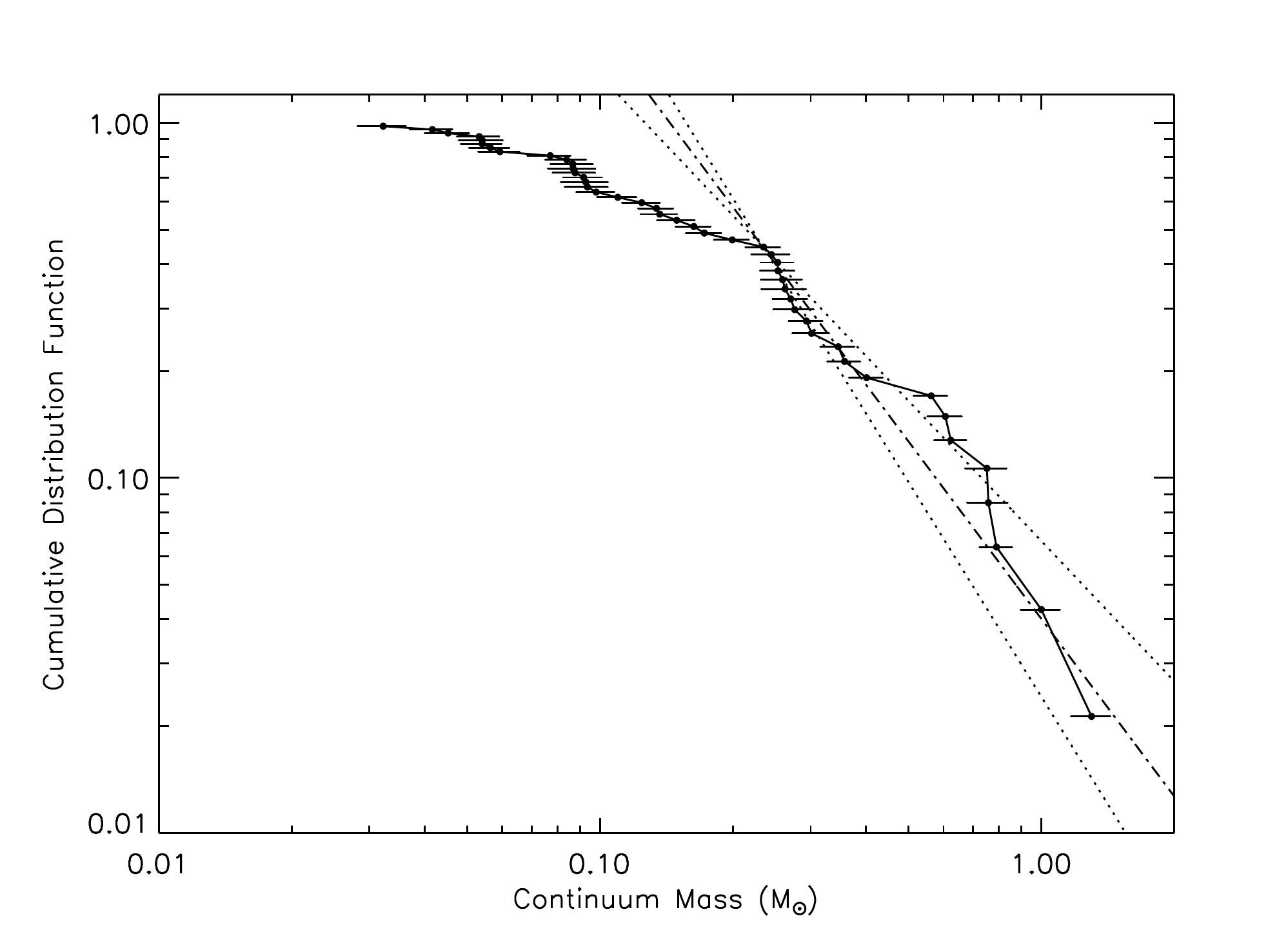}
\caption{Cumulative mass distribution function, with unbiased maximum likelihood estimator power-law index $\alpha_{uml}=2.7$ for cores with masses $>0.2\,$M$_{\odot}$ plotted as a dot-dashed line, and its $1\sigma$ $\pm 0.4$ error limits plotted as dotted lines.}
\label{fig:mass_cumulative}
\end{figure}

The traditional method of determining the power-law index of the CMF by fitting to binned data is liable to lead to a loss of accuracy in the fitted model.  We attempted to ameliorate this issue by also analysing the cumulative distribution function of core masses using the maximum likelihood estimator for an infinite power-law distribution \citep{koen2006,maschberger2009}, calculated over the same mass range ($M\ge 0.2 M_{\odot}$).  The cumulative distribution and fits are shown in Figure~\ref{fig:mass_cumulative}.  The empirical cumulative distribution function $\hat{F}$ is given, for the $i$\th\ data point in our sample, by
\begin{equation}
 \hat{F}(X_{i})\equiv \frac{i}{n+1},
\label{eq:cdf}
\end{equation}
where $n$ is the number of data points $X$. The maximum likelihood (ML) estimator for the exponent $\alpha$ of an infinite power-law distribution is
\begin{equation}
\alpha_{ml}=1+\frac{n}{\left(\sum_{i=1}^{n}X_{i}\right)-n{\rm ln}\left({\rm min}(X)\right)}.
\label{eq:ml_estimator}
\end{equation}
The unbiased maximum likelihood (UML) estimator, $\alpha_{uml}$ is then
\begin{equation}
\alpha_{uml}=1+\frac{n-1}{n}(\alpha_{ml}-1).
\label{eq:uml_estimator}
\end{equation}

The CMF power-law index found by this method was $\alpha_{uml}=2.7\pm0.4$.  Uncertainties were estimated by performing a set of Monte Carlo experiments, drawing a set of data points randomly from our distribution of masses, from which $\alpha_{\rm ml}$ was recalculated.  The error quoted is the standard deviation of the distribution of $\alpha_{uml}$ which results from this procedure.

In both cases, the power-law index is consistent with the high-mass power-law tail of the IMF, $\alpha=2.3$ \citep{salpeter,kroupa}.  That our two estimators for the power-law index only marginally agree with one another is likely a result of low number statistics.

Previous studies of the starless core population of Ophiuchus have found similar slopes for the high-mass distribution of core masses.  MAN98 found a slope of $\alpha\sim 1.5$ in the mass range 0.1--0.5\,M$_{\odot}$ and $\alpha\sim 2.5$ in the mass range 0.5--3.0\,M$_{\odot}$.  \citet{johnstone2000} found a similar behaviour: $\alpha=1.5$ for $M\leq 0.6\,$M$_{\odot}$ and $\alpha=2.0-2.5$ for $M>0.6\,$M$_{\odot}$.  \citet{sadavoy2010} found a power-law slope of $\alpha=2.26\pm0.20$ in the mass range $0.3\,$M$_{\odot}<M<5\,$M$_{\odot}$.  Our mass functions are consistent with the high-mass behaviour found by MAN98 and \citet{johnstone2000}, and with \citet{sadavoy2010} at all masses considered.

We conclude that our CMF is consistent with having a high-mass slope similar to that of the IMF, and with the CMFs found by previous studies of the same region.  The similarity between the CMF and IMF has been noted in many recent studies of molecular clouds \citep[e.g.][]{nutter2007}, leading to suggestions that the form of the IMF is caused by cloud fragmentation prior to the prestellar core stage of star formation \citep[see, e.g.,][and references therein]{PPVI}.

\begin{table*}
\centering
\caption{Properties derived from SCUBA-2 and Herschel data (see text for discussion).}
\label{tab:continuum}
\begin{tabular}{c c ccc ccc ccc ccc}
\hline
Source & Source & \multicolumn{3}{c}{Temperature} & \multicolumn{3}{c}{850\um\ Mass} & \multicolumn{3}{c}{Column Density} & \multicolumn{3}{c}{Density} \\
Index & ID & \multicolumn{3}{c}{(K)} & \multicolumn{3}{c}{(M$_{\odot}$)} & \multicolumn{3}{c}{($\times 10^{22}$ cm$^{-2}$)} & \multicolumn{3}{c}{($\times 10^{6}$ cm$^{-3}$)} \\
\hline
1 & SM1 & 17.2 & $\pm$ & 0.6 & 1.298 & $\pm$ & 0.134 & 30.520 & $\pm$ & 3.161 & 9.609 & $\pm$ & 0.995 \\
2 & SM1N & 17.3 & $\pm$ & 0.6 & 0.999 & $\pm$ & 0.104 & 29.477 & $\pm$ & 3.077 & 10.398 & $\pm$ & 1.085 \\
3 & SM2 & 18.5 & $\pm$ & 0.7 & 0.758 & $\pm$ & 0.082 & 7.758 & $\pm$ & 0.837 & 1.612 & $\pm$ & 0.174 \\
4 & VLA 1623 & 16.4 & $\pm$ & 0.5 & 1.158 & $\pm$ & 0.117 & 19.788 & $\pm$ & 2.006 & 5.311 & $\pm$ & 0.538 \\
5 & A-MM5 & 18.6 & $\pm$ & 0.7 & 0.259 & $\pm$ & 0.028 & 1.761 & $\pm$ & 0.192 & 0.298 & $\pm$ & 0.032 \\
6 & A-MM6 & 18.8 & $\pm$ & 0.8 & 0.752 & $\pm$ & 0.083 & 4.810 & $\pm$ & 0.529 & 0.790 & $\pm$ & 0.087 \\
7 & A-MM7 & 21.7 & $\pm$ & 1.0 & 0.262 & $\pm$ & 0.031 & 2.348 & $\pm$ & 0.278 & 0.456 & $\pm$ & 0.054 \\
8 & A-MM8 & 18.4 & $\pm$ & 0.7 & 0.276 & $\pm$ & 0.030 & 3.003 & $\pm$ & 0.322 & 0.643 & $\pm$ & 0.069 \\
9 & A-MM1 & 16.6 & $\pm$ & 0.6 & 0.087 & $\pm$ & 0.010 & 0.860 & $\pm$ & 0.096 & 0.176 & $\pm$ & 0.020 \\
10 & A-MM4 & 16.3 & $\pm$ & 0.5 & 0.110 & $\pm$ & 0.011 & 1.194 & $\pm$ & 0.125 & 0.256 & $\pm$ & 0.027 \\
11 & A-MM4a & 15.9 & $\pm$ & 0.5 & 0.042 & $\pm$ & 0.005 & 4.918 & $\pm$ & 0.555 & 3.468 & $\pm$ & 0.392 \\
12 & A-MM9 & 10.2 & $\pm$ & 0.2 & 0.268 & $\pm$ & 0.025 & 11.071 & $\pm$ & 1.053 & 4.621 & $\pm$ & 0.440 \\
13 & A-MM10 & 19.5 & $\pm$ & 0.9 & 0.144 & $\pm$ & 0.017 & 3.137 & $\pm$ & 0.360 & 0.950 & $\pm$ & 0.109 \\
14 & A-MM15 & 13.6 & $\pm$ & 0.4 & 0.069 & $\pm$ & 0.008 & 2.812 & $\pm$ & 0.311 & 1.167 & $\pm$ & 0.129 \\
15 & A-MM18 & 14.8 & $\pm$ & 0.4 & 0.301 & $\pm$ & 0.029 & 1.988 & $\pm$ & 0.194 & 0.332 & $\pm$ & 0.032 \\
16 & A-MM19 & 9.3 & $\pm$ & 0.3 & 0.382 & $\pm$ & 0.042 & 12.759 & $\pm$ & 1.399 & 4.788 & $\pm$ & 0.525 \\
17 & A-MM24 & 15.5 & $\pm$ & 0.5 & 0.070 & $\pm$ & 0.008 & 2.858 & $\pm$ & 0.309 & 1.186 & $\pm$ & 0.128 \\
18 & A-MM25 & 16.2 & $\pm$ & 0.6 & 0.030 & $\pm$ & 0.004 & 1.206 & $\pm$ & 0.172 & 0.501 & $\pm$ & 0.071 \\
19 & A-MM30 & 16.0 & $\pm$ & 0.5 & 0.053 & $\pm$ & 0.006 & 1.245 & $\pm$ & 0.139 & 0.391 & $\pm$ & 0.044 \\
20 & A-MM31 & 22.9 & $\pm$ & 1.2 & 0.087 & $\pm$ & 0.011 & 0.616 & $\pm$ & 0.077 & 0.106 & $\pm$ & 0.013 \\
21 & A-MM32 & 15.5 & $\pm$ & 0.5 & 0.069 & $\pm$ & 0.008 & 1.740 & $\pm$ & 0.197 & 0.568 & $\pm$ & 0.064 \\
22 & A-MM33 & 16.0 & $\pm$ & 0.5 & 0.047 & $\pm$ & 0.006 & 1.144 & $\pm$ & 0.134 & 0.365 & $\pm$ & 0.043 \\
23 & A-MM34 & 22.8 & $\pm$ & 1.2 & 0.093 & $\pm$ & 0.012 & 0.680 & $\pm$ & 0.085 & 0.119 & $\pm$ & 0.015 \\
24 & A-MM35 & 10.0 & $\pm$ & 0.2 & 0.033 & $\pm$ & 0.006 & 1.368 & $\pm$ & 0.238 & 0.568 & $\pm$ & 0.099 \\
25 & A-MM36 & 14.7 & $\pm$ & 0.5 & 0.078 & $\pm$ & 0.009 & 3.188 & $\pm$ & 0.366 & 1.324 & $\pm$ & 0.152 \\
26 & A2-MM1 & 15.8 & $\pm$ & 0.5 & 0.054 & $\pm$ & 0.006 & 0.724 & $\pm$ & 0.085 & 0.172 & $\pm$ & 0.020 \\
27 & A2-MM2 & 15.0 & $\pm$ & 0.4 & 0.032 & $\pm$ & 0.004 & 1.429 & $\pm$ & 0.183 & 0.618 & $\pm$ & 0.079 \\
28 & A3-MM1 & 17.6 & $\pm$ & 0.7 & 0.088 & $\pm$ & 0.010 & 0.687 & $\pm$ & 0.077 & 0.125 & $\pm$ & 0.014 \\
29 & B1-MM3 & 12.2 & $\pm$ & 0.3 & 0.270 & $\pm$ & 0.025 & 2.618 & $\pm$ & 0.240 & 0.529 & $\pm$ & 0.049 \\
30 & B1-MM4a & 11.8 & $\pm$ & 0.2 & 0.293 & $\pm$ & 0.026 & 2.972 & $\pm$ & 0.267 & 0.614 & $\pm$ & 0.055 \\
31 & B1-MM4b & 11.9 & $\pm$ & 0.3 & 0.059 & $\pm$ & 0.006 & 3.431 & $\pm$ & 0.375 & 1.695 & $\pm$ & 0.185 \\
32 & B1-MM5 & 12.1 & $\pm$ & 0.3 & 0.163 & $\pm$ & 0.015 & 2.066 & $\pm$ & 0.193 & 0.477 & $\pm$ & 0.045 \\
33 & B1B2-MM2 & 15.8 & $\pm$ & 0.5 & 0.071 & $\pm$ & 0.008 & 0.580 & $\pm$ & 0.066 & 0.107 & $\pm$ & 0.012 \\
34 & B1B2-MM3 & 16.4 & $\pm$ & 0.8 & 0.032 & $\pm$ & 0.005 & 0.693 & $\pm$ & 0.107 & 0.208 & $\pm$ & 0.032 \\
35 & B2-MM2a & 11.4 & $\pm$ & 0.2 & 0.172 & $\pm$ & 0.016 & 1.795 & $\pm$ & 0.169 & 0.376 & $\pm$ & 0.035 \\
36 & B2-MM2b & 11.6 & $\pm$ & 0.2 & 0.199 & $\pm$ & 0.019 & 1.941 & $\pm$ & 0.180 & 0.393 & $\pm$ & 0.037 \\
37 & B2-MM4 & 11.8 & $\pm$ & 0.3 & 0.134 & $\pm$ & 0.013 & 15.826 & $\pm$ & 1.481 & 11.162 & $\pm$ & 1.045 \\
38 & B2-MM6 & 11.3 & $\pm$ & 0.2 & 0.562 & $\pm$ & 0.050 & 4.477 & $\pm$ & 0.399 & 0.820 & $\pm$ & 0.073 \\
39 & B2-MM8a & 13.5 & $\pm$ & 0.4 & 0.243 & $\pm$ & 0.024 & 2.975 & $\pm$ & 0.289 & 0.676 & $\pm$ & 0.066 \\
40 & B2-MM8b & 13.8 & $\pm$ & 0.4 & 0.258 & $\pm$ & 0.025 & 1.613 & $\pm$ & 0.159 & 0.262 & $\pm$ & 0.026 \\
41 & B2-MM9 & 11.6 & $\pm$ & 0.3 & 0.606 & $\pm$ & 0.055 & 3.651 & $\pm$ & 0.334 & 0.582 & $\pm$ & 0.053 \\
42 & B2-MM10 & 15.8 & $\pm$ & 0.5 & 0.345 & $\pm$ & 0.035 & 2.676 & $\pm$ & 0.268 & 0.484 & $\pm$ & 0.049 \\
43 & B2-MM13 & 10.3 & $\pm$ & 0.2 & 0.623 & $\pm$ & 0.053 & 6.408 & $\pm$ & 0.546 & 1.334 & $\pm$ & 0.114 \\
44 & B2-MM14 & 10.7 & $\pm$ & 0.2 & 0.791 & $\pm$ & 0.068 & 4.875 & $\pm$ & 0.421 & 0.786 & $\pm$ & 0.068 \\
45 & B2-MM15 & 11.8 & $\pm$ & 0.3 & 0.346 & $\pm$ & 0.031 & 4.763 & $\pm$ & 0.431 & 1.147 & $\pm$ & 0.104 \\
46 & B2-MM16 & 10.4 & $\pm$ & 0.2 & 0.252 & $\pm$ & 0.022 & 29.785 & $\pm$ & 2.584 & 21.008 & $\pm$ & 1.823 \\
47 & B2-MM17 & 13.4 & $\pm$ & 0.3 & 0.272 & $\pm$ & 0.026 & 1.756 & $\pm$ & 0.168 & 0.290 & $\pm$ & 0.028 \\
48 & C-MM3 & 12.3 & $\pm$ & 0.3 & 0.244 & $\pm$ & 0.025 & 2.083 & $\pm$ & 0.210 & 0.395 & $\pm$ & 0.040 \\
49 & C-MM6a & 12.8 & $\pm$ & 0.4 & 0.077 & $\pm$ & 0.009 & 1.564 & $\pm$ & 0.179 & 0.457 & $\pm$ & 0.052 \\
50 & C-MM6b & 13.2 & $\pm$ & 0.4 & 0.094 & $\pm$ & 0.011 & 0.837 & $\pm$ & 0.095 & 0.163 & $\pm$ & 0.018 \\
51 & C-MM11 & 13.5 & $\pm$ & 0.4 & 0.078 & $\pm$ & 0.008 & 1.694 & $\pm$ & 0.180 & 0.513 & $\pm$ & 0.054 \\
52 & C-MM13 & 15.0 & $\pm$ & 0.5 & 0.019 & $\pm$ & 0.003 & 0.720 & $\pm$ & 0.129 & 0.291 & $\pm$ & 0.052 \\
53 & E-MM2d & 13.6 & $\pm$ & 0.4 & 0.149 & $\pm$ & 0.015 & 1.879 & $\pm$ & 0.187 & 0.433 & $\pm$ & 0.043 \\
54 & E-MM6 & 20.1 & $\pm$ & 0.9 & 0.074 & $\pm$ & 0.009 & 0.933 & $\pm$ & 0.111 & 0.215 & $\pm$ & 0.025 \\
55 & E-MM7 & 16.1 & $\pm$ & 0.6 & 0.068 & $\pm$ & 0.008 & 0.999 & $\pm$ & 0.113 & 0.249 & $\pm$ & 0.028 \\
56 & E-MM9 & 15.0 & $\pm$ & 0.5 & 0.045 & $\pm$ & 0.005 & 1.433 & $\pm$ & 0.173 & 0.528 & $\pm$ & 0.064 \\
57 & E-MM10 & 16.3 & $\pm$ & 0.6 & 0.032 & $\pm$ & 0.004 & 1.306 & $\pm$ & 0.172 & 0.542 & $\pm$ & 0.071 \\
58 & F-MM1 & 15.3 & $\pm$ & 0.5 & 0.054 & $\pm$ & 0.006 & 6.381 & $\pm$ & 0.691 & 4.501 & $\pm$ & 0.488 \\
59 & F-MM2b & 15.6 & $\pm$ & 0.5 & 0.028 & $\pm$ & 0.004 & 3.307 & $\pm$ & 0.427 & 2.333 & $\pm$ & 0.301 \\
60 & F-MM3 & 16.7 & $\pm$ & 0.6 & 0.101 & $\pm$ & 0.011 & 2.294 & $\pm$ & 0.246 & 0.711 & $\pm$ & 0.076 \\
61 & F-MM4 & 20.0 & $\pm$ & 0.9 & 0.055 & $\pm$ & 0.007 & 0.952 & $\pm$ & 0.114 & 0.257 & $\pm$ & 0.031 \\
62 & F-MM5 & 11.1 & $\pm$ & 0.3 & 0.057 & $\pm$ & 0.008 & 2.344 & $\pm$ & 0.321 & 0.973 & $\pm$ & 0.133 \\
63 & F-MM10 & 12.9 & $\pm$ & 0.3 & 0.031 & $\pm$ & 0.005 & 1.066 & $\pm$ & 0.155 & 0.403 & $\pm$ & 0.059 \\
\end{tabular}
\end{table*}
\addtocounter{table}{-1}
\begin{table*}
\centering
\caption{\emph{- continued}}
\begin{tabular}{c c ccc ccc ccc ccc}
\hline
Source & Source & \multicolumn{3}{c}{Temperature} & \multicolumn{3}{c}{850\um\ Mass} & \multicolumn{3}{c}{Column Density} & \multicolumn{3}{c}{Density} \\
Index & ID & \multicolumn{3}{c}{(K)} & \multicolumn{3}{c}{(M$_{\odot}$)} & \multicolumn{3}{c}{($\times 10^{22}$ cm$^{-2}$)} & \multicolumn{3}{c}{($\times 10^{6}$ cm$^{-3}$)} \\
\hline
64 & F-MM11 & 8.7 & $\pm$ & 0.2 & 0.055 & $\pm$ & 0.010 & 2.093 & $\pm$ & 0.367 & 0.836 & $\pm$ & 0.146 \\
65 & F-MM12 & 13.5 & $\pm$ & 0.4 & 0.021 & $\pm$ & 0.004 & 0.788 & $\pm$ & 0.147 & 0.315 & $\pm$ & 0.059 \\
66 & J-MM1 & 8.3 & $\pm$ & 0.2 & 0.161 & $\pm$ & 0.020 & 6.099 & $\pm$ & 0.756 & 2.435 & $\pm$ & 0.302 \\
67 & J-MM7 & 8.9 & $\pm$ & 0.3 & 0.116 & $\pm$ & 0.015 & 4.741 & $\pm$ & 0.628 & 1.968 & $\pm$ & 0.261 \\
68 & J-MM8 & 10.3 & $\pm$ & 0.3 & 0.212 & $\pm$ & 0.022 & 8.656 & $\pm$ & 0.911 & 3.593 & $\pm$ & 0.378 \\
69 & J-MM9 & 11.8 & $\pm$ & 0.3 & 0.155 & $\pm$ & 0.017 & 6.519 & $\pm$ & 0.700 & 2.747 & $\pm$ & 0.295 \\
70 & H-MM1 & 11.0 & $\pm$ & 0.2 & 0.358 & $\pm$ & 0.031 & 3.214 & $\pm$ & 0.282 & 0.626 & $\pm$ & 0.055 \\
71 & H-MM2 & 11.5 & $\pm$ & 0.3 & 0.062 & $\pm$ & 0.007 & 2.340 & $\pm$ & 0.282 & 0.934 & $\pm$ & 0.113 \\
72 & H-MM3 & 12.5 & $\pm$ & 0.3 & 0.156 & $\pm$ & 0.015 & 2.216 & $\pm$ & 0.213 & 0.542 & $\pm$ & 0.052 \\
73 & D/H-MM1 & 10.5 & $\pm$ & 0.3 & 0.065 & $\pm$ & 0.009 & 2.234 & $\pm$ & 0.296 & 0.850 & $\pm$ & 0.113 \\
74 & 88N SMM1 & 8.2 & $\pm$ & 0.3 & 0.053 & $\pm$ & 0.011 & 2.800 & $\pm$ & 0.605 & 1.322 & $\pm$ & 0.286 \\
75 & SMM 8 & 11.3 & $\pm$ & 0.2 & 0.253 & $\pm$ & 0.023 & 2.263 & $\pm$ & 0.207 & 0.440 & $\pm$ & 0.040 \\
76 & SMM 9 & 19.0 & $\pm$ & 0.8 & 0.080 & $\pm$ & 0.009 & 2.125 & $\pm$ & 0.245 & 0.712 & $\pm$ & 0.082 \\
77 & SMM 11 & 14.6 & $\pm$ & 0.4 & 0.131 & $\pm$ & 0.013 & 1.126 & $\pm$ & 0.114 & 0.215 & $\pm$ & 0.022 \\
78 & SMM 12 & 14.3 & $\pm$ & 0.4 & 0.098 & $\pm$ & 0.010 & 2.292 & $\pm$ & 0.230 & 0.720 & $\pm$ & 0.072 \\
79 & SMM 13 & 12.8 & $\pm$ & 0.3 & 0.056 & $\pm$ & 0.006 & 6.654 & $\pm$ & 0.712 & 4.693 & $\pm$ & 0.502 \\
80 & SMM 16a & 12.3 & $\pm$ & 0.3 & 0.124 & $\pm$ & 0.013 & 1.116 & $\pm$ & 0.112 & 0.217 & $\pm$ & 0.022 \\
81 & SMM 16b & 12.5 & $\pm$ & 0.3 & 0.045 & $\pm$ & 0.005 & 5.341 & $\pm$ & 0.624 & 3.767 & $\pm$ & 0.440 \\
82 & SMM 16c & 11.7 & $\pm$ & 0.3 & 0.136 & $\pm$ & 0.013 & 1.664 & $\pm$ & 0.162 & 0.377 & $\pm$ & 0.037 \\
83 & SMM 17 & 10.5 & $\pm$ & 0.3 & 0.070 & $\pm$ & 0.009 & 2.357 & $\pm$ & 0.295 & 0.886 & $\pm$ & 0.111 \\
84 & SMM 19 & 11.8 & $\pm$ & 0.3 & 0.402 & $\pm$ & 0.036 & 47.367 & $\pm$ & 4.264 & 33.409 & $\pm$ & 3.007 \\
85 & SMM 20 & 17.4 & $\pm$ & 0.7 & 3.555 & $\pm$ & 0.393 & 53.954 & $\pm$ & 5.961 & 13.649 & $\pm$ & 1.508 \\
86 & SMM 22 & 11.5 & $\pm$ & 0.3 & 0.235 & $\pm$ & 0.022 & 5.274 & $\pm$ & 0.485 & 1.624 & $\pm$ & 0.149 \\
87 & SMM 23 & 12.8 & $\pm$ & 0.3 & 0.015 & $\pm$ & 0.005 & 0.128 & $\pm$ & 0.040 & 0.025 & $\pm$ & 0.008 \\
88 & SMM 24 & 13.6 & $\pm$ & 0.4 & 0.084 & $\pm$ & 0.009 & 1.200 & $\pm$ & 0.130 & 0.295 & $\pm$ & 0.032 \\
89 & SMM 25 & 9.6 & $\pm$ & 0.2 & 0.090 & $\pm$ & 0.011 & 3.253 & $\pm$ & 0.406 & 1.269 & $\pm$ & 0.158 \\
90 & SMM 26 & 11.0 & $\pm$ & 0.3 & 0.046 & $\pm$ & 0.007 & 1.950 & $\pm$ & 0.293 & 0.820 & $\pm$ & 0.123 \\
91 & 1709 SMM1 & 12.9 & $\pm$ & 0.4 & 0.240 & $\pm$ & 0.025 & 5.436 & $\pm$ & 0.573 & 1.681 & $\pm$ & 0.177 \\
92 & 1709 SMM2 & 11.0 & $\pm$ & 0.2 & 0.092 & $\pm$ & 0.009 & 2.684 & $\pm$ & 0.275 & 0.942 & $\pm$ & 0.096 \\
93 & 1712 SMM1 & 5.8 & $\pm$ & 0.1 & 0.716 & $\pm$ & 0.089 & 20.386 & $\pm$ & 2.534 & 7.064 & $\pm$ & 0.878 \\
\hline
\end{tabular}
\end{table*}

\subsection{Source characterisation from spectral data}
\label{sec:linedata}

The typical column densities, masses and velocity dispersions derived from \nh\ and \co\ data were estimated for each core for which data were available.  For each core, the velocity dispersion was taken to be the average of the velocity dispersions in each good pixel covered by the aperture used for source photometry, while the mass was taken to be the average of the masses in the good pixels in the aperture, multiplied by the total number of pixels in the aperture.  The starless core properties derived from \nh\ and \co\ data are listed in Table~\ref{tab:line}.

Of the emission from the three isotopologues of CO mapped by HARP, that of \co\ was chosen as it has the lowest optical depth, typically $<0.5$, but reaching $\sim\!2$ in high-density regions \citep{white2015}.  \co\ emission can only probe the outer envelopes of starless cores; the freeze-out of heavy molecules onto dust grains at high densities and low temperatures means that CO (or its isotopologues) cannot be considered a reliable tracer for densities $n$(H$_{2}$)$>\!10^{5}$\,cm$^{-3}$ \citep[see, e.g.][and references therein]{PPVch2}.  Although Ophiuchus is known to have low average levels of CO depletion \citep{christie2012}, C$^{18}$O linewidths can only be used as a conservative measure of the bound state of a core, providing information on the behaviour of the moderately dense cloud material.

\nh\ emission is a better tracer of the bound state of the densest parts of starless cores than \co, with significant depletion not occurring until core densities exceed $\sim\!10^{6}$\,cm$^{-3}$ \citep[][and references therein]{PPVch2} .  However, due to the low abundance of \nh\ relative to H$_{2}$ ($X({\rm N_{2}H^{+}})=5.2\pm0.5\times10^{-10}$ -- \citealt{pirogov2003}), it is only detectable in regions of the highest H$_{2}$ column density.

Each pixel was fitted using an IDL routine utilising \emph{mpfit} \citep{mpfit}.  For \co, a single Gaussian was fitted to each pixel, and fits with signal-to-noise ratio (SNR) $\ge 5$ were accepted.  For \nh, a seven-component set of Gaussians were fitted to the multiplet, and fits were accepted for pixels where the weakest component had SNR $\ge 2$.

Column densities, and hence masses, were calculated for each core, using the CO and \nh\ data sets.  Column densities were calculated following \citet{garden1991}: 
\begin{equation}
  N=\frac{3k_{\textsc{b}}}{8\pi^{3}B\mu_{\textsc{d}}^{2}}\frac{e^{\nicefrac{hBJ(J+1)}{k_{\textsc{b}}T_{\rm{ex}}}}}{J+1}\frac{T_{\rm{ex}}+\frac{hB}{3k_{\textsc{b}}}}{1-e^{-\nicefrac{h\nu}{k_{\textsc{b}}T_{\rm{ex}}}}}\int\tau\mathrm{d}v
  \label{eq:coldensity}
\end{equation}
where $N$ is the column density of the species in question, $B$ and $\mu_{\textsc{d}}$ are the rotational constant and permanent dipole moment of the molecule respectively, and $J$ is the lower rotational level of the transition.  The excitation temperature, $T_{\rm ex}$, can be calculated as follows \citep[see, e.g.,][]{pineda2008}
\begin{equation}
  T_{\rm ex}=\frac{T_{0}}{{\rm ln}\left(1+T_{0}\left(\frac{T_{\textsc{r}}}{1-e^{-\tau}}+\frac{T_{0}}{e^{\nicefrac{T_{0}}{T_{\rm bg}}}-1}\right)^{-1}\right)}
  \label{eq:tex}
\end{equation}
where $T_{0}=h\nu/k_{\textsc{b}}$, $T_{\rm bg}$ is the cosmic microwave background temperature, 2.73\,K, and $T_{\textsc{r}}$ is the radiation temperature of the spectral line.

The integral in Equation~\ref{eq:coldensity} can be written as \citep[see, e.g.,][]{buckle2010}:
\begin{align}
  \int\tau(v)\mathrm{d}v & = & \frac{1}{J(T_{\rm ex})-J(T_{\rm bg})}\int\frac{\tau(v)}{1-e^{-\tau(v)}}T_{\textsc{mb}}\mathrm{d}v \\
   & \approx & \frac{1}{J(T_{\rm ex})-J(T_{\rm bg})}\frac{\tau(v_{0})}{1-e^{-\tau(v_{0})}}\int T_{\textsc{mb}}\mathrm{d}v
  \label{eq:tau_integral}
\end{align}
where $v_{0}$ is the central velocity of the line, $T_{\textsc{mb}}$ is the observed main beam temperature and $J(T)$ is the source function,
\begin{equation}
  J(T)=\frac{T_{0}}{e^{\nicefrac{T_{0}}{T}}-1}
  \label{eq:source_function}
\end{equation}
with $T_{0}$ defined as above.

Excitation temperatures and optical depths for \co\ were calculated under the assumption that the $^{13}$CO and \co\ emission trace material with the same excitation temperature, and that $^{13}$CO is optically thick everywhere.  The excitation temperature is calculated using Equation~\ref{eq:tex} in the limit $\tau_{^{13}\rm CO}\gg 1$, with $T_{\textsc{r}}=T_{\rm max,^{13}CO}$.  The optical depth of \co\ is determined using the relation
\begin{equation}
  \frac{T_{{\rm max,C^{18}O}}}{T_{{\rm max,^{13}CO}}}=\frac{1-e^{-\tau_{{\rm C^{18}O}}}}{1-e^{-\tau_{{\rm^{13}CO}}}},
  \label{eq:co_relative_strength}
\end{equation}
and the abundance ratio [$^{13}$CO/\co]$=$5.5 \citep{frerking1982}, i.e. $\tau_{{\rm^{13}CO}}=5.5\tau_{{\rm C^{18}O}}$.

For \co, $B$ and $\mu_{\textsc{d}}$ were taken from the NIST database \citep{cccbdb}: $B=5.79384\times10^{10}\,{\rm s}^{-1}$, and $\mu_{\textsc{d}}=0.112$\,D.  Thus, Equation~\ref{eq:coldensity} becomes
\begin{multline}
  N({\rm C}^{18}{\rm O})=7.94\times10^{8}e^{\nicefrac{16.88}{T_{\rm ex}}}\frac{T_{\rm ex}+0.927}{1-e^{-\nicefrac{16.88}{T_{\rm ex}}}}\times \\ \frac{1}{J(T_{\rm ex})-J(2.73\,{\rm K})}\frac{\tau}{1-e^{-\tau}}\Delta v\sum\limits_{i}T_{i}\,{\rm cm}^{-2},
  \label{eq:c18o_coldensity}
\end{multline}
where $\Delta v$ is the velocity channel width in cm\,s$^{-1}$, and $T_{i}$ is the best-fit main beam temperature in the $i^{\rm th}$ velocity channel.  The equivalent H$_{2}$ column density is found using the conversion factor $X({\rm C^{18}O})=2.635\times10^{-7}$.  This value of $X({\rm C^{18}O})$ was determined from the relations $N({\rm H}_{2})/A_{\textsc{v}}=9.4\times10^{20}\,{\rm cm}^{2}\,{\rm mag}^{-1}$ \citep[][and references therein]{pineda2010}, and $N(^{12}{\rm CO})/A_{\textsc{v}}=1.01\times 10^{17}\,{\rm cm}^{2}\,{\rm mag}^{-1}$ \citep{pineda2010}, i.e. $N({\rm H}_{2})/N(^{12}{\rm CO})=1.1\times10^{4}$.   For the abundance ratios [$^{13}$CO/\co]$=$5.5 \citep{frerking1982} and [$^{12}$CO/$^{13}$CO]=69 \citep{wilson1999}, this leads to the value of $X({\rm C^{18}O})$ given above.  The accuracy of the H$_{2}$ column densities calculated using this abundance ratio depends on all of the above relations being valid in Ophiuchus and consistent across all of our cores.  The total uncertainty resulting from all of these relations is difficult to quantify, but we state conservatively that our column densities determined from \co\ emission are likely to be accurate to within a factor of a few.

\begin{figure*}
\centering
\includegraphics[width=0.99\textwidth]{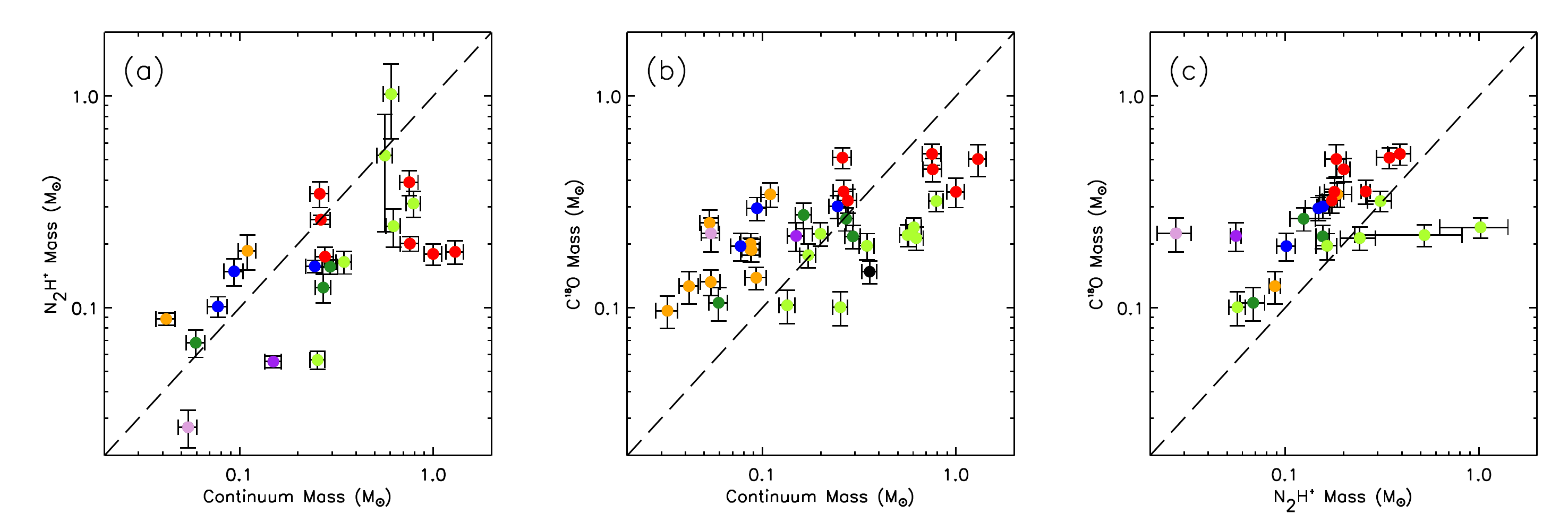}
\caption{Comparison of masses calculated from continuum, \nh\ and \co\ emission.  Panel (a) compares continuum- and \nh-derived masses, for the 23 cores for which \nh\ data are available. Panel (b) compares continuum- and \co-derived masses, for the 35 cores for which \co\ data are available.  Panel (c) compares \nh- and \co-derived masses, for the 23 \nh\ cores.  Colour coding is as in Figure~\ref{fig:m/r_plot}.  The dashed line is the line of unity.}
\label{fig:mass_comparison}
\end{figure*}

The hyperfine splitting of the \nh\ multiplet allows for the direct calculation of optical depth.  The optical depths of any pair of hyperfine transitions $j\to i$ and $m\to l$ are related to one another by their hyperfine statistical weights and Einstein A coefficients \citep[see, e.g.,][p. 308]{emerson1999}:
\begin{equation}
  \frac{\tau_{ji}}{\tau_{ml}}=\frac{g_{j}A_{ji}}{g_{m}A_{ml}}.
  \label{eq:relative_tau}
\end{equation}
Neglecting any background contribution, the relative strengths of the two lines will be
\begin{equation}
  \frac{T_{{\rm max},ji}}{T_{{\rm max},ml}}=\frac{T_{{\rm ex},ji}}{T_{{\rm ex},ml}}\frac{1-e^{-\tau_{ji}}}{1-e^{-\tau_{ml}}}.
  \label{eq:relative_strength}
\end{equation}
Assuming that the excitation temperature is the same for all of the hyperfine transitions, the relative strengths of each of the hyperfine components can be expressed as a function of optical depth, and hence optical depth can be fitted as a free parameter.  The excitation temperature can then be calculated using Equation~\ref{eq:tex}.  For each of the 15 hyperfine components, Equation~\ref{eq:coldensity} becomes
\begin{multline}
  N_{i}=3.10\times10^{6}\frac{T_{\rm ex}+0.745}{1-e^{-\nicefrac{h\nu_{i}}{k_{\textsc{b}}T_{\rm ex}}}}\times \\
  \frac{1}{J(T_{\rm ex})-J(2.73\,{\rm K})}\frac{\tau_{i}}{1-e^{-\tau_{i}}}\Delta v\sum\limits_{j}T_{j}\,{\rm cm}^{-2},
  \label{eq:n2hplus_component_coldensity}
\end{multline}
 where $T_{j}$ is the best-fit model main beam temperature of the of the $i^{\rm th}$ hyperfine component in the $j^{\rm th}$ velocity channel.  The frequencies and Einstein A coefficients of the hyperfine transitions are taken from \citet{daniel2006}, while the parameters $B=4.65869\times10^{10}\,$s$^{-1}$ and $\mu_{\textsc{d}}=3.40\,$D are taken from the CDMS database \citep{cdms}.  Summing over all hyperfine components, the total \nh\ column density is
\begin{equation}
  N({\rm N}_{2}{\rm H}^{+})=\sum\limits_{i=1}^{15}N_{i}.
  \label{eq:n2hplus_coldensity}
\end{equation}
The equivalent H$_{2}$ column density is found using the conversion factor $X({\rm N_{2}H^{+}})=5.2\times10^{-10}$ \citep{pirogov2003}.  We note that \citet{pirogov2003} determined this value of $X({\rm N_{2}H^{+}})$ by considering the mean \nh\ abundance across 36 massive molecular cloud cores; the applicability of this abundance to a low-to-intermediate mass star forming region such as Ophiuchus is not certain.  \citet{friesen2010} find \nh\ abundances in the range $2.5-17\times10^{-10}$ in Oph B, while \citet{difrancesco2004} find a mean \nh\ abundance of $1.3\times10^{-10}$ in Oph A, indicating \nh\ depletion in the Oph A region.  These results suggest that the \citet{pirogov2003} value of $X({\rm N_{2}H^{+}})$ is applicable to our cores, but that a wide scatter about this abundance is to be expected, and hence our H$_{2}$ column densities determined using this abundance are likely to be accurate to within a factor of $2-3$ in regions without significant \nh\ depletion.

Figure~\ref{fig:mass_comparison} compares the masses derived from each of our tracers, and shows that the masses of cores measured in \nh\ and in continuum emission correlate fairly well, although with significant scatter about the line of unity, whereas those from \co\ do not.  This correlation indicates that \nh\ and dust are tracing the same material.  The excess in continuum mass over \nh\ mass in the most massive cores in Oph A indicates that \nh\ is not tracing the very innermost regions of the densest cores.  As discussed above, depletion of \nh\ in the densest regions of Oph A has been previously noted by \citet{difrancesco2004}.  There is also considerable variation in core mass from region to region, as shown by the coloured symbols.  We return to a discussion of this variation in Section~\ref{sec:regional}.

It should be noted that different subsets of our set of starless cores are shown in each panel of Figure~\ref{fig:mass_comparison}.  \co\ data are available at the positions of 35 of the 46 starless cores which we are analysing (as shown in Figure~\ref{fig:mass_comparison}b).  \nh\ data are available for 23 of these 35 cores (shown in Figure~\ref{fig:mass_comparison}a).  There are no cores for which \nh\ data are available and \co\ data are not (i.e. the samples shown in Figures \ref{fig:mass_comparison}a and \ref{fig:mass_comparison}c are identical, and are a subset of those in Figure~\ref{fig:mass_comparison}b).  The \co\ and \nh\ masses of all cores for which data are available are listed in Table~\ref{tab:line}.  The virial analysis in Section 5 is performed only on those 23 cores for which continuum, \co\ and \nh\ data are all available.

\section{Energy balance and stability}

We estimate the magnitude of each of the terms in the virial equation in order to determine the energy balance, and hence the stability against collapse, of the cores in our sample.  We consider the virial equation in the form
\begin{equation}
\frac{1}{2}\ddot{\mathcal{I}}=2\Omega_{\textsc{k}}+\Omega_{\textsc{g}}+\Omega_{\textsc{m}}+\Omega_{\textsc{p}}
\label{eq:full_virial_theorem}
\end{equation}
where $\ddot{\mathcal{I}}$ is the second derivative of the moment of inertia, $\Omega_{\textsc{k}}$ is the internal energy, $\Omega_{\textsc{g}}$ is the gravitational potential energy, $\Omega_{\textsc{m}}$ is the magnetic energy, and $\Omega_{\textsc{p}}$ is the energy due to external pressure acting on the core.  If $\ddot{\mathcal{I}}<0$, a core's net energy is negative, and hence the core is collapsing.  Conversely, a core with $\ddot{\mathcal{I}}>0$ will be dispersing, and the virially stable mass of a core is the mass at which $\ddot{\mathcal{I}}=0$.

\subsection{Gravitational and internal energy}

The first two terms on the right-hand side of Equation~\ref{eq:full_virial_theorem} can be estimated from directly measured quantities.  The internal kinetic energy of a core of mass $M$ and one-dimensional velocity dispersion $\sigma$ is given by the relation
\begin{equation}
 \Omega_{\textsc{k}}=\frac{3}{2}M\sigma^{2}
\label{eq:kinetic_energy}
\end{equation}
where $\sigma$ is the velocity dispersion for the mean gas particle, related to the velocity dispersion in the tracer molecule ($\sigma_{\textsc{n}_{2}\textsc{h}^{+}}$) by
\begin{equation}
\sigma^{2}=\sigma^{2}_{\textsc{n}_{2}\textsc{h}^{+}}+k_{\textsc{b}}T_{gas}\left(\frac{1}{\mu m_{\textsc{h}}}-\frac{1}{m_{\textsc{n}_{2}\textsc{h}^{+}}}\right)
\label{eq:velocity_correction}
\end{equation}
where $T_{gas}$ is the typical gas temperature of the material traced by \nh\ \citep[see][]{fuller1992}.  We assume that \nh\ traces material at $T_{gas}\approx 7\,$K \citep{stamatellos2007}.  We apply a similar correction to the \co\ linewidths, there taking $T_{gas}$ to be the mean line-of-sight temperature of the core.  However, as discussed below, \co\ linewidths are significantly supersonic, making the effect of this correction minimal.

\begin{figure}
\centering
\includegraphics[width=0.47\textwidth]{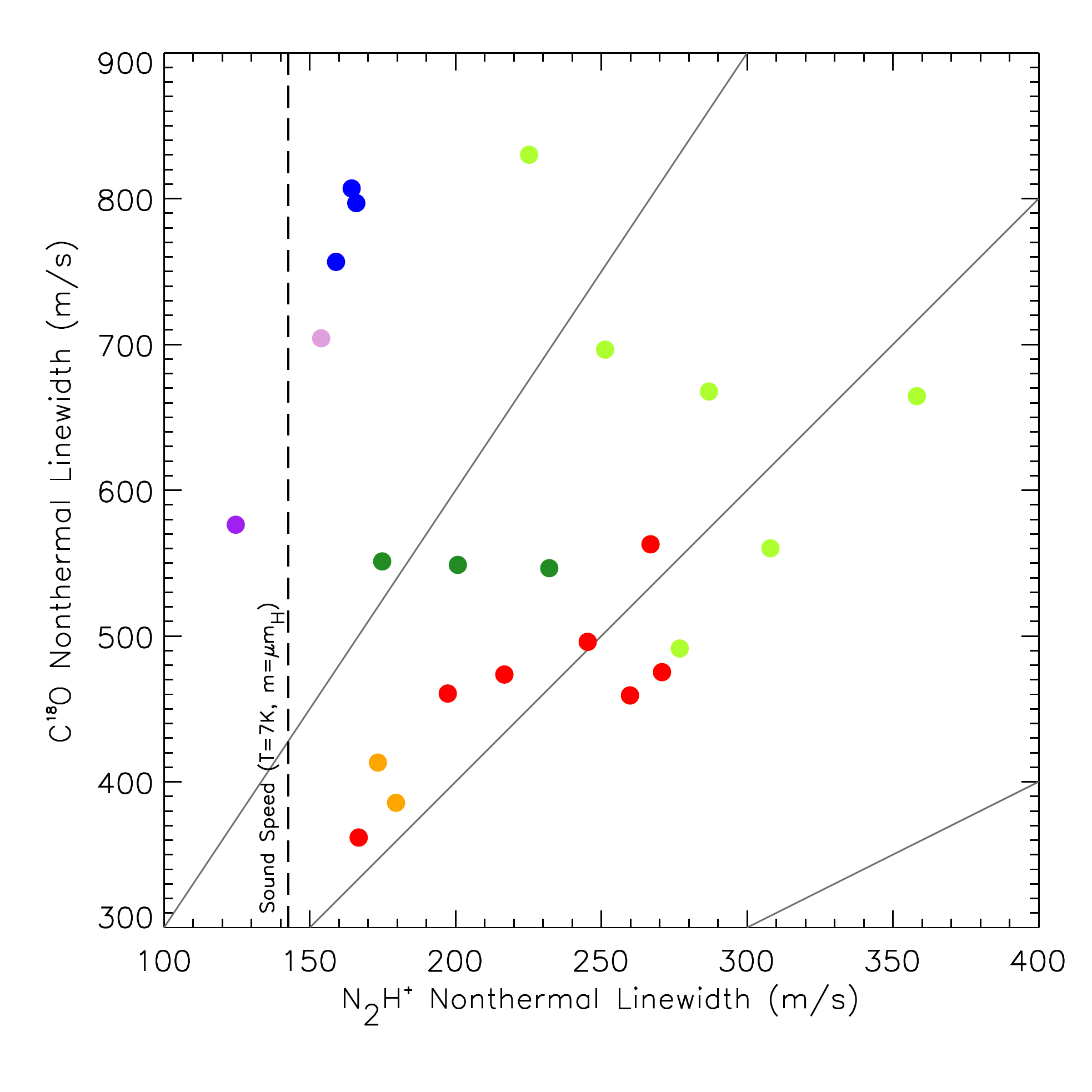}
\caption{Comparison of mean non-thermal linewidths for the 23 cores for which \nh\ data are available, as measured in \co\ and \nh.  The dashed line shows the mean gas sound speed at a temperature of 7\,K.  Grey lines show the 1:1, 2:1 and 3:1 \co:\nh\ linewidth ratios.  Colour coding is as in Figure~\ref{fig:m/r_plot}.}
\label{fig:therm_nontherm}
\end{figure}

The non-thermal component of the linewidth, $\sigma_{\textsc{nt}}$, can be derived using the gas temperature $T_{gas}$, and the relation $\sigma^{2}=\sigma^{2}_{\textsc{t}}+\sigma^{2}_{\textsc{nt}}$, where the sound speed, $\sigma_{\textsc{t}}$, is given by $\sqrt{k_{\textsc{b}}T_{gas}/m}$, and $m$ is the mass of the molecule being considered ($m_{\textsc{c}^{18}\textsc{o}}=30$ atomic mass units (amu); $m_{\textsc{n}_{2}\textsc{h}^{+}}=29$ amu).  Figure~\ref{fig:therm_nontherm} compares the non-thermal \nh\ and \co\ linewidths of our cores, with the sound speed in gas at 7\,K marked as a vertical line.

All of our cores have supersonic non-thermal velocity dispersions in \co.  The non-thermal velocity dispersions in \nh\ are consistently smaller than those measured in \co, typically being transonic or mildly supersonic.  This indicates a loss of turbulence between the material traced by \co\ and the denser material traced by \nh.  Transitions from supersonic turbulence at low densities to coherence at high densities have been observed in dense cores both in molecular clouds (e.g. \citealt{goodman1998}; \citealt{caselli2002}; \citealt{pineda2010}) and in isolation \citep{quinn2013}.  This behaviour is consistent with models of turbulent dissipation (e.g. \citealt{klessen2005}; \citealt{offner2008}).  The ratio between the non-thermal velocity dispersion in \co\ and the non-thermal velocity dispersion in \nh\ varies from region to region: in Oph B, $\sigma_{\textsc{nt}}($\co$)/\sigma_{\textsc{nt}}($\nh$)\sim 2.5$ while in Oph C, the ratio is $\sim 5$, suggesting that turbulence has been dissipated more in Oph C than in Oph B.

In keeping with the model used to characterise our sources, the gravitational potential energy is that of a spherically symmetric Gaussian density distribution, $\rho(r)=\rho_{0}{\rm e}^{-r^{2}/2\alpha^{2}}$ ($\alpha={\rm FWHM}/\sqrt{8{\rm ln}2}$):
\begin{equation}
  \Omega_{\textsc{g}}=-\frac{1}{2\sqrt{\pi}}\frac{GM^{2}}{\alpha}
\label{eq:gpe}
\end{equation}
(see Appendix C for a derivation).  We take $\alpha$ to be the geometric mean of the deconvolved Gaussian widths of each of our cores.

For our mean core mass, 0.27\,M\sun, and deconvolved size, FWHM\,$=0.01$\,pc, the gravitational potential energy $|\langle\Omega_{\textsc{g}}\rangle |\approx 4\times10^{41}$\,erg, and for our mean one-dimensional \nh\ velocity dispersion, 225\,ms$^{-1}$ (equivalent to $\sigma=262\,$ms$^{-1}$), the kinetic energy term in the virial equation is $2\langle\Omega_{\textsc{k}}\rangle\approx 11\times 10^{41}$\,erg.  Hence, these two terms are of similar order to one another, with the kinetic term slightly dominant.

\subsection{External gas pressure}
\label{sec:external_pressure_main}

Previous studies of starless cores in Ophiuchus have suggested that external gas pressure might be instrumental in confining dense cores.  \citet{maruta2010} estimate a typical surface pressure on cores in Ophiuchus of $\langle P_{\textsc{ext}}\rangle/k_{\textsc{b}}\approx 3\times 10^{6}\,$K\,cm$^{-2}$, sufficient to influence the energy balance of the cores.  Similarly, \citet{johnstone2000} estimate core surface pressures $P_{\textsc{ext}}/k_{\textsc{b}}\sim10^{6-7}\,$K\,cm$^{-3}$ by treating the starless cores they identify in Ophiuchus as pressure-confined Bonnor-Ebert spheres.

We consider the gas pressure in material traced by \co\ to be the external pressure acting on our starless cores, since CO becomes significantly depleted through freeze-out onto dust grains at densities $\gtrsim 10^{5}\,$cm$^{-3}$ \citep{PPVch2}, and as such is expected to trace the outer layers, or envelopes, of starless cores.  Higher-density tracers such as \nh\ are expected to trace the denser inner material of the cores themselves.

The external pressure term in the virial equation, $\Omega_{\textsc{p}}$, is given by
\begin{equation}
\Omega_{\textsc{p}}=-3P_{\textsc{ext}}V=-4\pi P_{\textsc{ext}}R^{3}
\label{eq:pressure_term}
\end{equation}
for a core of volume $V$ being acted on by an external pressure $P_{\textsc{ext}}$.  $P_{\textsc{ext}}$ can be estimated from the ideal gas law:
\begin{equation}
P_{\textsc{ext}}\approx\rho_{\textsc{c}^{18}\textsc{o}}\langle\sigma_{gas,\textsc{c}^{18}\textsc{o}}^{2}\rangle,
\label{eq:ideal_gas}
\end{equation}
where $\rho_{\textsc{c}^{18}\textsc{o}}$ is the density at which the transition between \co\ and \nh\ being effective tracers occurs, and $\langle\sigma_{gas,\textsc{c}^{18}\textsc{o}}^{2}\rangle$ is the mean gas velocity dispersion in material traced by \co.  We assume that \co\ does not trace densities higher than $10^{5}\,$cm$^{-3}$.  We must estimate a radius at which core density drops to $10^{5}\,$cm$^{-3}$ in order to determine the volume over which this surface pressure acts.  We continue to assume that our cores are characterised by Gaussian density distributions, in which case the radius at which the density drops to $\rho_{\textsc{c}^{18}\textsc{o}}$ is given by 
\begin{equation}
r_{\textsc{c}^{18}\textsc{o}}=\alpha\sqrt{2\,{\rm ln}\,\left(\frac{\rho_{0}}{\rho_{\textsc{c}^{18}\textsc{o}}}\right)}.
\label{eq:c18o_radius}
\end{equation}
The peak core density $\rho_{0}$ can be estimated from the measured mean density $\langle\rho_{\textsc{fwhm}}\rangle$ of each core (listed in Table~\ref{tab:continuum}), which is determined over an area of radius $1\times$FWHM:
\begin{equation}
\rho_{0}=\frac{\langle\rho_{\textsc{fwhm}}\rangle}{3}(8\,{\rm ln}\,2)^{3/2}\left(\sqrt{\frac{\pi}{2}}\,{\rm erf}\left(2\sqrt{\,{\rm ln}\,2}\right)-\frac{\sqrt{\,2{\rm ln}\,2}}{8}\right)^{-1}.
\label{eq:peak_density}
\end{equation}
These equations give typical $r_{\textsc{c}^{18}\textsc{o}}$ values in the range $\sim 0.7-1.5$ FWHM.

The mean energy due to external gas pressure on the material traced by \nh\ estimated from this method is $9\times 10^{41}\,$erg, roughly the same order of magnitude as the mean internal kinetic energy of our cores.  This is equivalent to $\langle P_{\textsc{ext}}\rangle/k_{\textsc{b}}\approx 1.8\times 10^{7}\,$K\,cm$^{-3}$, an order of magnitude higher pressure than that found by \citet{maruta2010}, but similar to the total pressure in Ophiuchus $P/k_{\textsc{b}}\sim 2\times 10^{7}\,$K\,cm$^{-3}$ estimated by \citet{johnstone2000}.

\subsection{External pressure from ionising photons}
\label{sec:external_pressure}

In Ophiuchus, the effects of the B2V star HD 147889 dominate the effects of the interstellar radiation field \citep{stamatellos2007}.  According to the cloud geometry model of \citet{liseau1999}, Oph A is the region of the cloud closest to HD 147889, at a distance of 1.1\,pc.  Furthermore, the B3-B5 star S1 appears to be influencing Oph A.  We estimate the pressure on cores in Oph A from ionising photons from these B stars, as being indicative of the maximum external pressure acting on any of the cores in our sample.

\begin{table}
\caption{Adopted B star properties}
\label{tab:b_stars}
\centering
\begin{tabular}{ccccc}
\hline
Star & Luminosity & T$_{{\rm eff}}$ & Radius & log$_{10}(\dot{N}^{0}_{{\rm LyC}})$ \\
 & (L\sun) & (K) & (R\sun) & (cm$^{-2}$s$^{-1}$) \\
\hline
HD 147889 & 4700 & 22300 & 4.6 & 20.4 \\
S1 & 1500 & 17200 & 4.4 & 18.5 \\
\hline
\end{tabular}
\end{table}

The pressure term of the virial equation due to ionising photons from an early-type star irradiating one side of a starless core is given by \citet{wardthompson2006} as
\begin{equation}
  \Omega_{\textsc{p}}\approx 2\pi R^{3}P_{{\rm ext}} \sim \frac{4R^{2}k_{\textsc{b}}T_{\textsc{ii}}}{D}\left(\frac{3\pi\dot{N}_{{\rm LyC}}R}{\alpha_{\ast}}\right)^{1/2}
  \label{pressure_term}
\end{equation}
where $R$ is the radius of the core; $D$ is the distance from the core to the exciting star; $T_{\textsc{ii}} \sim 10^{4}$\,K is the canonical temperature for gas in an H$_{{\textsc{ii}}}$ region; $\alpha_{\ast}\approx 2\times 10^{-13}$\,cm$^{3}$s$^{-1}$ is the recombination coefficient for atomic hydrogen into excited states at $T_{\textsc{ii}}$ and $\dot{N}_{{\rm LyC}}$ is the rate at which Lyman continuum photons are emitted from the exciting star.

We take the number of Lyman continuum photons emitted per unit surface area of the star, $\dot{N}^{0}_{{\rm LyC}}$, from \citet{dottori1980}, assuming in both cases log\,$g\sim 4.25$ \citep{strom1968}.  The total rate of ionising photons is then $\dot{N}_{{\rm LyC}}=4\pi R_{{\rm star}}^{2}\dot{N}^{0}_{{\rm LyC}}$, where the stellar radii are listed in Table~\ref{tab:b_stars}.  We take the distance to HD 147889 to be 1.1\,pc, and the distance to S1 to be 0.06\,pc, the plane-of-sky distance between the star S1 and the core SM1 at our assumed distance to Ophiuchus.  For a core radius equal to our mean deconvolved core FWHM, 0.01\,pc, the external pressure terms for a core in Oph A in close proximity to S1 will be
\begin{align}
\Omega_{\textsc{p,hd}} \sim\, & 3.4\times 10^{40} \,{\rm erg} \\
\Omega_{\textsc{p,s1}} \sim\, & 6.6\times 10^{40} \,{\rm erg} 
\label{eq:chis}
\end{align}
Hence, the maximum value we expect the ionising photon pressure term to take anywhere in Oph A is $\Omega_{\textsc{p}}\lesssim 10^{41}$\,erg, and outside Oph A, where the effect of HD 147889 will be lessened, and the effect of S1 will be minimal, we expect $\Omega_{\textsc{p}}\sim 10^{40}$\,erg.  Hence, we conclude that ionising photon pressure represents a small correction to the virial balance of our cores, typically being one to two orders of magnitude smaller than the gravitational and kinetic energy terms, and that we are justified in neglecting it in our virial analysis.

\subsection{Magnetic energy density}

Neither the strength nor the relative importance of magnetic fields in Ophiuchus are well known.  There have to date been only a few reliable measurements of magnetic fields in the cloud (\citealt{goodman1994}; \citealt{crutcher1993}; \citealt{troland1996}). The magnetic field at intermediate densities, measured through Zeeman splitting in OH \citep{crutcher1993,troland1996}, is what we consider in the subsequent analysis, as more representative of the magnetic field in the molecular gas.  \citet{troland1996} find the line of sight magnetic field strength $|B_{\rm los}|$ to be 10\,$\upmu$G at a density of $10^{3.2}\,$cm$^{-3}$, and find a 1D velocity dispersion in OH of $\sim 0.57\,$km\,s$^{-1}$.

\begin{table*}
\centering
\caption{Properties of starless cores, derived from spectral line data and from virial arguments and the Bonnor-Ebert criterion.}
\label{tab:line}
\setlength{\tabcolsep}{4pt}
\begin{tabular}{c c ccc ccc ccc c c c c}
\hline
Source & Source & \multicolumn{3}{c}{\nh\ Mass} & \multicolumn{3}{c}{\co\ Mass} & \multicolumn{3}{c}{Bonnor-Ebert Mass} & $-\Omega_{\textsc{g}}$ & $\Omega_{\textsc{k}}$ & $-\Omega_{\textsc{p}}$ & $\frac{1}{2}\ddot{\mathcal{I}}$\\
Index & Name & \multicolumn{3}{c}{(M$_{\odot}$)} & \multicolumn{3}{c}{(M$_{\odot}$)} & \multicolumn{3}{c}{(M$_{\odot}$)} & ($\times10^{41}$ erg) & ($\times10^{41}$ erg) & ($\times10^{41}$ erg) & ($\times10^{41}$ erg) \\
\hline
1 & SM1 & 0.184 & $\pm$ & 0.023 & 0.503 & $\pm$ & 0.086 & 0.261 & $\pm$ & 0.196 & 124.2 & 36.3 & 5.9 & $-57.4$ \\
2 & SM1N & 0.179 & $\pm$ & 0.020 & 0.353 & $\pm$ & 0.056 & 0.221 & $\pm$ & 0.168 & 82.4 & 27.3 & 5.9 & $-33.7$ \\
3 & SM2 & 0.201 & $\pm$ & 0.016 & 0.450 & $\pm$ & 0.057 & 0.308 & $\pm$ & 0.244 & 27.9 & 13.4 & 11.3 & $-12.4$ \\
5 & A-MM5 & 0.345 & $\pm$ & 0.048 & 0.511 & $\pm$ & 0.056 & 0.305 & $\pm$ & 0.242 & 2.7 & 5.2 & 10.1 & $-2.3$ \\
6 & A-MM6 & 0.391 & $\pm$ & 0.053 & 0.532 & $\pm$ & 0.061 & 0.297 & $\pm$ & 0.239 & 21.7 & 18.1 & 19.9 & $-5.5$ \\
7 & A-MM7 & 0.260 & $\pm$ & 0.013 & 0.354 & $\pm$ & 0.045 & 0.425 & $\pm$ & 0.374 & 3.1 & 6.9 & 8.0 & $+2.6$ \\
8 & A-MM8 & 0.174 & $\pm$ & 0.019 & 0.321 & $\pm$ & 0.042 & 0.387 & $\pm$ & 0.303 & 3.8 & 4.0 & 4.4 & $-0.3$ \\
9 & A-MM1 & \multicolumn{3}{c}{-} & 0.188 & $\pm$ & 0.024 & 0.180 & $\pm$ & 0.136 & 0.4 & - & 7.2 & - \\
10 & A-MM4 & 0.186 & $\pm$ & 0.035 & 0.343 & $\pm$ & 0.045 & 0.288 & $\pm$ & 0.207 & 0.6 & 1.7 & 3.0 & $-0.2$ \\
11 & A-MM4a & 0.088 & $\pm$ & 0.006 & 0.126 & $\pm$ & 0.022 & 0.255 & $\pm$ & 0.179 & 0.3 & 0.6 & 0.3 & $+0.7$ \\
19 & A-MM30 & \multicolumn{3}{c}{-} & 0.251 & $\pm$ & 0.038 & 0.151 & $\pm$ & 0.107 & 0.2 & - & 4.2 & - \\
20 & A-MM31 & \multicolumn{3}{c}{-} & 0.200 & $\pm$ & 0.023 & 0.580 & $\pm$ & 0.530 & 0.3 & - & 2.7 & - \\
23 & A-MM34 & \multicolumn{3}{c}{-} & 0.138 & $\pm$ & 0.017 & 0.624 & $\pm$ & 0.569 & 0.4 & - & 2.5 & - \\
26 & A2-MM1 & \multicolumn{3}{c}{-} & 0.132 & $\pm$ & 0.018 & 0.202 & $\pm$ & 0.145 & 0.2 & - & 3.0 & - \\
27 & A2-MM2 & \multicolumn{3}{c}{-} & 0.097 & $\pm$ & 0.017 & 0.143 & $\pm$ & 0.098 & 0.1 & - & 1.7 & - \\
28 & A3-MM1 & \multicolumn{3}{c}{-} & 0.187 & $\pm$ & 0.023 & 0.220 & $\pm$ & 0.169 & 0.3 & - & 6.6 & - \\
29 & B1-MM3 & 0.124 & $\pm$ & 0.019 & 0.263 & $\pm$ & 0.033 & 0.113 & $\pm$ & 0.067 & 3.5 & 4.1 & 10.8 & $-6.1$ \\
30 & B1-MM4a & 0.156 & $\pm$ & 0.012 & 0.217 & $\pm$ & 0.028 & 0.107 & $\pm$ & 0.062 & 4.2 & 6.5 & 10.7 & $-1.9$ \\
31 & B1-MM4b & 0.068 & $\pm$ & 0.010 & 0.105 & $\pm$ & 0.019 & 0.109 & $\pm$ & 0.064 & 0.4 & 1.1 & 1.2 & $+0.5$ \\
32 & B1-MM5 & \multicolumn{3}{c}{-} & 0.274 & $\pm$ & 0.038 & 0.121 & $\pm$ & 0.071 & 1.4 & - & 5.8 & - \\
35 & B2-MM2a & \multicolumn{3}{c}{-} & 0.177 & $\pm$ & 0.023 & 0.116 & $\pm$ & 0.066 & 1.5 & - & 6.0 & - \\
36 & B2-MM2b & \multicolumn{3}{c}{-} & 0.223 & $\pm$ & 0.028 & 0.092 & $\pm$ & 0.054 & 1.9 & - & 11.3 & - \\
37 & B2-MM4 & \multicolumn{3}{c}{-} & 0.103 & $\pm$ & 0.018 & 0.101 & $\pm$ & 0.060 & 3.0 & - & 0.8 & - \\
38 & B2-MM6 & 0.523 & $\pm$ & 0.296 & 0.220 & $\pm$ & 0.026 & 0.081 & $\pm$ & 0.047 & 13.6 & 25.0 & 25.8 & $+10.6$ \\
41 & B2-MM9 & 1.021 & $\pm$ & 0.394 & 0.240 & $\pm$ & 0.026 & 0.085 & $\pm$ & 0.051 & 13.7 & 18.6 & 33.8 & $-10.4$ \\
43 & B2-MM13 & 0.242 & $\pm$ & 0.049 & 0.213 & $\pm$ & 0.027 & 0.079 & $\pm$ & 0.043 & 18.9 & 21.4 & 15.2 & $+8.7$ \\
44 & B2-MM14 & 0.310 & $\pm$ & 0.044 & 0.319 & $\pm$ & 0.035 & 0.069 & $\pm$ & 0.038 & 23.6 & 19.7 & 40.7 & $-24.9$ \\
45 & B2-MM15 & 0.164 & $\pm$ & 0.020 & 0.196 & $\pm$ & 0.028 & 0.071 & $\pm$ & 0.042 & 6.8 & 7.3 & 20.3 & $-12.3$ \\
46 & B2-MM16 & 0.057 & $\pm$ & 0.005 & 0.100 & $\pm$ & 0.018 & 0.092 & $\pm$ & 0.050 & 10.5 & 7.3 & 0.7 & $+3.4$ \\
48 & C-MM3 & 0.157 & $\pm$ & 0.011 & 0.301 & $\pm$ & 0.038 & 0.084 & $\pm$ & 0.055 & 2.6 & 3.3 & 21.2 & $-17.2$ \\
49 & C-MM6a & 0.101 & $\pm$ & 0.011 & 0.195 & $\pm$ & 0.029 & 0.085 & $\pm$ & 0.058 & 0.4 & 1.1 & 7.1 & $-5.3$ \\
50 & C-MM6b & 0.148 & $\pm$ & 0.022 & 0.295 & $\pm$ & 0.037 & 0.091 & $\pm$ & 0.063 & 0.4 & 1.3 & 12.2 & $-10.0$ \\
53 & E-MM2d & 0.056 & $\pm$ & 0.004 & 0.218 & $\pm$ & 0.034 & 0.134 & $\pm$ & 0.088 & 1.2 & 1.6 & 7.2 & $-5.2$ \\
58 & F-MM1 & 0.027 & $\pm$ & 0.005 & 0.224 & $\pm$ & 0.041 & 0.139 & $\pm$ & 0.097 & 0.5 & 0.7 & 0.9 & $+0.0$ \\
70 & H-MM1 & \multicolumn{3}{c}{-} & 0.148 & $\pm$ & 0.018 & 0.223 & $\pm$ & 0.124 & 5.8 & - & 2.2 & - \\
\hline
\end{tabular}
\end{table*}

The magnetic field strength in the turbulent ISM is commonly related to the nonthermal velocity dispersion and density of the ISM \citep[see, e.g.][and references therein]{basu2000}:
\begin{equation}
B=B_{0}\frac{\sigma_{\textsc{nt}}}{\sigma_{\textsc{nt},0}}\left(\frac{n}{n_{0}}\right)^{1/2},
\label{eq:b_field}
\end{equation}
where the subscript `0' indicates the reference (measured) value of each quantity.  We note that this relation implies a constant ratio between turbulent and magnetic energy.  The magnetic energy can be expressed as
\begin{equation}
\Omega_{\textsc{m}}=\frac{B^{2}V}{2\mu_{0}}=\frac{1}{2\mu_{0}}\left(\frac{B_{0}^{2}}{\rho_{0}\sigma_{0,\textsc{nt}}^{2}}\right)M\sigma_{\textsc{nt}}^{2},
\label{eq:mag_energy}
\end{equation}
while the nonthermal component of the kinetic energy, $\Omega_{\textsc{k,nt}}$, is given by $1.5M\sigma_{\textsc{nt}}^{2}$ (see Equation~\ref{eq:kinetic_energy}).  Thus, the ratio between turbulent and magnetic energy is given by
\begin{equation}
\frac{\Omega_{\textsc{m}}}{\Omega_{\textsc{k,nt}}}=\frac{1}{3\mu_{0}}\frac{B_{0}^{2}}{\rho_{0}\sigma_{0,\textsc{nt}}^{2}},
\label{eq:mag_turb_ratio}
\end{equation}
which, for the values of $B_{0}$, $\rho_{0}$ and $\sigma_{0,\textsc{nt}}$ given by \citet{crutcher1993} and \citet{troland1996}, gives a ratio of $\Omega_{\textsc{m}}/\Omega_{\textsc{k,nt}}=0.11$ for the Ophiuchus molecular cloud.  Therefore, for our cores (if Equation~\ref{eq:b_field} holds) the magnetic energy of a core cannot exceed $\sim 10$\% of the core's internal energy.  In the case of transonic or subsonic motions within the core, the fraction will be smaller still.  Furthermore, the internal energy term in the virial equation is $2\Omega_{\textsc{k}}$, while the magnetic energy term is merely $\Omega_{\textsc{m}}$.  Consequently, the contribution of magnetic energy to core stability will in this case be $\sim 5$\% that of the turbulent kinetic energy.  Therefore, we also neglect the magnetic energy term in our virial analysis.  We note the need for further measurements of magnetic field strengths in high-density regions, in order to test the validity of analyses of this kind.

\subsection{Core stability}

\begin{figure}
\centering
\includegraphics[width=0.47\textwidth]{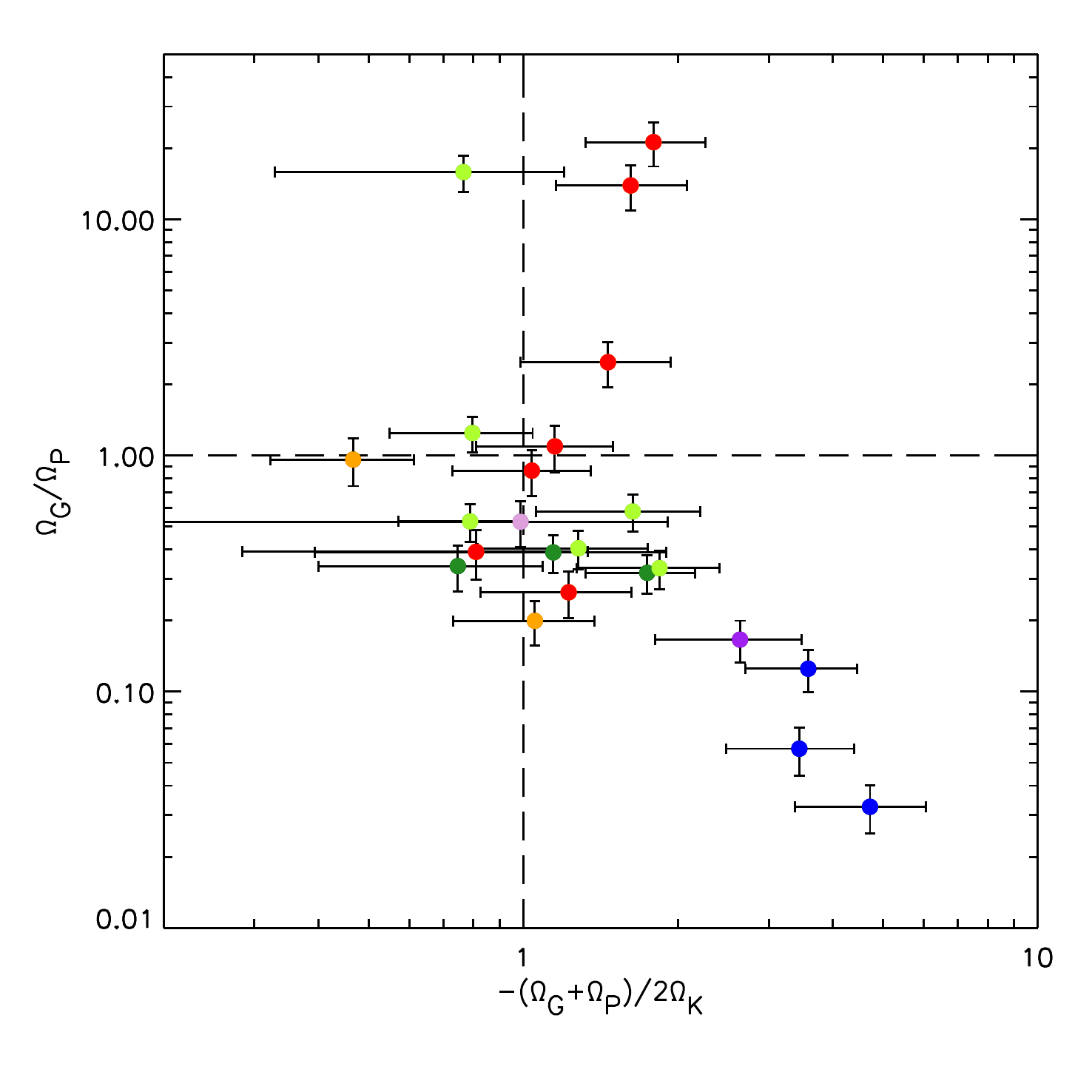}
\caption{Virial stability of the 23 cores in our catalogue for which \nh\ data are available, compared to the ratio of gravitational energy and external pressure terms in the virial equation.  The vertical dashed line indicates the line of virial stability, with the right-hand side of the plot being bound and the left-hand side being unbound.  The horizontal dashed line marks equipartition between external pressure energy and gravitational potential energy; cores above the line are gravitationally bound, while cores below the line are pressure-confined.  Colour coding is as in Figure~\ref{fig:m/r_plot}.}
\label{fig:energy_balance}
\end{figure}

On average, for those cores in our sample for which \nh\ and \co\ data are available, the gravitational potential energy and the external pressure energy are of similar magnitude, and together slightly dominate over the internal energy.  However, there is a wide variation from core to core.  Table~\ref{tab:line} lists the values of gravitational potential energy, internal energy, external pressure energy and the virial parameter for all those cores for which data are available.

Figure~\ref{fig:energy_balance} shows the ratio of $\Omega_{\textsc{g}}$ to $\Omega_{\textsc{p}}$ plotted against $-(\Omega_{\textsc{g}}+\Omega_{\textsc{p}})/2\Omega_{\textsc{k}}$, the virial stability criterion.  The vertical dashed line marks the locus of virial stability.  It can be seen that the majority of our cores lie to the right of this line, indicating that they are virially bound.  Of the 23 cores for which \nh\ data are available, 22 are either bound or virialised, having a virial ratio $-(\Omega_{\textsc{g}}+\Omega_{\textsc{p}})/2\Omega_{\textsc{k}}\geq 1$.  However, as can be seen in Figure~\ref{fig:energy_balance}, 1 core, in Oph A$^{\prime}$, is marginally unbound, with virial ratio $<$ 1, and with uncertainty on this ratio such that a ratio of 1 is consistent.

The horizontal dashed line on Figure~\ref{fig:energy_balance} marks the division between those cores that are gravitationally bound (above the line) and those that are pressure-confined (below the line).   There is a wide variation from region to region, with Oph A being the most gravitationally bound and Oph C being the most highly pressure-confined.  These differences are discussed further in Section~\ref{sec:regional}.  It should be noted that a full virial analysis has only been performed on those cores located in regions targeted for \nh\ observations, i.e. the regions of highest column density.  The results of this analysis cannot necessarily be generalised to the cores for which \nh\ data are not available.

\subsection{Bonnor-Ebert critical mass}

The Bonnor-Ebert (BE) model of a starless core \citep{bonnor1956,ebert1955} is frequently used as a measure of the stability of starless cores \citep[e.g.][]{alves2001}.  The BE model treats a core as an isothermal, self-gravitating, polytropic sphere bounded by external pressure.  The mass at which a BE sphere at temperature $T$, with sound speed $c_{s}(T)$, and bounded by external pressure $P_{\textsc{ext}}$, is critically stable against gravitational collapse is given by
\begin{equation}
M_{\textsc{BE},crit}=1.18\frac{c_{s}(T)^{4}}{P_{\textsc{ext}}^{1/2}G^{3/2}}.
\label{eq:BE}
\end{equation}
The critical BE mass is often considered a useful proxy for virial mass as, if the critically-stable BE model is appropriate and the radius at which cores are bounded by external pressure can be estimated, the stability of a core can be estimated without velocity dispersion data, as cores with observed masses greater than their critically-stable BE mass (i.e. $M/M_{\textsc{BE},crit}>1$) will be undergoing gravitational collapse. 

\begin{figure}
\centering
\includegraphics[width=0.47\textwidth]{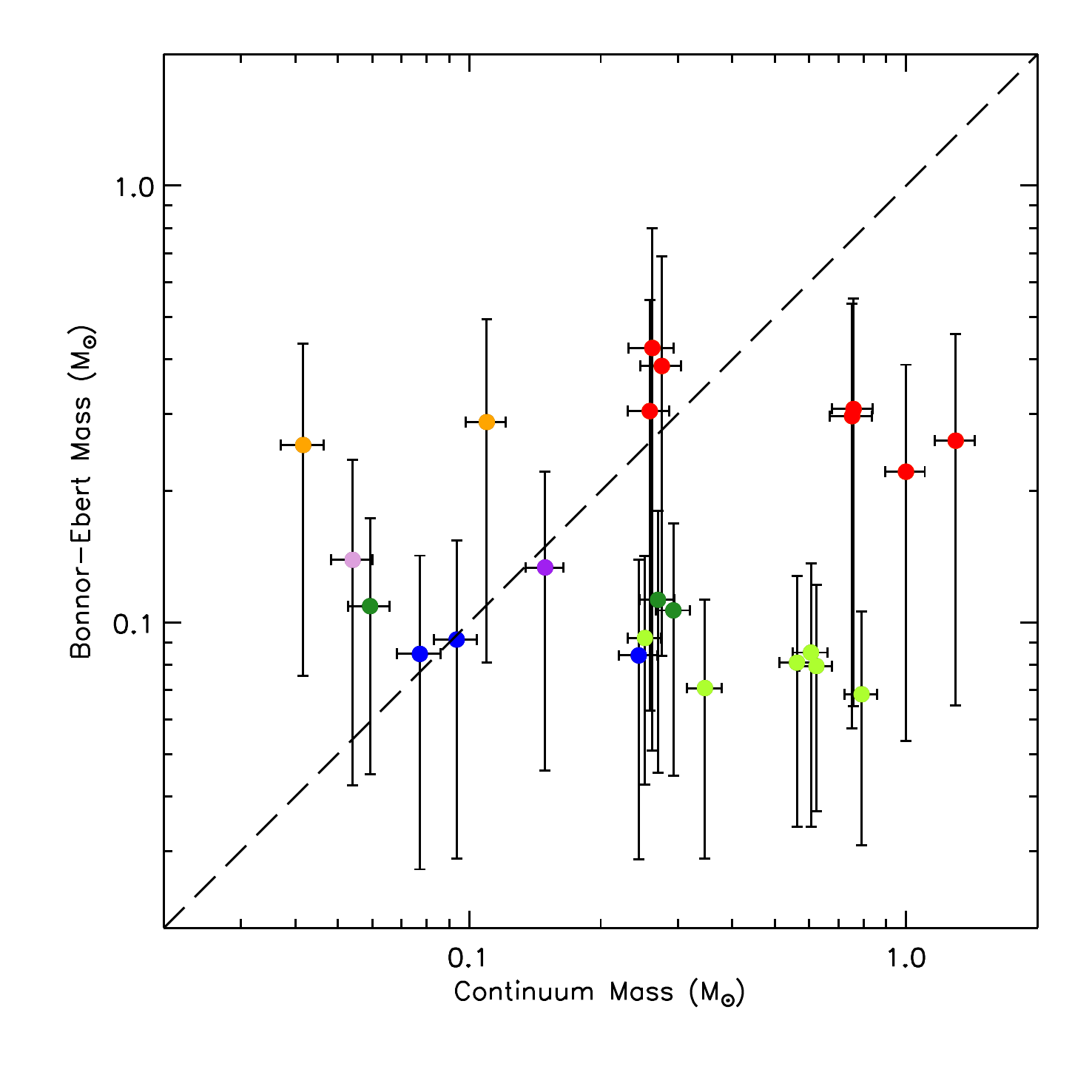}
\caption{Comparison of continuum-derived mass and Bonnor-Ebert critical mass for the 23 cores for which \nh\ data are available.  Cores to the right of the dashed line are collapsing according to the critical Bonner-Ebert criterion.  Colour coding is as in Figure~\ref{fig:m/r_plot}.}
\label{fig:be_plot}
\end{figure}

We investigated whether the critical BE stability criterion ($M/M_{\textsc{BE},crit}$) can accurately predict the virial balance of starless cores in L1688, and hence whether it can be reliably used as a proxy for virial mass in regions for which line data are not available.  We determined the critically stable masses of our cores by considering the external pressure, $P_{\textsc{ext}}$, on our cores to be the gas pressure in \co.  The critically stable BE masses and continuum masses of the subset of our cores for which \nh\ data are available are compared in Figure~\ref{fig:be_plot}.  Critically stable BE masses for the remainder of the set of cores for which \co\ data are are available are listed in Table~\ref{tab:line}, but are excluded from Figure~\ref{fig:be_plot} in order to aid comparison with Figure~\ref{fig:energy_balance}.

Figure~\ref{fig:be_plot} shows that there is no correlation between observed mass and critical BE mass, indicating that, as would be expected for a set of virially unstable cores, our cores cannot be modelled as static, critically-stable, BE spheres.  Moreover, the critical BE stability criterion does not reliably predict the either virially bound state or the energy balance of the \nh\ cores. A core lying to the right of the line of unity on Figure~\ref{fig:be_plot}  has no stable BE solution and must, according to BE analysis, be collapsing under its own gravity, while a core lying to the left of the line of unity may be modelled as a stable, pressure-confined BE sphere.

We find that the BE criterion typically over-predicts the degree to which our cores are gravitationally unstable. Of the 15 cores predicted to be collapsing under gravity according to Figure~\ref{fig:be_plot}, 9 are in fact found to pressure-confined. However, there are no cases where the BE analysis suggests a core is pressure-confined and it is found to in fact be gravitationally bound. The degree to which cores are virially bound is in many cases also overestimated. For example, the BE criterion predicts that all 6 of the cores in Oph B2 will be strongly gravitationally bound, whereas Figure~\ref{fig:energy_balance}  shows that of these 6 cores, 4 are approximately virialised, and the other 2, while virially bound, are confined by external pressure.

A possible explanation for these discrepancies is that in this analysis we have used the standard BE mass formula (Equation~\ref{eq:BE}), which does not account for the contribution of non-thermal motion to internal support. However, as shown in Figure~\ref{fig:therm_nontherm}, our cores typically have transonic or mildly supersonic internal motions at the radii traced by \nh, and hence assuming all support against collapse is thermal is likely to overestimate the degree to which our cores are both gravitationally unstable and virially bound. An accurate parameterisation of the effect of non-thermal internal motion on core support might improve the accuracy of the BE analysis.

Another important consideration is that while in principle the 8 cores lying to the left of the line of unity in Figure~\ref{fig:be_plot} can be modelled as stable, pressure-confined BE spheres, Figure~\ref{fig:energy_balance}  shows that many of our cores, whether confined by pressure or by gravity, are not in virial equilibrium. Caution must be exercised when applying an equilibrium model such as a BE sphere to a non-equilibrium set of objects such as the cores in this sample.


\section{Regional variations in core properties}
\label{sec:regional}

Figure~\ref{fig:energy_balance} shows that most of the cores in our sample for which \nh\ data are available are either bound or virialised.  Figure~\ref{fig:m/r_plot} shows that our cores occupy the part of the mass/size plane in which prestellar cores are expected to lie.  However, whether our cores are gravitationally bound (i.e. prestellar) or pressure-confined varies from region to region.  Gravity strongly dominates over external pressure in the most massive cores in Oph A, the well-known prestellar cores SM1, SM1N and SM2.  Cores in Oph A$^\prime$ and B are typically in approximate equipartition between gravitational and pressure energy or marginally pressure-dominated.  However, cores in Oph C and E are strongly pressure-dominated and virially bound.

It is noticeable from all of the above that the properties of the starless cores, including the degree to which cores are bound, as well as whether they are gravitationally bound or pressure-confined, and the extent to which turbulence is dissipated, varies more between regions than within them.  This suggests that the local environment has a significant effect on the nature of the starless cores.  \citet{enoch2009} provide a catalogue of deeply embedded Class 0 and Class I protostars in L1688 and L1689, marked as yellow stars on Figure~\ref{fig:850data}.  We refer to this catalogue in the following discussion.

\subsection{Oph A}
Oph A is the only region in L1688 within which substantially gravitationally bound cores are found (see Figure~\ref{fig:energy_balance}).  Temperatures in Oph A are higher than in other parts of the cloud.  The Oph A region is also the part of the cloud most clearly being influenced by stars that have already formed: the B2 protostar HD 147889 drives a PDR at the western edge of Oph A, while on the eastern side of Oph A there is a reflection nebula associated with the B4 protostar S1, both of which can be seen in Figure~\ref{fig:rgb}.  This suggests a morphology in the region in which the dense gas that makes up the central, submillimetre-bright cores of Oph A is being influenced by its local environment.  However, as shown in Figure~\ref{fig:energy_balance}, cores in the densest regions of Oph A do not appear to be dominated by external pressure.  \citet{enoch2009} list only one protostar embedded in Oph A: the Class 0 protostar VLA 1623 (the only Class 0 source in L1688).  This is consistent with star formation in this dense clump being in its early stages.
\subsection{Oph A$^{\prime}$}
The cores in Oph A$^{\prime}$ are at similar temperatures to those in Oph A, but are among the least bound of the cores in our sample.  Gravity and external pressure appear to be contributing approximately equally to the confinement of these cores.  This region is confused, particularly along its western edge, where much of the emission is from the PDR associated with HD 147889.  \citet{enoch2009} list three embedded Class I protostars in Oph A$^{\prime}$.
\subsection{Oph B}
The Oph B region appears to be relatively quiescent: it is the coldest of the regions; there are few embedded protostars; and the cores are typically virialised or marginally bound. \citet{enoch2009} list four embedded Class I protostars in Oph B: none in Oph B1; one in Oph B1B2; and three in Oph B2, of which one is the outflow-driving source IRS 47 \citep{white2015}.  Cores in Oph B1 and B2 typically show similar behaviour, although the ratio of gravitational to pressure energy is consistently in the range 0.3--0.4 in B1, and more varied in B2.  As shown in Figure~\ref{fig:therm_nontherm}, cores in B2 have the highest non-thermal linewidths measured in \nh, suggesting that turbulence is not being effectively dissipated in this region.  We hypothesise that this could be due to the influence of the outflow from IRS 47, as protostellar outflows have been shown to inject and sustain turbulence on small scales in molecular clouds \citep{duartecabral2012}.

We note that the pre-brown dwarf candidate Oph B-11 \citep{pound1995,greaves2003,andre2012}, located between Oph B1 and B2, is detected in SCUBA-2 850-\um\ emission.  Oph B-11 is discussed in detail in Appendix~D.

\subsection{Oph C}

\begin{figure*}
  \centering
  \includegraphics[width=0.7\textwidth]{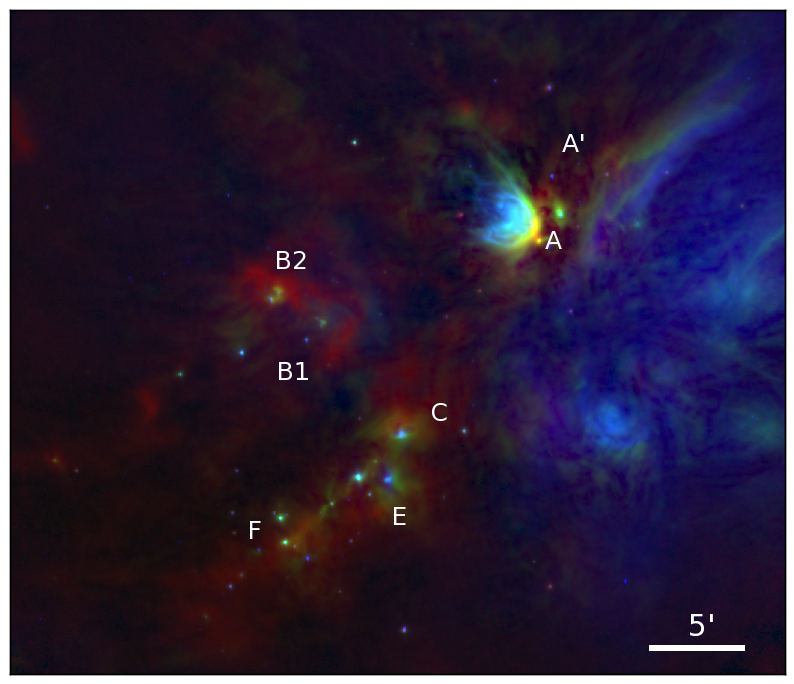}
  \caption{Three-colour image of L1688, with regions labelled.  Red channel: SCUBA-2 850-\um\ data (this work).  Green channel: \emph{Herschel} 100-\um\ data (Ladjalate et al. 2015).  Blue channel: \emph{Spitzer} 8-\um\ data (\citealt{evans2003}).}
  \label{fig:rgb3}
\end{figure*}

Oph C appears to be extremely quiescent, and substantially less evolved than the rest of the Oph C-E-F `filament' of which it appears to be a part.  The three cores we identify within Oph C are all substantially bound and pressure-confined, with broad \co\ linewidths, as shown in Figure~\ref{fig:therm_nontherm}.  The \nh\ linewidths, however, indicate that the cores in Oph C are among the least supersonic in \nh.  The reason for this apparently very effective dissipation of turbulence is not clear, although the lack of embedded sources driving outflows might be a possibility.  The lack of embedded sources in Oph C, along with its considerably lower aspect ratio than its neighbours Oph E and Oph F, lead us to suggest that Oph C may have a slightly different line-of-sight distance than other regions, possibly being further from influences such as HD 147889.  There are no embedded protostars in Oph C listed by \citet{enoch2009}.
\subsection{Oph E and F}
We consider Oph E and Oph F together, due to the low number of cores detected in these regions, along with the similarities between the two regions.  These appear to be the most evolved regions of L1688, with a high ratio of embedded sources to starless cores: Oph E has four embedded Class I sources, while Oph F has six.   Cores in Oph F are at a similar temperature to those in Oph A and A$^{\prime}$, although without any obvious external heating.  The core in Oph F for which an energy balance can be determined appears to be gravitationally bound, while the core in Oph E is pressure-confined.  \co\ linewidths show substantial turbulence, similarly to Oph C, while these cores are the least supersonic in \nh.  Again, we hypothesise that this effective dissipation of turbulence may be the result of a lack of outflows in either of these regions.
\subsection{L1689 and L1709}
The starless cores we find in L1689 and L1709 are typically of similar mass to those in Oph B, C and E.  We find six starless cores in L1689S; four in L1689; and one in L1709.  \citet{enoch2009} list four embedded Class I protostars in L1689S; one Class 0 source in L1689N; and two Class I sources in L1709.  The low number of cores relative to L1688, the low ratio of embedded sources to starless cores, and the presence of the Class 0 source IRAS 16293-2422 suggests that L1689 and L1709 are likely to be less evolved than, or forming stars less efficiently than, L1688.  This was explained by NWA06 as due to L1689 being further from the Sco OB2 association than L1688, and hence less active.

\subsection{Gradients across the cloud}

It is clear from the discussion above that the different regions of the L1688 cloud do not show the same properties or evolutionary stage, despite being in close proximity both to one another and to HD 147889.  There is a marked variation in temperature across the cloud, with Oph A and A$^{\prime}$ being the warmest regions, followed by Oph F, E, C, B1, and B2, in that order.  Oph A and A$^{\prime}$ are clearly being influenced by the nearby B stars.  As discussed in Section~\ref{sec:external_pressure}, the flux of ionising photons from the two B stars is not a dominant term in the virial equation in Oph A.  However, these stars will be heating the gas and dust within Oph A.

Figure~\ref{fig:rgb3} shows in blue the warm dust traced by \emph{Spitzer} 8-\um\ emission (\citealt{evans2003}; \citealt{enoch2009}), which surrounds Oph A and A$^{\prime}$ on two sides.  It should be noted that while the relative influence of HD 147889 on L1688 as a whole must be much greater than that of S1, the flux of ionising photons from S1 on Oph A is approximately twice that of HD 147889; the S1 reflection nebula is likely to have at least as much influence on Oph A as the PDR driven by HD 147889, even though the former is much smaller.

Oph A and Oph B appear to be at similar evolutionary stages, despite their marked difference in temperature.  Both regions have embedded sources driving outflows, which may be hindering the dissipation of turbulence within the region.  However, while Oph A shows the influence of local effects, Oph B appears to be evolving in a more quiescent location: it is the coldest of the regions, and Figure~\ref{fig:rgb3} shows no sign of it being bordered by PDRs or reflection nebulae.

Cores in Oph A and Oph B are typically of similar mass (see Figure~\ref{fig:mass_comparison}).  However, as shown in Figure~\ref{fig:energy_balance}, while some of the cores in Oph A are strongly gravitationally dominated, the cores in Oph B are close to equipartition between gravitational potential energy and pressure.  It is possible that material in Oph A might have been swept up by the PDR and the reflection nebula, increasing local density and hence leading to the strongly gravitationally bound prestellar cores in this region.

Oph E and F appear to be at a later evolutionary stage than Oph A and B, with a high ratio of protostars to starless cores, several embedded sources, and no embedded sources young enough to be driving outflows.  Those starless cores that are found are among the least massive in L1688 (see Figure~\ref{fig:mass_comparison}).  These regions are both at an intermediate temperature.  There is no obvious source of external heating, similarly to Oph B, suggesting that the embedded sources in Oph E and F might be heating their surroundings.  What might have led these regions to begin forming stars earlier than Oph A and B is not clear.

Oph C is noticeably different from the other regions in L1688, being an apparently entirely quiescent region, with only a few low-mass, pressure-confined cores and no embedded sources.  As discussed above, this leads us to suggest that Oph C might be at a slightly different line-of-sight distance than the neighbouring regions.

There appears to be a general gradient in evolutionary stage from southwest to northeast across the cloud (except for Oph C).  This could be due to the influence of the Sco OB2 association, located behind and to the southwest of Ophiuchus \citep{mamajek2008}; HD 147889, also behind Ophiuchus \citep{liseau1999}, appears to be primarily of importance in Oph A, and to have relatively limited influence elsewhere.

While a global southwest/northeast gradient in evolutionary stage can be inferred, and is consistent with previous studies (\citealt{loren1989a}; NWA06), it must be emphasised that the properties of regions within L1688 appear to be determined substantially by local effects.  In particular, the differences in temperature and energy balance between cores in Oph A and Oph B, two regions apparently at similar evolutionary stages, but with different immediate local environments (Oph A being heavily influenced by two B stars, and Oph B evolving in a less disturbed location), indicate the importance of local effects in determining the properties of starless cores.

\section{Conclusions}

In this paper, we have extracted a set of sources from the SCUBA-2 850-\um\ map of the Ophiuchus molecular cloud, and have characterised the properties of these cores using SCUBA-2, \emph{Herschel}, IRAM and HARP data sets.

We identified sources using the CuTEx curvature-based soure extraction algorithm, which gave us a catalogue of 93 sources, 70 of which were in the central region of the L1688 sub-cloud.  Of these 93 sources, 46 were identified as protostellar, and 47 were identified as starless cores.  Of the 70 sources in L1688, 47 were uniquely identified with a source in the S08 catalogue.

We determined the dust temperature of each source by SED fitting, which allowed an accurate mass determination to be made for each source.  The distribution of masses of the starless cores is consistent with the expected shape of the core mass function.  The low counting statistics of our sample did not allow us to accurately determine the power-law index of our core mass function, although the two slope values determined, $\alpha=2.0\pm0.4$ and $\alpha=2.7\pm0.4$ are both consistent with the expected behaviour of the high-mass Initial Mass Function.

We calculated the masses of our cores from \nh\ and \co\ emission.   We found that the mass of a core determined from 850-\um\ continuum emission and the mass determined from \nh\ emission correlate well, indicating that \nh\ and continuum emission are tracing the same material.  The most massive cores, those in Oph A, have consistently higher continuum masses than \nh\ masses, indicating that as expected, \nh\ emission does not trace the very densest material in prestellar cores.

We performed full virial stability analyses for the 23 cores for which both \co\ and \nh\ data were available, estimating the contributions of gravitational energy, internal pressure (both thermal and non-thermal) and external pressure to the energy balance of the cores.  Existing measurements of the magnetic field strength in Ophiuchus suggest that magnetic energy is unlikely to significantly alter the energy balance of our cores.  We found that most of our cores are bound or virialised, with a virial ratio $\geq\,1$.

We calculated the Bonnor-Ebert critically stable masses for each of the 23 cores for which \nh\ data are available.  We found that our cores cannot be modelled as critically stable Bonnor-Ebert spheres, and that the Bonnor-Ebert critically stable mass is not a good estimator of the bound state of the cores for which we can perform a full virial analysis, typically overestimating the degree to which cores are gravitationally bound.

We found that whether our cores are gravitationally bound or pressure confined depends strongly on the region in which they are located.  Cores in the centre of Oph A are gravitationally bound, while cores in Oph C and E are pressure-confined.  Cores in Oph A$^{\prime}$, B and F are in approximate equipartition between gravitational potential energy and external pressure energy, with pressure typically slightly dominating.

We see a loss of turbulence between core linewidths measured in \co\ and core linewidths measured in \nh.  This supports a picture in which dissipation of turbulence occurs in the dense centres of starless cores.  At the radii traced by \nh\ emission, turbulence is dissipating, but is not yet fully dissipated, with a transonic or mildly supersonic non-thermal component to the core linewidth still present even when the core is on the brink of gravitational collapse.  The degree to which which turbulence is dissipated varies between regions, with turbulence being dissipated more within Oph C, E and F than within Oph A, A$^{\prime}$ and B.

These results show that starless cores in the Ophiuchus molecular cloud are non-equilibrium objects with complex relationships with their local environments, and that a detailed analysis of their energy balance, of the sort we have carried out here, is required in order to accurately determine their virial state.  In particular, we have shown that external pressure is of key importance to the energy balance of most of the densest starless cores in Ophiuchus, and thus cannot be neglected in a virial analysis.  The wealth of continuum and kinematic data now available for many galactic star-forming regions now allows for detailed analyses of the virial balance of starless cores in other regions to be performed, and a thorough understanding of their behaviour and relationship with their environments to be developed.  In future papers we will carry out such studies in other Gould Belt star-forming regions.

\section{Acknowledgements}

K.P. wishes to thank STFC for studentship support while this research was carried out.  The James Clerk Maxwell Telescope has historically been operated by the Joint Astronomy Centre on behalf of the Science and Technology Facilities Council of the United Kingdom, the National Research Council of Canada and the Netherlands Organisation for Scientific Research.  Additional funds for the construction of SCUBA-2 were provided by the Canada Foundation for Innovation.  \emph{Herschel} is an ESA space observatory with science instruments provided by European-led Principal Investigator consortia and with important participation from NASA.  IRAM is supported by INSU/CNRS (France), MPG (Germany) and IGN (Spain).

\bibliographystyle{mn2e_fix}

\begin{thebibliography}{102}
\expandafter\ifx\csname natexlab\endcsname\relax\def\natexlab#1{#1}\fi

\bibitem[{{Alves}, {Lada} \& {Lada}(2001){Alves}, {Lada}, \&
  {Lada}}]{alves2001}
{Alves} J.~F., {Lada} C.~J., {Lada} E.~A., 2001, Nature, 409, 159

\bibitem[{{Andr{\'e}} {et~al}\mbox{.}(2007){Andr{\'e}}, {Belloche}, {Motte}, \&
  {Peretto}}]{andre2007}
{Andr{\'e}} P., {Belloche} A., {Motte} F., {Peretto} N., 2007, A\&A, 472, 519

\bibitem[{{Andr{\'e}} {et~al}\mbox{.}(2014){Andr{\'e}}, {Di Francesco},
  {Ward-Thompson}, {Inutsuka}, {Pudritz}, \& {Pineda}}]{PPVI}
{Andr{\'e}} P., {Di Francesco} J., {Ward-Thompson} D., {Inutsuka} S.-I.,
  {Pudritz} R.~E., {Pineda} J., 2014, Protostars and Planets VI, 27

\bibitem[{{Andr{\'e}} {et~al}\mbox{.}(2010){Andr{\'e}}, {Men'shchikov},
  {Bontemps}, {K{\"o}nyves}, {Motte}, {Schneider}, {Didelon}, {Minier},
  {Saraceno}, {Ward-Thompson}, {Di Francesco}, {White}, {Molinari}, {Testi},
  {Abergel}, {Griffin}, {Henning}, {Royer}, {Mer{\'{\i}}n}, {Vavrek}, {Attard},
  {Arzoumanian}, {Wilson}, {Ade}, {Aussel}, {Baluteau}, {Benedettini},
  {Bernard}, {Blommaert}, {Cambr{\'e}sy}, {Cox}, {di Giorgio}, {Hargrave},
  {Hennemann}, {Huang}, {Kirk}, {Krause}, {Launhardt}, {Leeks}, {Le Pennec},
  {Li}, {Martin}, {Maury}, {Olofsson}, {Omont}, {Peretto}, {Pezzuto}, {Prusti},
  {Roussel}, {Russeil}, {Sauvage}, {Sibthorpe}, {Sicilia-Aguilar}, {Spinoglio},
  {Waelkens}, {Woodcraft}, \& {Zavagno}}]{andre2010}
{Andr{\'e}} P. {et~al.}, 2010, {A\&A}, 518, {L102}

\bibitem[{{Andr\'{e}} \& {Montmerle}(1994)}]{andre1994}
{Andr\'{e}} P., {Montmerle} T., 1994, ApJ, 420, 837

\bibitem[{Andr\'{e}, {Ward-Thompson} \& Barsony(1993)Andr\'{e},
  {Ward-Thompson}, \& Barsony}]{andre1993}
Andr\'{e} P., {Ward-Thompson} D., Barsony M., 1993, {ApJ}, 406, 122

\bibitem[{{Andr{\'e}}, {Ward-Thompson} \& {Greaves}(2012){Andr{\'e}},
  {Ward-Thompson}, \& {Greaves}}]{andre2012}
{Andr{\'e}} P., {Ward-Thompson} D., {Greaves} J., 2012, Science, 337, 69

\bibitem[{{Aniano} {et~al}\mbox{.}(2011){Aniano}, {Draine}, {Gordon}, \&
  {Sandstrom}}]{aniano2011}
{Aniano} G., {Draine} B.~T., {Gordon} K.~D., {Sandstrom} K., 2011, PASP, 123,
  1218

\bibitem[{{Basu}(2000)}]{basu2000}
{Basu} S., 2000, ApJL, 540, L103

\bibitem[{{Beckwith} {et~al}\mbox{.}(1990){Beckwith}, {Sargent}, {Chini}, \&
  {Guesten}}]{beckwith1990}
{Beckwith} S.~V.~W., {Sargent} A.~I., {Chini} R.~S., {Guesten} R., 1990, AJ,
  99, 924

\bibitem[{{Bintley} {et~al}\mbox{.}(2014){Bintley}, {Holland}, {MacIntosh},
  {Friberg}, {Bell}, {Berke}, {Berry}, {Berthold}, {Cookson}, {Coulson},
  {Currie}, {Dempsey}, {Gibb}, {Gorges}, {Graves}, {Jenness}, {Johnstone},
  {Parsons}, {Thomas}, {Walther}, \& {Wouterloot}}]{bintley2014}
{Bintley} D. {et~al.}, 2014, in Society of Photo-Optical Instrumentation
  Engineers (SPIE) Conference Series, Vol. 9153, {Millimeter, Submillimeter,
  and Far-Infrared Detectors and Instrumentation for Astronomy VII}, {Holland}
  W.~S., {Zmuidzinas} J., eds., p. 915303

\bibitem[{{Bonnor}(1956)}]{bonnor1956}
{Bonnor} W.~B., 1956, MNRAS, 116, 351

\bibitem[{{Bontemps} {et~al}\mbox{.}(2001){Bontemps}, {Andr{\'e}}, {Kaas},
  {Nordh}, {Olofsson}, {Huldtgren}, {Abergel}, {Blommaert}, {Boulanger},
  {Burgdorf}, {Cesarsky}, {Cesarsky}, {Copet}, {Davies}, {Falgarone},
  {Lagache}, {Montmerle}, {P{\'e}rault}, {Persi}, {Prusti}, {Puget}, \&
  {Sibille}}]{bontemps2001}
{Bontemps} S. {et~al.}, 2001, A\&A, 372, 173

\bibitem[{{Buckle} {et~al}\mbox{.}(2010){Buckle}, {Curtis}, {Roberts}, {White},
  {Hatchell}, {Brunt}, {Butner}, {Cavanagh}, {Chrysostomou}, {Davis},
  {Duarte-Cabral}, {Etxaluze}, {Di Francesco}, {Friberg}, {Friesen}, {Fuller},
  {Graves}, {Greaves}, {Hogerheijde}, {Johnstone}, {Matthews}, {Matthews},
  {Nutter}, {Rawlings}, {Richer}, {Sadavoy}, {Simpson}, {Tothill}, {Tsamis},
  {Viti}, {Ward-Thompson}, {Wouterloot}, \& {Yates}}]{buckle2010}
{Buckle} J.~V. {et~al.}, 2010, MNRAS, 401, 204

\bibitem[{{Buckle} {et~al}\mbox{.}(2009){Buckle}, {Hills}, {Smith}, {Dent},
  {Bell}, {Curtis}, {Dace}, {Gibson}, {Graves}, {Leech}, {Richer},
  {Williamson}, {Withington}, {Yassin}, {Bennett}, {Hastings}, {Laidlaw},
  {Lightfoot}, {Burgess}, {Dewdney}, {Hovey}, {Willis}, {Redman}, {Wooff},
  {Berry}, {Cavanagh}, {Davis}, {Dempsey}, {Friberg}, {Jenness}, {Kackley},
  {Rees}, {Tilanus}, {Walther}, {Zwart}, {Klapwijk}, {Kroug}, \&
  {Zijlstra}}]{buckle2009}
{Buckle} J.~V. {et~al.}, 2009, {MNRAS}, 399, 1026

\bibitem[{{Caselli} {et~al}\mbox{.}(2002){Caselli}, {Walmsley}, {Zucconi},
  {Tafalla}, {Dore}, \& {Myers}}]{caselli2002}
{Caselli} P., {Walmsley} C.~M., {Zucconi} A., {Tafalla} M., {Dore} L., {Myers}
  P.~C., 2002, ApJ, 565, 344

\bibitem[{Chabrier(2003)}]{chabrier2003}
Chabrier G., 2003, {PASP}, 115, 763

\bibitem[{{Chapin} {et~al}\mbox{.}(2013){Chapin}, {Berry}, {Gibb}, {Jenness},
  {Scott}, {Tilanus}, {Economou}, \& {Holland}}]{chapin2013}
{Chapin} E.~L., {Berry} D.~S., {Gibb} A.~G., {Jenness} T., {Scott} D.,
  {Tilanus} R.~P.~J., {Economou} F., {Holland} W.~S., 2013, MNRAS, 430, 2545

\bibitem[{{Christie} {et~al}\mbox{.}(2012){Christie}, {Viti}, {Yates},
  {Hatchell}, {Fuller}, {Duarte-Cabral}, {Sadavoy}, {Buckle}, {Graves},
  {Roberts}, {Nutter}, {Davis}, {White}, {Hogerheijde}, {Ward-Thompson},
  {Butner}, {Richer}, \& {Di Francesco}}]{christie2012}
{Christie} H. {et~al.}, 2012, MNRAS, 422, 968

\bibitem[{{Crutcher} {et~al}\mbox{.}(1993){Crutcher}, {Troland}, {Goodman},
  {Heiles}, {Kazes}, \& {Myers}}]{crutcher1993}
{Crutcher} R.~M., {Troland} T.~H., {Goodman} A.~A., {Heiles} C., {Kazes} I.,
  {Myers} P.~C., 1993, ApJ, 407, 175

\bibitem[{{Daniel}, {Cernicharo} \& {Dubernet}(2006){Daniel}, {Cernicharo}, \&
  {Dubernet}}]{daniel2006}
{Daniel} F., {Cernicharo} J., {Dubernet} M.-L., 2006, ApJ, 648, 461

\bibitem[{{Dempsey} {et~al}\mbox{.}(2013){Dempsey}, {Friberg}, {Jenness},
  {Tilanus}, {Thomas}, {Holland}, {Bintley}, {Berry}, {Chapin}, {Chrysostomou},
  {Davis}, {Gibb}, {Parsons}, \& {Robson}}]{dempsey2013}
{Dempsey} J.~T. {et~al.}, 2013, MNRAS, 430, 2534

\bibitem[{{Di Francesco}, {Andr{\'e}} \& {Myers}(2004){Di Francesco},
  {Andr{\'e}}, \& {Myers}}]{difrancesco2004}
{Di Francesco} J., {Andr{\'e}} P., {Myers} P.~C., 2004, ApJ, 617, 425

\bibitem[{{Di Francesco} {et~al}\mbox{.}(2007){Di Francesco}, {Evans},
  {Caselli}, {Myers}, {Shirley}, {Aikawa}, \& {Tafalla}}]{PPVch2}
{Di Francesco} J., {Evans} N. J.~I., {Caselli} P., {Myers} P.~C., {Shirley} Y.,
  {Aikawa} Y., {Tafalla} M., 2007, in {Protostars and Planets V}, {Reipurth}
  B., {Jewitt} D., {Tucson} K.~K., eds., {University of Arizona Press}, pp.
  {17--32}

\bibitem[{{Dolidze} \& {Arakelyan}(1959)}]{dolidze1959}
{Dolidze} M.~V., {Arakelyan} M.~A., 1959, AZH, 36, 444

\bibitem[{{Dottori}(1980)}]{dottori1980}
{Dottori} H.~A., 1980, Ap\&SS, 73, 175

\bibitem[{{Drabek} {et~al}\mbox{.}(2012){Drabek}, {Hatchell}, {Friberg},
  {Richer}, {Graves}, {Buckle}, {Nutter}, {Johnstone}, \& {Di
  Francesco}}]{drabek2012}
{Drabek} E. {et~al.}, 2012, {MNRAS}, 426, 23

\bibitem[{{Duarte-Cabral} {et~al}\mbox{.}(2012){Duarte-Cabral}, {Chrysostomou},
  {Peretto}, {Fuller}, {Matthews}, {Schieven}, \& {Davis}}]{duartecabral2012}
{Duarte-Cabral} A., {Chrysostomou} A., {Peretto} N., {Fuller} G.~A., {Matthews}
  B., {Schieven} G., {Davis} G.~R., 2012, A\&A, 543, A140

\bibitem[{{Ebert}(1955)}]{ebert1955}
{Ebert} R., 1955, Z.Ap., 37, 217

\bibitem[{{Elmegreen} \& {Falgarone}(1996)}]{elmegreen1996}
{Elmegreen} B.~G., {Falgarone} E., 1996, ApJ, 471, 816

\bibitem[{{Emerson}(1999)}]{emerson1999}
{Emerson} D., 1999, {Interpreting Astronomical Spectra}. Wiley-VCH

\bibitem[{Enoch {et~al}\mbox{.}(2008)Enoch, Evans, Sargent, Glenn, Rosolowsky,
  \& Myers}]{enoch2008}
Enoch M.~L., Evans, N.~J. I., Sargent A.~I., Glenn J., Rosolowsky E., Myers P.,
  2008, {ApJ}, 684, 1240

\bibitem[{{Enoch} {et~al}\mbox{.}(2009){Enoch}, {Evans}, {Sargent}, \&
  {Glenn}}]{enoch2009}
{Enoch} M.~L., {Evans}, II N.~J., {Sargent} A.~I., {Glenn} J., 2009, ApJ, 692,
  973

\bibitem[{{Evans} {et~al}\mbox{.}(2003){Evans}, {Allen}, {Blake}, {Boogert},
  {Bourke}, {Harvey}, {Kessler}, {Koerner}, {Lee}, {Mundy}, {Myers}, {Padgett},
  {Pontoppidan}, {Sargent}, {Stapelfeldt}, {van Dishoeck}, {Young}, \&
  {Young}}]{evans2003}
{Evans}, II N.~J. {et~al.}, 2003, PASP, 115, 965

\bibitem[{{Evans} {et~al}\mbox{.}(2009){Evans}, {Dunham}, {J{\o}rgensen},
  {Enoch}, {Mer{\'{\i}}n}, {van Dishoeck}, {Alcal{\'a}}, {Myers},
  {Stapelfeldt}, {Huard}, {Allen}, {Harvey}, {van Kempen}, {Blake}, {Koerner},
  {Mundy}, {Padgett}, \& {Sargent}}]{evans2009}
{Evans}, II N.~J. {et~al.}, 2009, ApJS, 181, 321

\bibitem[{{Frerking}, {Langer} \& {Wilson}(1982){Frerking}, {Langer}, \&
  {Wilson}}]{frerking1982}
{Frerking} M.~A., {Langer} W.~D., {Wilson} R.~W., 1982, ApJ, 262, 590

\bibitem[{{Friesen} {et~al}\mbox{.}(2010){Friesen}, {Di Francesco},
  {Shimajiri}, \& {Takakuwa}}]{friesen2010}
{Friesen} R.~K., {Di Francesco} J., {Shimajiri} Y., {Takakuwa} S., 2010, ApJ,
  708, 1002

\bibitem[{{Fuller} \& {Myers}(1992)}]{fuller1992}
{Fuller} G.~A., {Myers} P.~C., 1992, ApJ, 384, 523

\bibitem[{{Garden} {et~al}\mbox{.}(1991){Garden}, {Hayashi}, {Hasegawa},
  {Gatley}, \& {Kaifu}}]{garden1991}
{Garden} R.~P., {Hayashi} M., {Hasegawa} T., {Gatley} I., {Kaifu} N., 1991,
  ApJ, 374, 540

\bibitem[{{Goodman} {et~al}\mbox{.}(1998){Goodman}, {Barranco}, {Wilner}, \&
  {Heyer}}]{goodman1998}
{Goodman} A.~A., {Barranco} J.~A., {Wilner} D.~J., {Heyer} M.~H., 1998, ApJ,
  504, 223

\bibitem[{{Goodman} \& {Heiles}(1994)}]{goodman1994}
{Goodman} A.~A., {Heiles} C., 1994, ApJ, 424, 208

\bibitem[{{Gould}(1879)}]{gould1879}
{Gould} B.~G., 1879, Resultados del Observatorio Nacional Argentino, 1, 0

\bibitem[{{Greaves}, {Holland} \& {Pound}(2003){Greaves}, {Holland}, \&
  {Pound}}]{greaves2003}
{Greaves} J.~S., {Holland} W.~S., {Pound} M.~W., 2003, MNRAS, 346, 441

\bibitem[{{Greene} {et~al}\mbox{.}(1994){Greene}, {Wilking}, {Andr\'{e}},
  {Young}, \& {Lada}}]{greene1994}
{Greene} T.~P., {Wilking} B.~A., {Andr\'{e}} P., {Young} E.~T., {Lada} C.~J.,
  1994, ApJ, 434, 614

\bibitem[{{Greene} \& {Young}(1992)}]{greene1992}
{Greene} T.~P., {Young} E.~T., 1992, ApJ, 395, 516

\bibitem[{{Griffin} {et~al}\mbox{.}(2010){Griffin}, {Abergel}, {Abreu}, {Ade},
  {Andr{\'e}}, {Augueres}, {Babbedge}, {Bae}, {Baillie}, {Baluteau}, {Barlow},
  {Bendo}, {Benielli}, {Bock}, {Bonhomme}, {Brisbin}, {Brockley-Blatt},
  {Caldwell}, {Cara}, {Castro-Rodriguez}, {Cerulli}, {Chanial}, {Chen},
  {Clark}, {Clements}, {Clerc}, {Coker}, {Communal}, {Conversi}, {Cox},
  {Crumb}, {Cunningham}, {Daly}, {Davis}, {de Antoni}, {Delderfield}, {Devin},
  {di Giorgio}, {Didschuns}, {Dohlen}, {Donati}, {Dowell}, {Dowell}, {Duband},
  {Dumaye}, {Emery}, {Ferlet}, {Ferrand}, {Fontignie}, {Fox}, {Franceschini},
  {Frerking}, {Fulton}, {Garcia}, {Gastaud}, {Gear}, {Glenn}, {Goizel},
  {Griffin}, {Grundy}, {Guest}, {Guillemet}, {Hargrave}, {Harwit}, {Hastings},
  {Hatziminaoglou}, {Herman}, {Hinde}, {Hristov}, {Huang}, {Imhof}, {Isaak},
  {Israelsson}, {Ivison}, {Jennings}, {Kiernan}, {King}, {Lange}, {Latter},
  {Laurent}, {Laurent}, {Leeks}, {Lellouch}, {Levenson}, {Li}, {Li},
  {Lilienthal}, {Lim}, {Liu}, {Lu}, {Madden}, {Mainetti}, {Marliani}, {McKay},
  {Mercier}, {Molinari}, {Morris}, {Moseley}, {Mulder}, {Mur}, {Naylor},
  {Nguyen}, {O'Halloran}, {Oliver}, {Olofsson}, {Olofsson}, {Orfei}, {Page},
  {Pain}, {Panuzzo}, {Papageorgiou}, {Parks}, {Parr-Burman}, {Pearce},
  {Pearson}, {P{\'e}rez-Fournon}, {Pinsard}, {Pisano}, {Podosek}, {Pohlen},
  {Polehampton}, {Pouliquen}, {Rigopoulou}, {Rizzo}, {Roseboom}, {Roussel},
  {Rowan-Robinson}, {Rownd}, {Saraceno}, {Sauvage}, {Savage}, {Savini},
  {Sawyer}, {Scharmberg}, {Schmitt}, {Schneider}, {Schulz}, {Schwartz},
  {Shafer}, {Shupe}, {Sibthorpe}, {Sidher}, {Smith}, {Smith}, {Smith},
  {Spencer}, {Stobie}, {Sudiwala}, {Sukhatme}, {Surace}, {Stevens}, {Swinyard},
  {Trichas}, {Tourette}, {Triou}, {Tseng}, {Tucker}, {Turner}, {Vaccari},
  {Valtchanov}, {Vigroux}, {Virique}, {Voellmer}, {Walker}, {Ward}, {Waskett},
  {Weilert}, {Wesson}, {White}, {Whitehouse}, {Wilson}, {Winter}, {Woodcraft},
  {Wright}, {Xu}, {Zavagno}, {Zemcov}, {Zhang}, \& {Zonca}}]{SPIRE}
{Griffin} M.~J. {et~al.}, 2010, A\&A, 518, L3

\bibitem[{{Herschel}(1847)}]{herschel1847}
{Herschel}, Sir J.~F.~W., 1847, {Results of astronomical observations made
  during the years 1834, 5, 6, 7, 8, at the Cape of Good Hope; being the
  completion of a telescopic survey of the whole surface of the visible
  heavens, commenced in 1825}

\bibitem[{Hildebrand(1983)}]{hildebrand1983}
Hildebrand R.~H., 1983, {Q. Jl R. astr. Soc.}, 24, 267

\bibitem[{{Holland} {et~al}\mbox{.}(2013){Holland}, {Bintley}, {Chapin},
  {Chrysostomou}, {Davis}, {Dempsey}, {Duncan}, {Fich}, {Friberg}, {Halpern},
  {Irwin}, {Jenness}, {Kelly}, {MacIntosh}, {Robson}, {Scott}, {Ade},
  {Atad-Ettedgui}, {Berry}, {Craig}, {Gao}, {Gibb}, {Hilton}, {Hollister},
  {Kycia}, {Lunney}, {McGregor}, {Montgomery}, {Parkes}, {Tilanus}, {Ullom},
  {Walther}, {Walton}, {Woodcraft}, {Amiri}, {Atkinson}, {Burger}, {Chuter},
  {Coulson}, {Doriese}, {Dunare}, {Economou}, {Niemack}, {Parsons},
  {Reintsema}, {Sibthorpe}, {Smail}, {Sudiwala}, \& {Thomas}}]{holland2013}
{Holland} W.~S. {et~al.}, 2013, MNRAS, 430, 2513

\bibitem[{{Johnson}(2013)}]{cccbdb}
{Johnson} R.~D.~I., 2013, {http://cccbdb.nist.gov/}, release 16a

\bibitem[{{Johnstone} {et~al}\mbox{.}(2000){Johnstone}, {Wilson},
  {Moriarty-Schieven}, {Joncas}, {Smith}, {Gregersen}, \&
  {Fich}}]{johnstone2000}
{Johnstone} D., {Wilson} C.~D., {Moriarty-Schieven} G., {Joncas} G., {Smith}
  G., {Gregersen} E., {Fich} M., 2000, ApJ, 545, 327

\bibitem[{{Kirk} {et~al}\mbox{.}(2013){Kirk}, {Ward-Thompson}, {Palmeirim},
  {Andr{\'e}}, {Griffin}, {Hargrave}, {K{\"o}nyves}, {Bernard}, {Nutter},
  {Sibthorpe}, {Di Francesco}, {Abergel}, {Arzoumanian}, {Benedettini},
  {Bontemps}, {Elia}, {Hennemann}, {Hill}, {Men'shchikov}, {Motte},
  {Nguyen-Luong}, {Peretto}, {Pezzuto}, {Rygl}, {Sadavoy}, {Schisano},
  {Schneider}, {Testi}, \& {White}}]{csar}
{Kirk} J.~M. {et~al.}, 2013, {MNRAS}, 432, 1424

\bibitem[{{Klessen} {et~al}\mbox{.}(2005){Klessen}, {Ballesteros-Paredes},
  {V{\'a}zquez-Semadeni}, \& {Dur{\'a}n-Rojas}}]{klessen2005}
{Klessen} R.~S., {Ballesteros-Paredes} J., {V{\'a}zquez-Semadeni} E.,
  {Dur{\'a}n-Rojas} C., 2005, ApJ, 620, 786

\bibitem[{{Koen}(2006)}]{koen2006}
{Koen} C., 2006, MNRAS, 365, 590

\bibitem[{{K{\"o}nyves} {et~al}\mbox{.}(2010){K{\"o}nyves}, {Andr{\'e}},
  {Men'shchikov}, {Schneider}, {Arzoumanian}, {Bontemps}, {Attard}, {Motte},
  {Didelon}, {Maury}, {Abergel}, {Ali}, {Baluteau}, {Bernard}, {Cambr{\'e}sy},
  {Cox}, {Di Francesco}, {di Giorgio}, {Griffin}, {Hargrave}, {Huang}, {Kirk},
  {Li}, {Martin}, {Minier}, {Molinari}, {Olofsson}, {Pezzuto}, {Russeil},
  {Roussel}, {Saraceno}, {Sauvage}, {Sibthorpe}, {Spinoglio}, {Testi},
  {Ward-Thompson}, {White}, {Wilson}, {Woodcraft}, \& {Zavagno}}]{konyves2010}
{K{\"o}nyves} V. {et~al.}, 2010, A\&A, 518, L106

\bibitem[{Kroupa(2001)}]{kroupa}
Kroupa P., 2001, {MNRAS}, 322, 231

\bibitem[{{Leous} {et~al}\mbox{.}(1991){Leous}, {Feigelson}, {Andr\'{e}}, \&
  {Montmerle}}]{leous1991}
{Leous} J.~A., {Feigelson} E.~D., {Andr\'{e}} P., {Montmerle} T., 1991, ApJ,
  379, 683

\bibitem[{{Liseau} {et~al}\mbox{.}(1999){Liseau}, {White}, {Larsson}, {Sidher},
  {Olofsson}, {Kaas}, {Nordh}, {Caux}, {Lorenzetti}, {Molinari}, {Nisini}, \&
  {Sibille}}]{liseau1999}
{Liseau} R. {et~al.}, 1999, A\&A, 344, 342

\bibitem[{Loren(1989)}]{loren1989a}
Loren R.~B., 1989, ApJ, 338, 902

\bibitem[{{Mamajek}(2008)}]{mamajek2008}
{Mamajek} E.~E., 2008, Astronomische Nachrichten, 329, 10

\bibitem[{{Markwardt}(2009)}]{mpfit}
{Markwardt} C.~B., 2009, in Astronomical Society of the Pacific Conference
  Series, Vol. 411, Astronomical Data Analysis Software and Systems XVIII,
  {Bohlender} D.~A., {Durand} D., {Dowler} P., eds., p. 251

\bibitem[{{Maruta} {et~al}\mbox{.}(2010){Maruta}, {Nakamura}, {Nishi}, {Ikeda},
  \& {Kitamura}}]{maruta2010}
{Maruta} H., {Nakamura} F., {Nishi} R., {Ikeda} N., {Kitamura} Y., 2010, ApJ,
  714, 680

\bibitem[{{Maschberger} \& {Kroupa}(2009)}]{maschberger2009}
{Maschberger} T., {Kroupa} P., 2009, MNRAS, 395, 931

\bibitem[{{Mizuno} {et~al}\mbox{.}(1990){Mizuno}, {Fukui}, {Iwata}, {Nozawa},
  \& {Takano}}]{mizuno1990}
{Mizuno} A., {Fukui} Y., {Iwata} T., {Nozawa} S., {Takano} T., 1990, ApJ, 356,
  184

\bibitem[{{Molinari} {et~al}\mbox{.}(2011){Molinari}, {Schisano}, {Faustini},
  {Pestalozzi}, {di Giorgio}, \& {Liu}}]{cutex}
{Molinari} S., {Schisano} E., {Faustini} F., {Pestalozzi} M., {di Giorgio}
  A.~M., {Liu} S., 2011, {A\&A}, 530, A133

\bibitem[{{Motte}, {Andr\'{e}} \& {Neri}(1998){Motte}, {Andr\'{e}}, \&
  {Neri}}]{motte1998}
{Motte} F., {Andr\'{e}} P., {Neri} R., 1998, A\&A, 336, 150

\bibitem[{{Motte} {et~al}\mbox{.}(2001){Motte}, {Andr{\'e}}, {Ward-Thompson},
  \& {Bontemps}}]{motte2001}
{Motte} F., {Andr{\'e}} P., {Ward-Thompson} D., {Bontemps} S., 2001, A\&A, 372,
  L41

\bibitem[{{M{\"u}ller} {et~al}\mbox{.}(2001){M{\"u}ller}, {Thorwirth}, {Roth},
  \& {Winnewisser}}]{cdms}
{M{\"u}ller} H.~S.~P., {Thorwirth} S., {Roth} D.~A., {Winnewisser} G., 2001,
  A\&A, 370, L49

\bibitem[{{Nutter} \& {Ward-Thompson}(2007)}]{nutter2007}
{Nutter} D., {Ward-Thompson} D., 2007, MNRAS, 374, 1413

\bibitem[{{Nutter}, {Ward-Thompson} \& {Andr{\'e}}(2006){Nutter},
  {Ward-Thompson}, \& {Andr{\'e}}}]{nutter2006}
{Nutter} D., {Ward-Thompson} D., {Andr{\'e}} P., 2006, MNRAS, 368, 1833

\bibitem[{{Offner}, {Klein} \& {McKee}(2008){Offner}, {Klein}, \&
  {McKee}}]{offner2008}
{Offner} S.~S.~R., {Klein} R.~I., {McKee} C.~F., 2008, ApJ, 686, 1174

\bibitem[{{Pilbratt} {et~al}\mbox{.}(2010){Pilbratt}, {Riedinger}, {Passvogel},
  {Crone}, {Doyle}, {Gageur}, {Heras}, {Jewell}, {Metcalfe}, {Ott}, \&
  {Schmidt}}]{herschel}
{Pilbratt} G.~L. {et~al.}, 2010, A\&A, 518, L1

\bibitem[{{Pineda}, {Caselli} \& {Goodman}(2008){Pineda}, {Caselli}, \&
  {Goodman}}]{pineda2008}
{Pineda} J.~E., {Caselli} P., {Goodman} A.~A., 2008, ApJ, 679, 481

\bibitem[{{Pineda} {et~al}\mbox{.}(2010){Pineda}, {Goodman}, {Arce}, {Caselli},
  {Foster}, {Myers}, \& {Rosolowsky}}]{pineda2010}
{Pineda} J.~E., {Goodman} A.~A., {Arce} H.~G., {Caselli} P., {Foster} J.~B.,
  {Myers} P.~C., {Rosolowsky} E.~W., 2010, ApJL, 712, L116

\bibitem[{{Pirogov} {et~al}\mbox{.}(2003){Pirogov}, {Zinchenko}, {Caselli},
  {Johansson}, \& {Myers}}]{pirogov2003}
{Pirogov} L., {Zinchenko} I., {Caselli} P., {Johansson} L.~E.~B., {Myers}
  P.~C., 2003, A\&A, 405, 639

\bibitem[{{Planck HFI Core Team} {et~al}\mbox{.}(2011){Planck HFI Core Team},
  {Ade}, {Aghanim}, {Ansari}, {Arnaud}, {Ashdown}, {Aumont}, {Banday},
  {Bartelmann}, {Bartlett}, {Battaner}, {Benabed}, {Beno{\^i}t}, {Bernard},
  {Bersanelli}, {Bock}, {Bond}, {Borrill}, {Bouchet}, {Boulanger}, {Bradshaw},
  {Bucher}, {Cardoso}, {Castex}, {Catalano}, {Challinor}, {Chamballu}, {Chary},
  {Chen}, {Chiang}, {Church}, {Clements}, {Colley}, {Colombi}, {Couchot},
  {Coulais}, {Cressiot}, {Crill}, {Crook}, {de Bernardis}, {Delabrouille},
  {Delouis}, {D{\'e}sert}, {Dolag}, {Dole}, {Dor{\'e}}, {Douspis}, {Dunkley},
  {Efstathiou}, {Filliard}, {Forni}, {Fosalba}, {Ganga}, {Giard}, {Girard},
  {Giraud-H{\'e}raud}, {Gispert}, {G{\'o}rski}, {Gratton}, {Griffin}, {Guyot},
  {Haissinski}, {Harrison}, {Helou}, {Henrot-Versill{\'e}},
  {Hern{\'a}ndez-Monteagudo}, {Hildebrandt}, {Hills}, {Hivon}, {Hobson},
  {Holmes}, {Huffenberger}, {Jaffe}, {Jones}, {Kaplan}, {Kneissl}, {Knox},
  {Kunz}, {Lagache}, {Lamarre}, {Lange}, {Lasenby}, {Lavabre}, {Lawrence}, {Le
  Jeune}, {Leroy}, {Lesgourgues}, {Mac{\'{\i}}as-P{\'e}rez}, {MacTavish},
  {Maffei}, {Mandolesi}, {Mann}, {Marleau}, {Marshall}, {Masi}, {Matsumura},
  {McAuley}, {McGehee}, {Melin}, {Mercier}, {Mitra}, {Miville-Desch{\^e}nes},
  {Moneti}, {Montier}, {Mortlock}, {Murphy}, {Nati}, {Netterfield},
  {N{\o}rgaard-Nielsen}, {North}, {Noviello}, {Novikov}, {Osborne}, {Pajot},
  {Patanchon}, {Peacocke}, {Pearson}, {Perdereau}, {Perotto}, {Piacentini},
  {Piat}, {Plaszczynski}, {Pointecouteau}, {Ponthieu}, {Pr{\'e}zeau}, {Prunet},
  {Puget}, {Reach}, {Remazeilles}, {Renault}, {Riazuelo}, {Ristorcelli},
  {Rocha}, {Rosset}, {Roudier}, {Rowan-Robinson}, {Rusholme}, {Saha}, {Santos},
  {Savini}, {Schaefer}, {Shellard}, {Spencer}, {Starck}, {Stolyarov},
  {Stompor}, {Sudiwala}, {Sunyaev}, {Sutton}, {Sygnet}, {Tauber}, {Thum},
  {Torre}, {Touze}, {Tristram}, {van Leeuwen}, {Vibert}, {Vibert}, {Wade},
  {Wandelt}, {White}, {Wiesemeyer}, {Woodcraft}, {Yurchenko}, {Yvon}, \&
  {Zacchei}}]{planckoffsets}
{Planck HFI Core Team} {et~al.}, 2011, A\&A, 536, A6

\bibitem[{{Poglitsch} {et~al}\mbox{.}(2010){Poglitsch}, {Waelkens}, {Geis},
  {Feuchtgruber}, {Vandenbussche}, {Rodriguez}, {Krause}, {Renotte}, {van
  Hoof}, {Saraceno}, {Cepa}, {Kerschbaum}, {Agn{\`e}se}, {Ali}, {Altieri},
  {Andreani}, {Augueres}, {Balog}, {Barl}, {Bauer}, {Belbachir}, {Benedettini},
  {Billot}, {Boulade}, {Bischof}, {Blommaert}, {Callut}, {Cara}, {Cerulli},
  {Cesarsky}, {Contursi}, {Creten}, {De Meester}, {Doublier}, {Doumayrou},
  {Duband}, {Exter}, {Genzel}, {Gillis}, {Gr{\"o}zinger}, {Henning},
  {Herreros}, {Huygen}, {Inguscio}, {Jakob}, {Jamar}, {Jean}, {de Jong},
  {Katterloher}, {Kiss}, {Klaas}, {Lemke}, {Lutz}, {Madden}, {Marquet},
  {Martignac}, {Mazy}, {Merken}, {Montfort}, {Morbidelli}, {M{\"u}ller},
  {Nielbock}, {Okumura}, {Orfei}, {Ottensamer}, {Pezzuto}, {Popesso},
  {Putzeys}, {Regibo}, {Reveret}, {Royer}, {Sauvage}, {Schreiber}, {Stegmaier},
  {Schmitt}, {Schubert}, {Sturm}, {Thiel}, {Tofani}, {Vavrek}, {Wetzstein},
  {Wieprecht}, \& {Wiezorrek}}]{PACS}
{Poglitsch} A. {et~al.}, 2010, A\&A, 518, L2

\bibitem[{{Pound} \& {Blitz}(1995)}]{pound1995}
{Pound} M.~W., {Blitz} L., 1995, ApJ, 444, 270

\bibitem[{{Quinn}(2013)}]{quinn2013}
{Quinn} C., 2013, PhD thesis, {Cardiff University}

\bibitem[{{Roy} {et~al}\mbox{.}(2014){Roy}, {Andr{\'e}}, {Palmeirim}, {Attard},
  {K{\"o}nyves}, {Schneider}, {Peretto}, {Men'shchikov}, {Ward-Thompson},
  {Kirk}, {Griffin}, {Marsh}, {Abergel}, {Arzoumanian}, {Benedettini}, {Hill},
  {Motte}, {Nguyen Luong}, {Pezzuto}, {Rivera-Ingraham}, {Roussel}, {Rygl},
  {Spinoglio}, {Stamatellos}, \& {White}}]{roy2014}
{Roy} A. {et~al.}, 2014, A\&A, 562, A138

\bibitem[{{Sadavoy} {et~al}\mbox{.}(2010){Sadavoy}, {Di Francesco}, {Bontemps},
  {Megeath}, {Rebull}, {Allgaier}, {Carey}, {Gutermuth}, {Hora}, {Huard},
  {McCabe}, {Muzerolle}, {Noriega-Crespo}, {Padgett}, \&
  {Terebey}}]{sadavoy2010}
{Sadavoy} S.~I. {et~al.}, 2010, ApJ, 710, 1247

\bibitem[{{Sadavoy} {et~al}\mbox{.}(2013){Sadavoy}, {Di Francesco},
  {Johnstone}, {Currie}, {Drabek}, {Hatchell}, {Nutter}, {Andr{\'e}},
  {Arzoumanian}, {Benedettini}, {Bernard}, {Duarte-Cabral}, {Fallscheer},
  {Friesen}, {Greaves}, {Hennemann}, {Hill}, {Jenness}, {K{\"o}nyves},
  {Matthews}, {Mottram}, {Pezzuto}, {Roy}, {Rygl}, {Schneider-Bontemps},
  {Spinoglio}, {Testi}, {Tothill}, {Ward-Thompson}, {White}, {JCMT}, \&
  {Herschel Gould Belt Survey Teams}}]{sadavoy2013}
{Sadavoy} S.~I. {et~al.}, 2013, ApJ, 767, 126

\bibitem[{Salpeter(1955)}]{salpeter}
Salpeter E.~E., 1955, {ApJ}, 121, {161}

\bibitem[{{Simpson} {et~al}\mbox{.}(2011){Simpson}, {Johnstone}, {Nutter},
  {Ward-Thompson}, \& {Whitworth}}]{simpson2011}
{Simpson} R.~J., {Johnstone} D., {Nutter} D., {Ward-Thompson} D., {Whitworth}
  A.~P., 2011, MNRAS, 417, 216

\bibitem[{{Simpson}, {Nutter} \& {Ward-Thompson}(2008){Simpson}, {Nutter}, \&
  {Ward-Thompson}}]{simpson2008}
{Simpson} R.~J., {Nutter} D., {Ward-Thompson} D., 2008, MNRAS, 391, 205

\bibitem[{{Stamatellos}, {Whitworth} \& {Ward-Thompson}(2007){Stamatellos},
  {Whitworth}, \& {Ward-Thompson}}]{stamatellos2007}
{Stamatellos} D., {Whitworth} A.~P., {Ward-Thompson} D., 2007, MNRAS, 379, 1390

\bibitem[{{Strom} \& {Peterson}(1968)}]{strom1968}
{Strom} S.~E., {Peterson} D.~M., 1968, ApJ, 152, 859

\bibitem[{{Swinyard} {et~al}\mbox{.}(2010){Swinyard}, {Ade}, {Baluteau},
  {Aussel}, {Barlow}, {Bendo}, {Benielli}, {Bock}, {Brisbin}, {Conley},
  {Conversi}, {Dowell}, {Dowell}, {Ferlet}, {Fulton}, {Glenn}, {Glauser},
  {Griffin}, {Griffin}, {Guest}, {Imhof}, {Isaak}, {Jones}, {King}, {Leeks},
  {Levenson}, {Lim}, {Lu}, {Makiwa}, {Naylor}, {Nguyen}, {Oliver}, {Panuzzo},
  {Papageorgiou}, {Pearson}, {Pohlen}, {Polehampton}, {Pouliquen},
  {Rigopoulou}, {Ronayette}, {Roussel}, {Rykala}, {Savini}, {Schulz},
  {Schwartz}, {Shupe}, {Sibthorpe}, {Sidher}, {Smith}, {Spencer}, {Trichas},
  {Triou}, {Valtchanov}, {Wesson}, {Woodcraft}, {Xu}, {Zemcov}, \&
  {Zhang}}]{swinyard2010}
{Swinyard} B.~M. {et~al.}, 2010, A\&A, 518, L4

\bibitem[{{Troland} {et~al}\mbox{.}(1996){Troland}, {Crutcher}, {Goodman},
  {Heiles}, {Kazes}, \& {Myers}}]{troland1996}
{Troland} T.~H., {Crutcher} R.~M., {Goodman} A.~A., {Heiles} C., {Kazes} I.,
  {Myers} P.~C., 1996, ApJ, 471, 302

\bibitem[{{Vrba}(1977)}]{vrba1977}
{Vrba} F.~J., 1977, AJ, 82, 198

\bibitem[{{Vrba} {et~al}\mbox{.}(1975){Vrba}, {Strom}, {Strom}, \&
  {Grasdalen}}]{vrba1975}
{Vrba} F.~J., {Strom} K.~M., {Strom} S.~E., {Grasdalen} G.~L., 1975, ApJ, 197,
  77

\bibitem[{Ward-Thompson {et~al}\mbox{.}(2007)Ward-Thompson, Andr\'{e},
  Crutcher, Johnstone, Onishi, \& Wilson}]{PPVch3}
Ward-Thompson D., Andr\'{e} P., Crutcher R., Johnstone D., Onishi T., Wilson
  C., 2007, in {Protostars and Planets V}, {Reipurth} B., {Jewitt} D., {Tucson}
  K.~K., eds., {University of Arizona Press}, pp. {33--46}

\bibitem[{{Ward-Thompson}, {Andr{\'e}} \& {Kirk}(2002){Ward-Thompson},
  {Andr{\'e}}, \& {Kirk}}]{wardthompson2002a}
{Ward-Thompson} D., {Andr{\'e}} P., {Kirk} J.~M., 2002, MNRAS, 329, 257

\bibitem[{{Ward-Thompson} {et~al}\mbox{.}(2007){Ward-Thompson}, {Di Francesco},
  {Hatchell}, {Hogerheijde}, {Nutter}, {Bastien}, {Basu}, {Bonnell}, {Bowey},
  {Brunt}, {Buckle}, {Butner}, {Cavanagh}, {Chrysostomou}, {Curtis}, {Davis},
  {Dent}, {van Dishoeck}, {Edmunds}, {Fich}, {Fiege}, {Fissel}, {Friberg},
  {Friesen}, {Frieswijk}, {Fuller}, {Gosling}, {Graves}, {Greaves}, {Helmich},
  {Hills}, {Holland}, {Houde}, {Jayawardhana}, {Johnstone}, {Joncas}, {Kirk},
  {Kirk}, {Knee}, {Matthews}, {Matthews}, {Matzner}, {Moriarty-Schieven},
  {Naylor}, {Padman}, {Plume}, {Rawlings}, {Redman}, {Reid}, {Richer},
  {Shipman}, {Simpson}, {Spaans}, {Stamatellos}, {Tsamis}, {Viti}, {Weferling},
  {White}, {Whitworth}, {Wouterloot}, {Yates}, \& {Zhu}}]{scuba2survey}
{Ward-Thompson} D. {et~al.}, 2007, {PASP}, 119, 855

\bibitem[{{Ward-Thompson} {et~al}\mbox{.}(2006){Ward-Thompson}, {Nutter},
  {Bontemps}, {Whitworth}, \& {Attwood}}]{wardthompson2006}
{Ward-Thompson} D., {Nutter} D., {Bontemps} S., {Whitworth} A., {Attwood} R.,
  2006, MNRAS, 369, 1201

\bibitem[{Ward-Thompson {et~al}\mbox{.}(1994)Ward-Thompson, Scott, Hills, \&
  Andr\'{e}}]{wardthompson1994}
Ward-Thompson D., Scott P.~F., Hills R.~E., Andr\'{e} P., 1994, {MNRAS}, 268,
  {276}

\bibitem[{{Wenger} {et~al}\mbox{.}(2000){Wenger}, {Ochsenbein}, {Egret},
  {Dubois}, {Bonnarel}, {Borde}, {Genova}, {Jasniewicz}, {Lalo{\"e}},
  {Lesteven}, \& {Monier}}]{simbad}
{Wenger} M. {et~al.}, 2000, A\&AS, 143, 9

\bibitem[{{White} {et~al}\mbox{.}(2015){White}, {Drabek-Maunder}, {Rosolowsky},
  {Ward-Thompson}, {Davis}, {Gregson}, {Hatchell}, {Etxaluze}, {Stickler},
  {Buckle}, {Johnstone}, {Friesen}, {Sadavoy}, {Natt}, {Currie}, {Richer},
  {Pattle}, {Spaans}, {Francesco}, \& {Hogerheijde}}]{white2015}
{White} G.~J. {et~al.}, 2015, MNRAS, 447, 1996

\bibitem[{{Wilking}, {Gagn{\'e}} \& {Allen}(2008){Wilking}, {Gagn{\'e}}, \&
  {Allen}}]{wilking2008}
{Wilking} B.~A., {Gagn{\'e}} M., {Allen} L.~E., 2008, {Star Formation in the
  {$\rho$} Ophiuchi Molecular Cloud}, {Reipurth} B., ed., p. 351

\bibitem[{{Wilking} \& {Lada}(1983)}]{wilking1983}
{Wilking} B.~A., {Lada} C.~J., 1983, ApJ, 274, 698

\bibitem[{{Wilson}(1999)}]{wilson1999}
{Wilson} T.~L., 1999, Reports on Progress in Physics, 62, 143

\bibitem[{{Young}, {Lada} \& {Wilking}(1986){Young}, {Lada}, \&
  {Wilking}}]{young1986}
{Young} E.~T., {Lada} C.~J., {Wilking} B.~A., 1986, ApJL, 304, L45

\end{thebibliography}

\setcounter{figure}{0}
\makeatletter 
\renewcommand{\thefigure}{A\@arabic\c@figure}
\makeatother

\section*{Appendix A: Data}

We show in full the regions within Ophiuchus observed using SCUBA-2. Figure A1 shows the 850-\um\ flux density data.  Figure A2 shows the 850-\um\ variance map.  Figure A3 shows the 450-\um\ flux density data.  Figure A4 shows the 450-\um\ variance map.  Figures A1--A4 all show the data in square-root scaling. Figure A5 shows the mask used to define areas of significant emission in both the 850-\um\ and 450-\um\ data.

\begin{figure*}
\centering
\includegraphics[height=0.9\textheight]{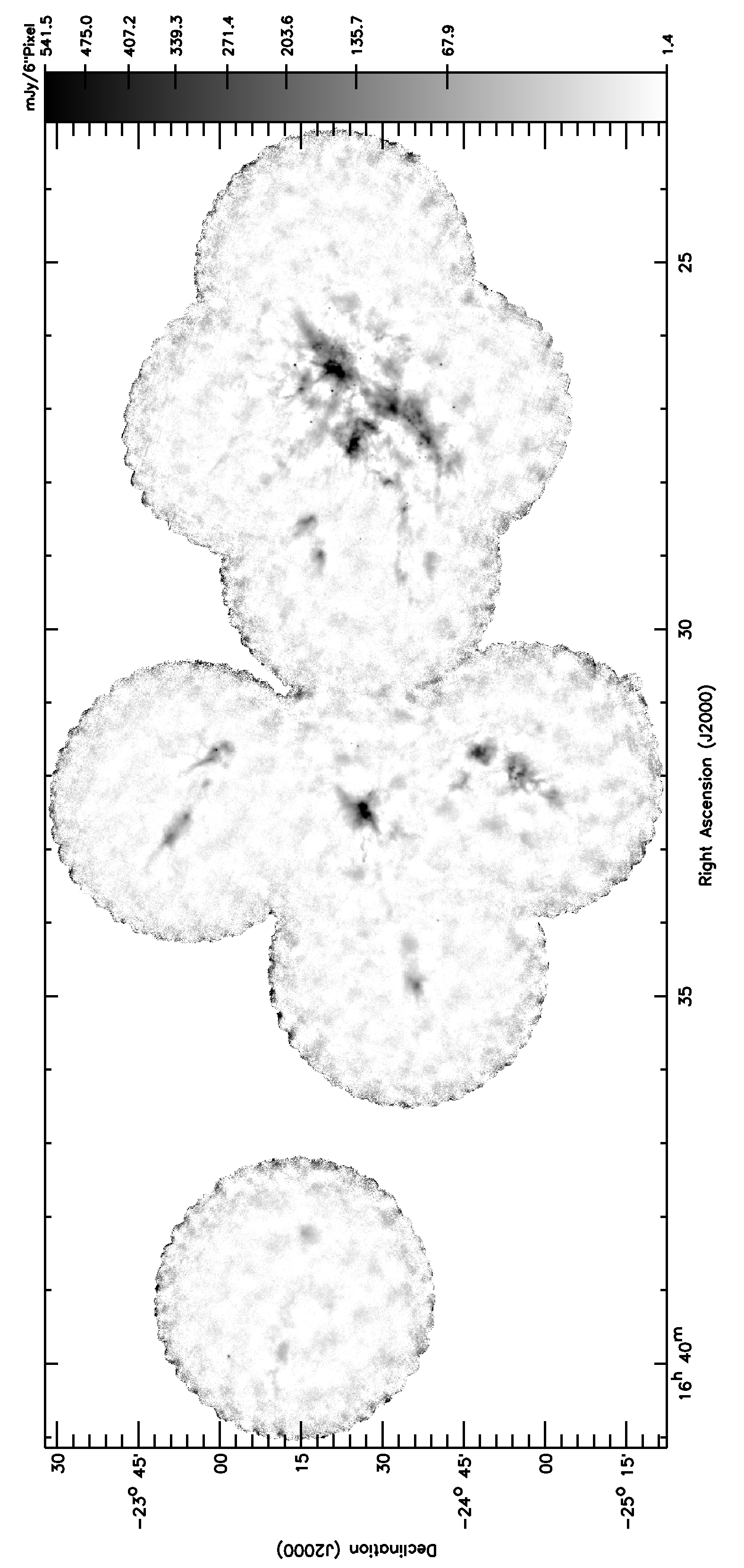}
\caption{The 850-\um\ flux density measured in Ophiuchus with SCUBA-2, shown in square root scaling.}
\label{fig:850appendix}
\end{figure*}

\begin{figure*}
\centering
\includegraphics[height=0.9\textheight]{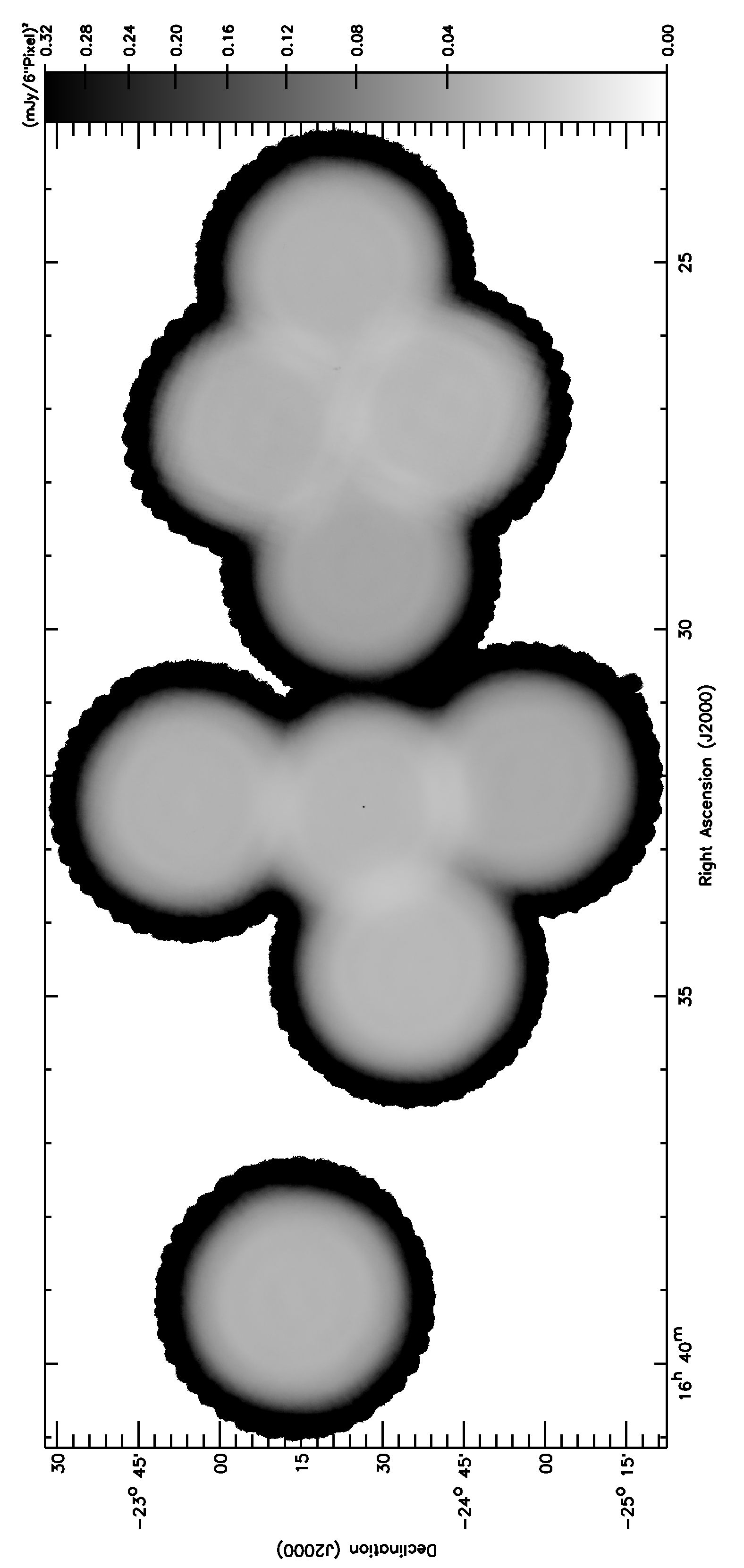}
\caption{The variance on the SCUBA-2 850-\um\ flux density data, shown in square root scaling.}
\label{fig:850varappendix}
\end{figure*}

\begin{figure*}
\centering
\includegraphics[height=0.9\textheight]{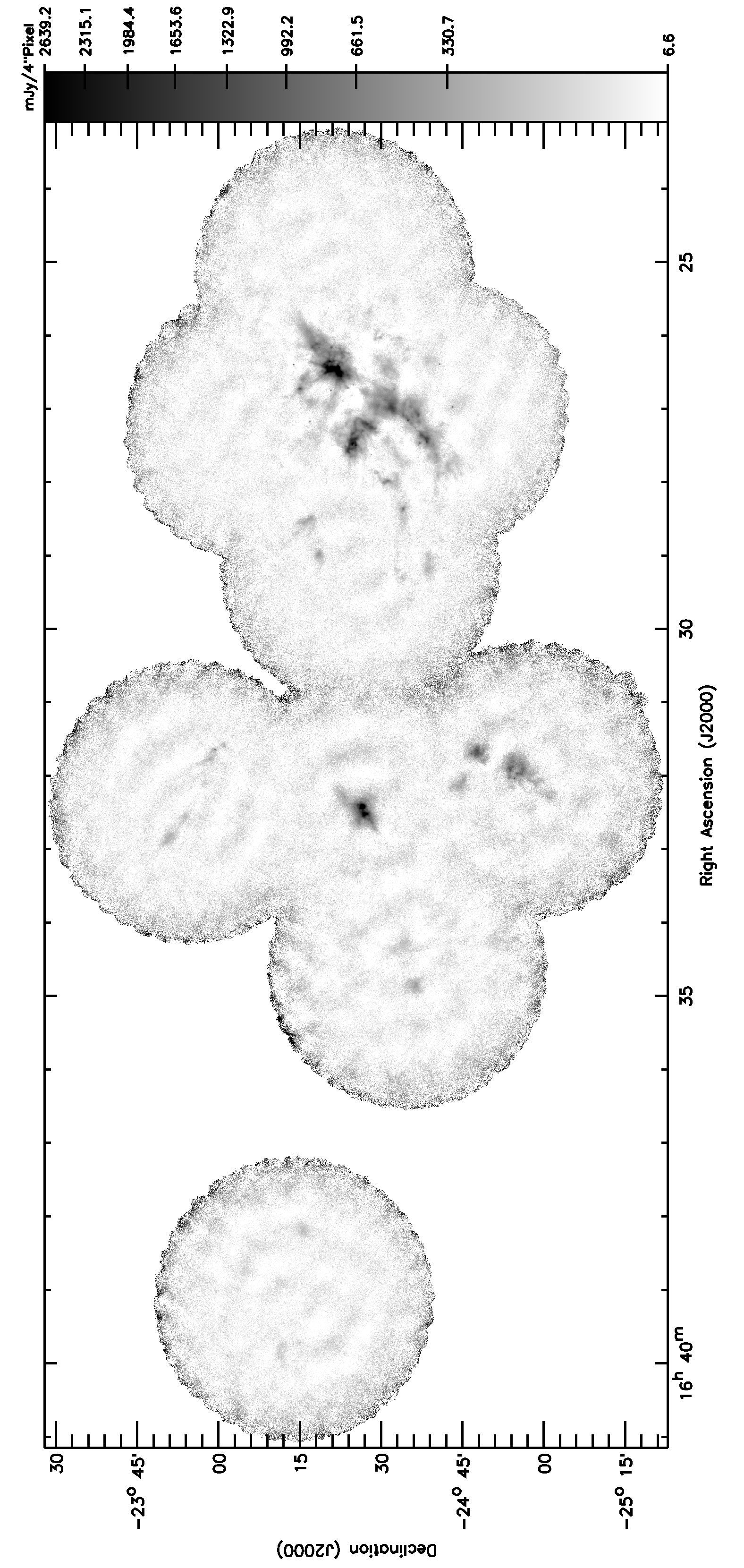}
\caption{The 450-\um\ flux density measured in Ophiuchus with SCUBA-2, shown in square root scaling.}
\label{fig:450appendix}
\end{figure*}

\begin{figure*}
\centering
\includegraphics[height=0.9\textheight]{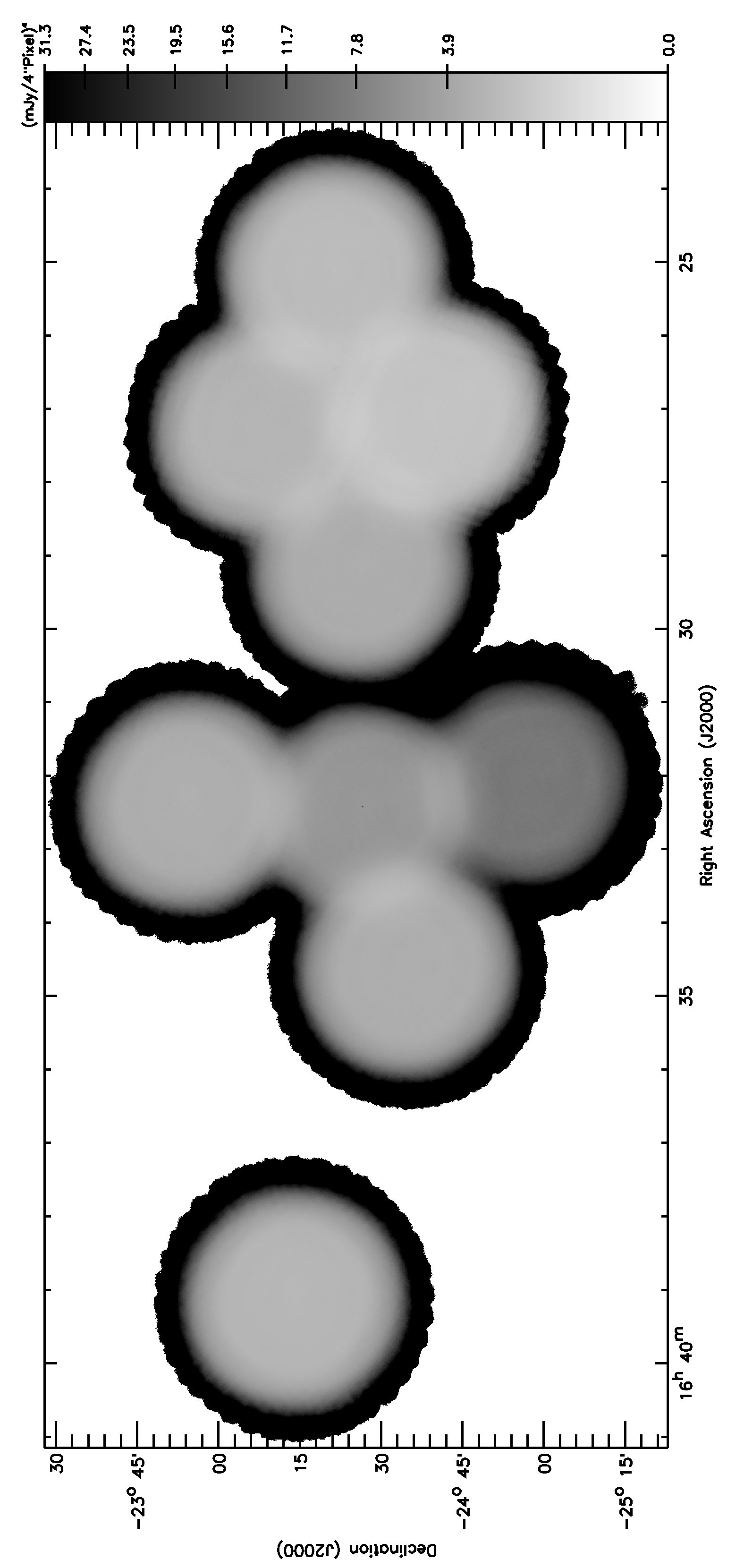}
\caption{The variance on the SCUBA-2 450-\um\ flux density data, shown in square root scaling.}
\label{fig:450varappendix}
\end{figure*}

\begin{figure*}
\centering
\includegraphics[height=0.9\textheight]{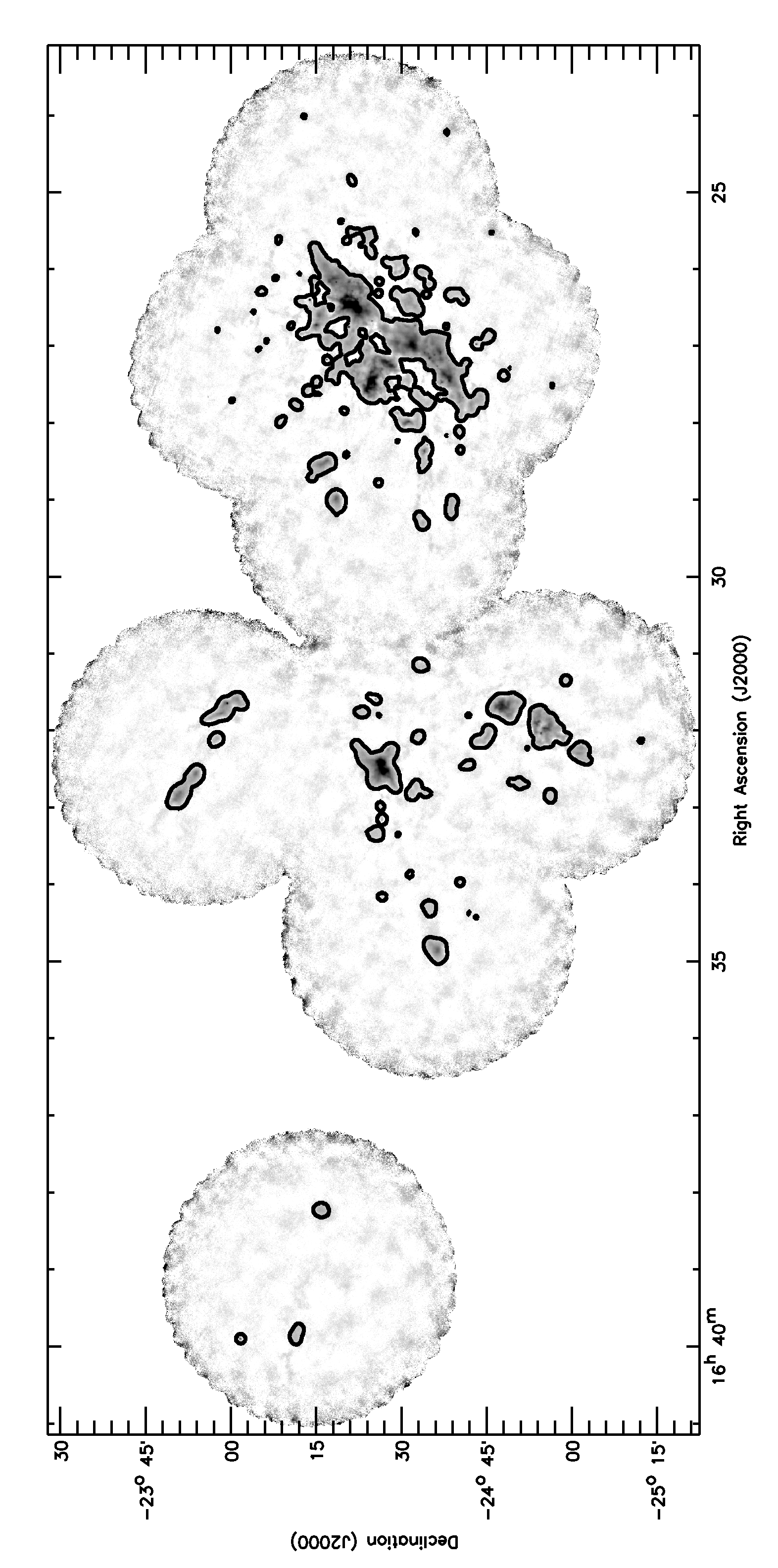}
\caption{The mask used in the data reduction process, enclosing regions of significant emission, shown as a thick contour, overlaying the 850-\um\ flux density data, shown as a greyscale.}
\label{fig:maskappendix}
\end{figure*}

\section*{Appendix B: RGB Images of L1688}

\setcounter{figure}{0}
\makeatletter 
\renewcommand{\thefigure}{B\@arabic\c@figure}
\makeatother

\vspace{5mm}

\noindent\begin{minipage}{\textwidth}
  \centering
  \includegraphics[width=0.57\textwidth]{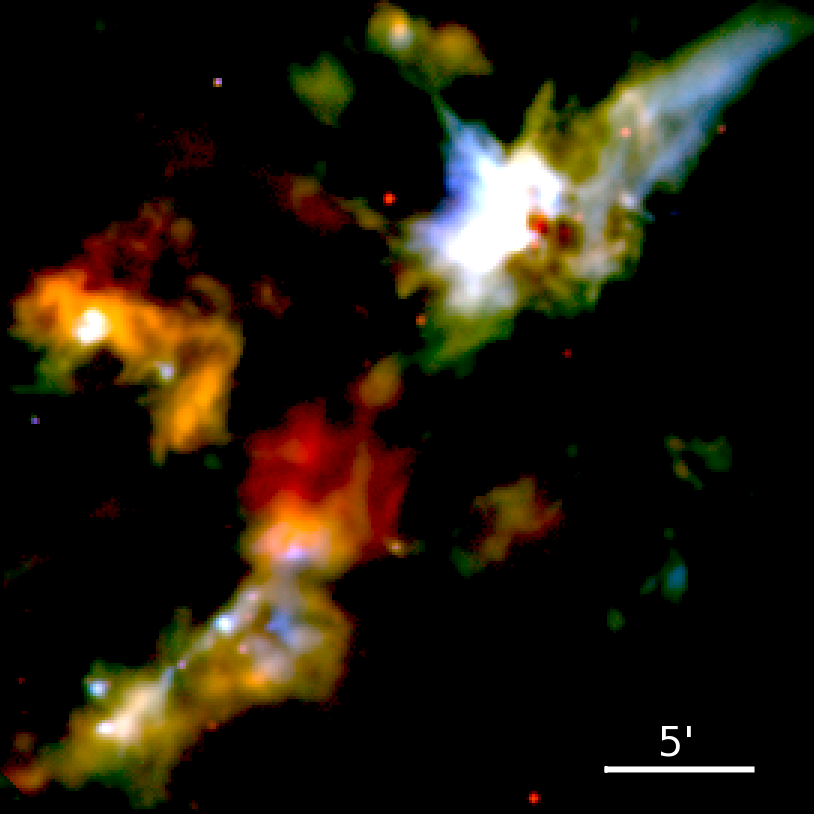}
  \captionof{figure}{Three-colour image of L1688.  Red channel: SCUBA-2 850-\um\ data.  Green channel: spatially filtered \emph{Herschel} 250-\um\ data.  Blue channel: spatially filtered \emph{Herschel} 160-\um\ data.}
  \label{fig:rgb}

\vspace{5mm}

  \includegraphics[width=0.57\textwidth]{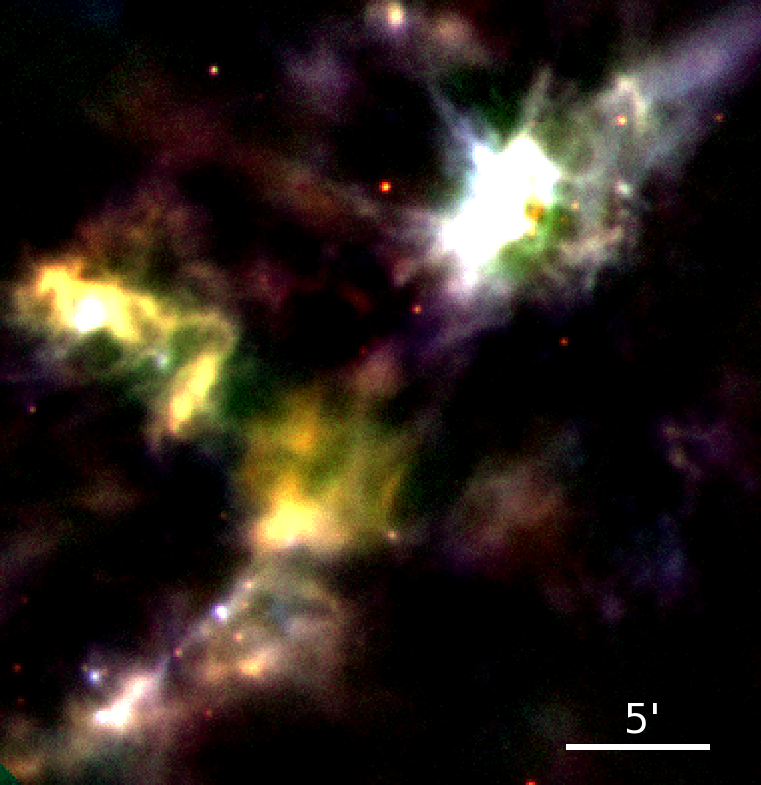}
 \captionof{figure}{Three-colour image of L1688.  Red channel: SCUBA-2 850-\um\ data.  Green channel: SCUBA-2 450-\um\ data. Blue channel: spatially filtered \emph{Herschel} 250-\um\ data.} 
  \label{fig:rgb2}

\end{minipage}

\clearpage

\section*{Appendix C: Gravitational Potential Energy of a Gaussian Distribution}

\setcounter{equation}{0}
\makeatletter 
\renewcommand{\theequation}{C\@arabic\c@equation}
\makeatother
We give here a brief derivation of the gravitational potential energy of a Gaussian distribution, as used in our virial analysis.  This is the first time that this has been shown.

For a radially symmetric potential, the gravitational potential energy $\Omega_{\textsc{g}}$ is given by
\begin{equation}
\Omega_{\textsc{g}}=-4\pi G\int_{0}^{\infty}{\rm d}r\,r\,\rho(r)M(r), 
\end{equation}
where $\rho(r)$ and $M(r)$ are the density and mass at radius $r$, respectively.  $M(r)$ is given by
\begin{equation}
M(r)=4\pi\int_{0}^{r}{\rm d}r^{\prime}\,r^{\prime2}\rho(r^{\prime}). 
\end{equation}
We assume a radially symmetric Gaussian density distribution
\begin{equation}
\rho(r)=\rho_{0}e^{-\nicefrac{r^{2}}{2\alpha^{2}}}, 
\end{equation}
for which the total mass at radius $r$ is given by
\begin{align}
M(r) & = 4\pi\rho_{0}\int_{0}^{r}{\rm d}r^{\prime}\,r^{\prime2}\,e^{-r^{\prime2}/{2\alpha^{2}}} \\ 
 & = 4\pi\rho_{0}\left[\alpha^{3}\sqrt{\frac{\pi}{2}}\,{\rm erf}\left(\frac{r}{\alpha\sqrt{2}}\right)-\alpha^{2}re^{\nicefrac{-r^{2}}{2\alpha^{2}}}\right], 
\end{align}
and the total mass summed over all radii is given by
\begin{align}
M & = 4\pi\rho_{0}\int_{0}^{\infty}{\rm d}r^{\prime}\,r^{\prime2}\,e^{-r^{\prime2}/{2\alpha^{2}}} \\ 
 & = 2\sqrt{2}\pi^{\nicefrac{3}{2}}\rho_{0}\alpha^{3}. 
\end{align}
Using Equations~B1 and B5, $\Omega_{\textsc{g}}$ is given by
\begin{align}
\Omega_{\textsc{g}} & = -16\pi^{2}G\rho_{0}^{2}\alpha^{2}\times \notag \\
 &  \quad\int_{0}^{\infty}{\rm d}r\,r\,e^{\nicefrac{-r^{2}}{2\alpha^{2}}}\left[\alpha\sqrt{\frac{\pi}{2}}\,{\rm erf}\left(\frac{r}{\alpha\sqrt{2}}\right)-re^{\nicefrac{-r^{2}}{2\alpha^{2}}}\right] \\ 
 & = -16\pi^{2}G\rho_{0}^{2}\alpha^{2}\times\left(\frac{\alpha^{3}\sqrt{\pi}}{4}\right) \\ 
 & = -4\pi^{\nicefrac{5}{2}}G\rho_{0}^{2}\alpha^{5}. 
\end{align}
Combining Equations B7 and B10, the gravitational potential energy of a Gaussian distribution of characteristic width $\alpha$ and total mass $M$ is
\begin{equation}
\Omega_{\textsc{g}} = -\frac{1}{2\sqrt{\pi}}\frac{GM^{2}}{\alpha}.
\end{equation}
This is used in Equation~\ref{eq:gpe} in Section~5.1 in the text.

\section*{Appendix D: Oph B-11}

\setcounter{figure}{0}
\makeatletter 
\renewcommand{\thefigure}{D\@arabic\c@figure}
\makeatother

We investigated whether the pre-brown dwarf source Oph B-11 was detectable in our 850-\um\ map of Ophiuchus.  Originally detected and identified as a pre-brown dwarf candidate in a DCO$^{+}$ search \citep{pound1995}, Oph B-11 was observed using SCUBA by \citet{greaves2003}, who classed the source as a very young `isolated planet'.  \citet{andre2012} observed Oph B-11 using the IRAM Plateau de Bure Interferometer (PdBI), determining that the source was in fact a gravitationally bound pre-brown dwarf, with mass 0.02-0.03\,M$_{\odot}$.

When observed by Greaves et al. (2003), the integration time for the 2.7\arcmin\ diameter field was 2 hours, resulting in a very sensitive SCUBA map with 1$\sigma$ RMS noise of 6\,mJy/15\arcsec\ beam.  The 1$\sigma$ RMS noise in our SCUBA-2 map of the same region is 6.3\,mJy/15\arcsec\ beam, almost identical.  This was achieved using 4$\times$PONG1800 observations, taking a total of 2 hours 40 minutes (i.e. essentially the same integration time as with SCUBA) to cover a field of 30\arcmin\ diameter, compared to 2.7\arcmin\ with SCUBA (i.e. roughly 120 times the area in the same time).

\begin{figure}
\centering
\includegraphics[width=0.47\textwidth]{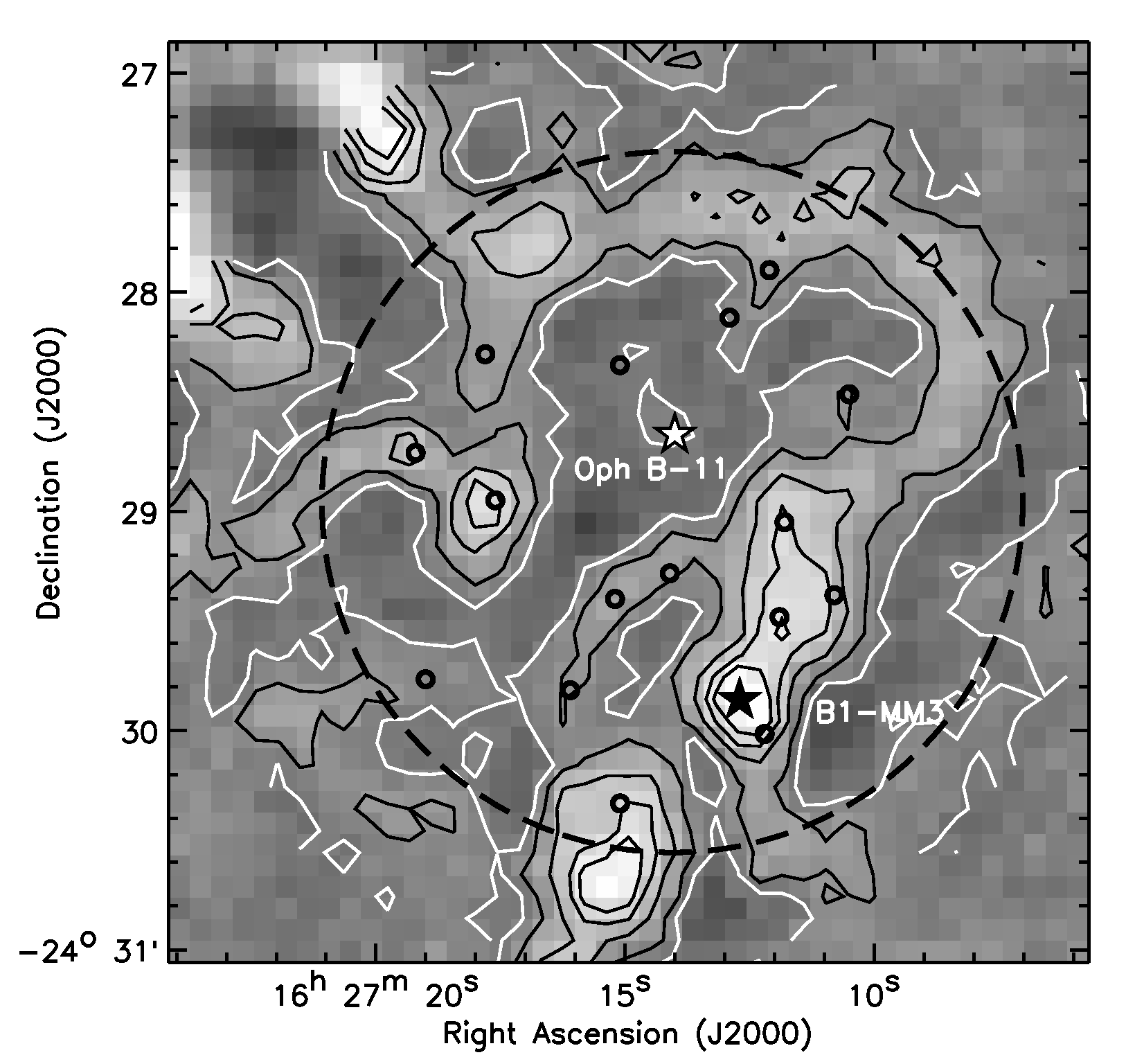}
\caption{Unsharp-masked SCUBA-2 850-\um\ image of the region surrounding Oph B-11, smoothed to 15\arcsec\ resolution.  Our peak position for Oph B-11 is marked as a white star.  B1-MM3 is marked as a black star.  Other sources identified by Greaves et al. (2003) are marked as black circles.  The approximate area observed by Greaves et al. (2003) is enclosed by the dashed line.  Contour levels are 28, 66, 94, 115 and 129 mJy/15\arcsec\ beam above the local minimum, to approximately match the contours of Greaves et al. (2003).}
\label{fig:oph_b11}
\end{figure}

In order to detect this extremely faint source, we repeated the unsharp masking process used by Greaves et al. (2003) on their SCUBA map of the region.  We smoothed the SCUBA-2 map with a 30\arcsec\ Gaussian filter, and subtracted the smoothed emission from the original map, removing all structure significantly more extended than the 14.1\arcsec\ beam.  Data were them smoothed to a 15\arcsec\ beam to match the SCUBA data of Greaves et al. (2003).

After removing the extended structure in this way from the SCUBA-2 map, we were able to detect Oph B-11.  The emission peaks at R.A. $=$ 16$^{\rm h}$:27$^{\rm m}$:$14^{\rm s}.0$, Dec. $=$ $-24^{\circ}$:28\arcmin:39\arcsec.  Greaves et al. (2003) found the source position to be R.A. $=$ 16$^{\rm h}$:27$^{\rm m}$:14$^{\rm s}$.0 Dec. $=$ $-24^{\circ}$:28\arcmin:31\arcsec, while Andr\'{e} et al. (2012) give the source position as R.A. $=$ 16$^{\rm h}$:27$^{\rm m}$:13$^{\rm s}$.96 Dec. $=$ $-24^{\circ}$:28\arcmin:29.3\arcsec.  All of these positions are consistent within the quoted errors.

We measure a peak flux density above the local background for Oph B-11 of $55\pm6$\,mJy/15\arcsec\ beam with an uncertainty on the local background of $\pm 9$\,mJy/15\arcsec\ beam.  Greaves et al. (2003) find a peak 850-\um\ flux density for Oph B-11 of $39\pm6\,$mJy/15\arcsec\ beam, with an uncertainty on their local background of $\pm 5$\,mJy/15\arcsec\ beam.  Thus, our measurement of the peak flux density of Oph B-11 is consistent with that of Greaves et al. (2003).  We converted our peak flux density to a mass using the Greaves et al. (2003) temperature estimate of $12-20$\,K, taking $\kappa_{850\upmu{\rm m}}=0.01$\,cm$^{2}$g$^{-1}$, and assuming a distance of 138\,pc.  We find a mass range for Oph B-11 of $0.011-0.024\,$M$_{\odot}$.  Thus, our data are consistent with the IRAM mass estimate (Andr\'{e} et al. 2012), and hence with the pre-brown dwarf interpretation of Oph B-11.

\label{lastpage}

\end{document}